\documentclass[12pt]{article}

\AtBeginDocument{%
   \setlength\abovedisplayskip{0.05in}
   \setlength\belowdisplayskip{0.08in}}

\usepackage[onehalfspacing]{setspace}

\makeatletter
\renewcommand\@footnotetext[1]{%
  \insert\footins{%
    \reset@font\footnotesize
    \interlinepenalty\interfootnotelinepenalty
    \splittopskip\footnotesep
    \splitmaxdepth \dp\strutbox \floatingpenalty \@MM
    \hsize\columnwidth \@parboxrestore
    {\setstretch{1.0}\protect\@makefntext{%
      \rule{\z@}{\footnotesep}\ignorespaces#1}}}}
\makeatother

\usepackage[top=50mm, bottom=-500mm, left=50mm, right=50mm]{geometry}
\usepackage{lineno}
\usepackage{amssymb}
\usepackage{amsthm}
\usepackage{amsmath}
\usepackage{cases}
\usepackage{caption}
\usepackage{epsfig}
\usepackage{graphicx}
\usepackage{graphics}
\usepackage{float}
\usepackage{subfigure}
\usepackage{multirow}
\usepackage{xcolor}
\usepackage{lineno}
\usepackage{fullpage}
\usepackage[normalem]{ulem} 
\usepackage{makeidx}
\usepackage{xspace}
\usepackage{wrapfig}
\usepackage{amsmath}
\usepackage{lscape}
\usepackage{mathrsfs}
\usepackage{adjustbox}
\usepackage{multirow}
\usepackage{varwidth}
\usepackage{rotating}
\usepackage{lscape}
\usepackage[bottom]{footmisc}
\usepackage[colorlinks,citecolor=darkblue,urlcolor=darkblue,linkcolor=darkblue]{hyperref} 
\usepackage{cleveref}
\makeindex
\newtheorem{theorem}{Theorem}
\newtheorem{assumption}{Assumption}
\newtheorem{corollary}{Corollary}

\newtheorem{lemma}{Lemma}
\newtheorem{ex}{Example}

\newtheorem{proposition}{Proposition}
\newtheorem{definition}{Definition}

\definecolor{purple}{rgb}{0.6, 0.4, 0.8}
\definecolor{darkred}{rgb}{1, 0.1, 0.3}
\definecolor{darkblue}{rgb}{0.0, 0.0, 0.55}
\definecolor{darkgreen}{rgb}{0,0.6,0.5}
\definecolor{forestgreen}{rgb}{0.0, 0.46, 0.37}
\definecolor{bittersweet}{rgb}{1.0, 0.44, 0.37}
\definecolor{navy}{rgb}{0.0, 0.0, 0.55}
\definecolor{brown}{rgb}{0.53, 0.18, 0.09}
\definecolor{Green}{rgb}{0.0, 0.47, 0.44}

\newcommand {\mm}[1] {\ifmmode{#1}\else{\mbox{\(#1\)}}\fi}

\newcommand\E{\mathbb{E}}
\newcommand\Proba{\mathbb{P}}
\newcommand\R{\mathbb{R}}

\usepackage{dsfont}
\newcommand{\m}[1]{\ensuremath{\boldsymbol {#1}}}
\newcommand{\mb}[1]{\textbf{#1}}

\usepackage{sectsty}
\sectionfont{\centering\Large}
\subsectionfont{\large}
\usepackage{fancyhdr}
\usepackage{natbib}

\newtheorem{remark}{Remark}

\usepackage{pst-node, pst-plot, pst-poly}
\usepackage{rotating}

\usepackage{tikz}
\usepackage{venndiagram}
\usepackage{pstricks}
\usepackage{tikz-cd}
\usepackage{siunitx}
\usepackage{pgfplots}
\usepgfplotslibrary{fillbetween}
\usepackage{algorithm}
\usepackage{algorithmic}
\usepackage{bbm}
\usepackage{csquotes}

\usepackage{scalerel}
\usepackage[T1]{fontenc}
\usepackage{titling} 
\usetikzlibrary{shapes,arrows,trees,calc}
\pgfplotsset{compat=1.12}

\allowdisplaybreaks 
\begin{document}

\setlength{\droptitle}{-0.85in}   
\title{\small\bf \mbox{\MakeUppercase{Collective Intelligence in Dynamic Networks}}\thanks{I am very thankful to Drew Fudenberg and Stephen Morris for their guidance and mentorship. I am also grateful to Abhijit Banerjee, Arun Chandrasekhar, Krishna Dasaratha, Ben Golub, Ani Gosh, Matthew Jackson, Ali Jadbabaie, Philippe Jehiel,  M. Ali Khan, Evan Sadler, Alireza Tahbaz-Salehi, Omer Tamuz, and Alex Wolitzky for all their insightful comments and discussions.  \textit{JEL codes}: D83, D85.}
\vspace{-0.2in}
}

\author{
\textsc{Florian Mudekereza}\thanks{Department of Economics, MIT, \href{mailto:florianm@mit.edu}{\texttt{\footnotesize florianm@mit.edu}}.}
}

\date{}
\maketitle
\thispagestyle{empty}
\setcounter{page}{0}
\vspace{-0.86in}
\begin{abstract}
We revisit DeGroot learning 
to examine the robustness of social learning in \textit{dynamic} networks---networks that evolve \textit{randomly} over time. Dynamics have double-edged effects depending on social structure: while they can foster consensus and \textit{boost} collective intelligence in ``sparse'' networks, they can have adverse effects such as slowing down the speed of learning and causing long-run \textit{disagreement} in ``well-connected'' networks.  
Collective intelligence arises in dynamic networks when average influence and trust remain \textit{balanced} as society grows. We also find that the initial social structure of a dynamic network plays a central role in shaping long-run beliefs. We then propose a robust measure of \textit{homophily} based on  the likelihood of the worst network fragmentation.
\end{abstract}
   {\noindent\small\textit{The question we have to ask in thinking about collective wisdom, then, is: Can people make collectively intelligent decisions even when they are in constant, if erratic, interaction with each other?}
    \par\hfill---\citet[][pp. 42--43]{wis05}}
\vspace{-0.1in}

\section{Introduction}

There is substantial evidence that social networks play a central role in people’s lives. They shape people's aspirations, boost economic mobility, facilitate job acquisitions, and foster \textit{collective intelligence}  \citep{moj19}. The latter, known as the ``wisdom of crowds'' \citep{wis05}, suggests that a large decentralized society of heterogeneous agents can learn the true state of the world through mutual communication. 
This phenomenon dates back to Aristotle, and perhaps since \citet{galton907}, it has served as the main justification for collective schemes such as crowdsourcing, prediction markets, opinion polls, and jury/voting systems. There is also growing interest among scientists to design collectively intelligent networks composed of humans and computer systems (e.g., artificial intelligence)  \citep{ai15}.
\par  \citet{moj10} find that a large society is collectively intelligent when no agent has disproportionate \textit{influence}. This result builds on the pioneering work of \citet{degroot74} and relies crucially on networks being exogenously fixed over time, which may be plausible in some cases. In most other cases, however, intelligent systems typically rely on environmental feedback and their ability to \textit{dynamically} reorganize, which result in random evolutionary dynamics \citep[][]{dynam20,moj23}. Hence, \citet{moj23} highlight ``the importance of tracking networks over time.'' Consider, for instance, the social network consisting of the low-income families that participated in the Moving to Opportunity (MTO) experiment in the mid-1990s \citep{mto01}.
Some of these families were randomly offered housing vouchers to move to lower-poverty neighborhoods. This 
random treatment assignment played the role of a random matching process because families that relocated had to create new social contacts and friendships at their destinations. In fact, \citet[][p. 613]{mto01} acknowledge that ``relocations could [...] disrupt social support networks that are important sources of informal child care, job referrals, and other labor market information.''   
Our introductory quote can therefore be rephrased as follows: what are the long/medium-run effects of random network dynamics on social learning outcomes?
\par \citet{bernd19} offer a very negative answer to this question: ``Even if the social network does not privilege any agent in terms of influence, a large society almost always fails to converge to the truth'' when there exist random network dynamics. In contrast, we offer a more complete and nuanced answer by highlighting both positive and negative effects of random dynamics on social learning outcomes. Our main contributions are outlined below. 
\par In Section \ref{sec:wisdom}, we show that a large dynamic network is collectively intelligent if and only if the average influence of the most influential agent vanishes as society grows. This result relies on the same assumptions as in \citet{moj10}: initial signals are independent and unbiased, but it is robust to signal correlations and systematic biases. We find that achieving collective intelligence in dynamic networks requires striking a \textit{balance} between the average influence agents exert on others and the average level of trust they place in others as society grows. Notably, introducing even very small amounts of random dynamics is often enough to promote consensus and \textit{boost} collective intelligence. 
This result is consistent with \citeauthor{dynam20}'s (\citeyear{dynam20}) recent experimental evidence and simulations. 
\par Our technical contribution is a novel framework that enables the analysis of social learning in dynamic networks. In Section \ref{sec:model}, we propose a general dynamical system that plays the role of a ``data generating process'' of the dynamic network.  
The key assumption on the dynamics is \textit{stationarity}---the joint distribution of the networks does not change over time. 
This allows us to build directly on random matrix theory: we adapt \citeauthor{hen97}'s (\citeyear{hen97}) results about convergence of products of nonnegative random matrices, which allows us to nest most existing frameworks \citep[][]{dem03,iff08,iff09,moj10,banerjee21}. Then, Section \ref{sec:consensus} proposes a simple condition that establishes existence of consensus in dynamic networks, which is noteworthy because: (1) it is a natural extension of the well-known ``primitive'' condition used in deterministic networks. (2) It is very easy to verify in many settings because it can be characterized in terms of social structure.  
More generally, this paper is the first to unify social learning theory and random matrix theory in a tractable framework. By leveraging stationarity, our framework serves as a natural intermediate model between the deterministic and independent-and-identically-distributed (iid) network models, thus enabling us to bridge the gap between the two most prominent models in the literature. This allows us to ask new questions such as what classes of dynamic networks are most conducive for social learning.  

\par All the results presented above suggest that the main insights from deterministic networks are \textit{robust} to random dynamics. The rest of the paper complements these results by providing novel insights that are unique to dynamic networks. In Section \ref{sec:fragile}, we focus on iid networks to characterize consensus in terms of \textit{early} social structure. 
We find that the initial \textit{social topology} (or ``skeleton'') of an iid network  reveals whether consensus exists. Two  societies have the same social topology at time $t$ if the zero entries in their respective interaction matrices are located in the same position at time $t$. Thus, whether consensus exists in an iid network can be inferred by simply examining the initial interaction matrix. This also simplifies the computation of agents' influence weights, which is otherwise an open problem in random matrix theory. 
 This result therefore suggest that using the DeGroot rule in dynamic networks yields a form of ``anchoring bias'' on early beliefs. This has policy implications: it highlights the importance of the \textit{timing} of interventions targeted to improve social learning outcomes. Specifically, early interventions can enhance collective intelligence by mitigating anchoring biases. This is reminiscent of \citeauthor{raj15}'s (\citeyear{raj15}) findings, showing the improved long-run outcomes of children whose families moved earlier in MTO. In general, these results can provide guidance to a planner who seeks to design and implement interventions in dynamic social networks \citep[][]{ban19,ban23,ban24}. 

\par Section \ref{sec:med} investigates the double-edged effects of random dynamics. 
 While analyzing the medium-run behavior of the learning process, we find that the effect of random dynamics on the speed of convergence is nuanced because it depends on the social structure. Introducing random dynamics in ``sparse'' networks tends to speed up convergence whereas it has the opposite effect in ``well-connected'' networks. That is, nudging agents to randomize their weights could be beneficial in networks where many agents are isolated.  
 Section \ref{sec:conv} proposes a robust probabilistic measure of \textit{homophily}\footnote{Homophily is the tendency of people to associate disproportionately with those having similar traits.} based on the likelihood of the worst possible network fragmentation.
We then show that convergence speed in stationary networks can be faster than in iid networks when beliefs evolve with moderate persistence. If a planner wishes to boost social learning, then designing a stationary network might be optimal.  
 Section \ref{sec:disag}  illustrates how random dynamics can cause long-run \textit{disagreement}. 
 
\par \noindent\textit{--- Related Work}: Although the  literature acknowledges random dynamics, dynamic networks have yet to receive enough attention in economics. 
This is perhaps due to the mathematical complexities introduced by randomness, and the fact that the tools needed to handle them are buried in random matrix theory.
 One of our main goals is therefore to gather and apply results from random matrix theory to analyze dynamic networks. \citet[][Appendix 1.A]{dem03} acknowledge the presence of random matching processes but circumvent their challenges by assuming that randomness vanishes in the aggregate.  
Notable exceptions to this literature are \citet{iff08,iff09} who study dynamic networks, and their frameworks are special cases of ours (Section \ref{sec:unique}).  
In general, this adjacent literature focuses only on consensus, whereas we explore when consensus is \textit{correct}, which is very relevant in economics. The main takeaway from this literature is that consensus is completely determined by the \textit{average} behavior of a dynamic network. Our generalization reveals that the average no longer suffices for consensus, so we introduce a condition that disciplines the \textit{tails} of a dynamic network. This is important because it highlights that analyzing the medium-run of a dynamic network requires an understanding of the behavior of its tails---not just its average as previously suggested in \citet[][p. 532]{golub17}.   
\par \noindent\textit{--- Organization}: Section \ref{sec:model} introduces our framework, Section \ref{sec:consensus} establishes consensus, and Section \ref{sec:wisdom} studies collective intelligence. Section \ref{sec:fragile} analyzes consensus, Section \ref{sec:med} explores the medium run, Section \ref{sec:dis} is a discussion, and Section \ref{sec:lit} is a more detailed literature review.

\section{Framework}\label{sec:model}

\subsection{Preliminaries}\label{sec:prelim}

Our framework consists of a finite set $N = \{1, 2,\dots, n\}$ of agents who interact according to a dynamic social network. The interaction patterns among the agents at time $t\in\{1,2,\dots\}$ are captured by an $n \times n$ nonnegative matrix $\m{X}_t$, where $(\m{X}_{t})_{i,j\in N}>0$ indicates that $i$ pays attention (or listens) to $j$ at time $t$. We refer to $\m{X}_t$ as the ``interaction  matrix at time $t$,'' which is assumed to be (row) \textit{stochastic}, i.e., each row sums to one. 
This setup is a dynamic extension of the standard DeGroot learning \citep[e.g.,][Section I.A]{moj10}. 
\par Let's first build some intuition for dynamic networks. Consider two agents (hereafter, Ann and Bob) with fixed interaction matrix $\big(\begin{smallmatrix}
    1&0\\0&1
\end{smallmatrix}\big)$. This network has no practical value within the standard DeGroot model except that it illustrates failure of consensus. However, this ``trivial'' network has a dynamic foundation. At each $t$, suppose the interaction matrix $\m{X}_t$ takes value in $\big\{\big(\begin{smallmatrix}
    1&0\\0&1
\end{smallmatrix}\big),\big(\begin{smallmatrix}
    1-\epsilon&\epsilon\\\epsilon&1-\epsilon
\end{smallmatrix}\big)\big\}$, and let Ann and Bob be two neighbors who may encounter each other randomly while walking in their neighborhood. In periods when they do not encounter each other, each agent can only give weight to their own belief. In periods when they encounter each other, each agent gives weight $\epsilon>0$ to the other agent's belief. Thus, the network $\big(\begin{smallmatrix}
    1&0\\0&1
\end{smallmatrix}\big)$ is the limit of this dynamic network when randomness vanishes $\epsilon\rightarrow0$. This simple example yields two preliminary insights: (1) random dynamics can provide real-life foundations for networks that may otherwise seem pathological, and (2) introducing even very small amounts of randomness may be enough to produce drastically different outcomes.

\subsection{Network Generating Process}\label{sec:dgp}
We build on \citet[][Section 1.1]{hen97} to propose a random process that generates the interaction matrices $\{\m{X}_t\}_{t\geq1}$. Let $\varSigma$ denote the set of all $n\times n$ stochastic matrices  
and  $(\varOmega,\mathscr{F},\Proba,\theta)$ denote an \textit{ergodic} dynamical system,\footnote{Ergodicity in this context refers to the property of the ``measure-preserving'' map $\theta:\varOmega\rightarrow\varOmega$, which is the main concept in ergodic theory. 
Appendix \hyperref[app:tech]{A} provides a comprehensive review of key technical details. 
} where a  stochastic matrix $\m{X}_0$ is a random variable on $\varOmega$ taking values in $\varSigma$. Elements in $\varOmega$ and $\varSigma$ are denoted $\omega$ and $\sigma$, respectively. 
Then, $\m{X}_t(\omega)$---the interaction matrix of $n$ agents at time $t\geq0$---is generated according to 
\begin{align}\label{eq:DGP}
    \m{X}_t(\omega)=\m{X}_0\circ\theta^t(\omega),
\end{align}
for $\omega\in\varOmega$. This process assumes \textit{stationarity}, so there exists an invariant probability measure $\mu$ on the Borel subsets $B$ of $\varSigma$, i.e., $\mu(B) = \Proba \big(\m{X}_{t}(\omega)\in B\big)$, for all $t$, with support $\varSigma_\mu\subset\varSigma$. 
\par\noindent \textit{--- Interpretation}: The pair $(\mu,\varSigma_{\mu})$ is the main primitive of our model. The set $\varSigma_{\mu}$ is a collection of all possible (or feasible) interaction patterns that may realize as a dynamic network evolves, and $\mu$ describes the frequency with which such patterns emerge over time.  
\par\noindent \textit{--- Special cases}$: \{\m{X}_t(\omega)\}_{t\geq1}$ can be iid but in general there is no restriction on their correlation structure. There is also no restriction on the correlation across the rows of each $\m{X}_{t}$, i.e., agents' weights can be correlated at every $t$. Section \ref{sec:disc} provides a microfoundation. 
\subsection{Evolution of Beliefs}
\par In standard DeGroot learning, each agent $i$ receives, at $t=0$, an initial signal, denoted $p^{(0)}_i\in\R$---normalized, without loss of generality, to lie in $[0,1]$. Each agent then updates her belief  by repeatedly taking the \textit{same} weighted averages of everyone's beliefs over time, and these weights are captured by an exogenously fixed interaction matrix $\m{T}$ \citep[][Section I.B]{moj10}. Here, we extend this as follows: the repeated weighted averaging process of all agents at $t\geq1$ is captured by a random (left) product $\m{X}^{(t)}(\omega)$ defined recursively as 
\begin{align}\label{eq:prod}
    \m{X}^{(1)}(\omega)=\m{X}_1(\omega), \quad\text{and}\quad \m{X}^{(t+1)}(\omega)=\m{X}_{t+1}(\omega)\m{X}^{(t)}(\omega),
\end{align}
for $\omega\in\varOmega$. This allows the weights that agents put on (themselves and) their neighbors to vary randomly over time. 
Then, for any vector of initial beliefs $\m{p}^{(0)}\in[0,1]^n$, the agents' updated beliefs at $t\geq1$, $\m{p}^{(t)}$, evolve according to the discrete-time linear dynamical system
\begin{align}\label{eq:evol}
    \m{p}^{(t)}(\omega)=\m{X}^{(t)}(\omega)\m{p}^{(0)}.
\end{align}
 When each $\m{X}_t(\omega)$ in eq. (\ref{eq:DGP}) is time-invariant, i.e., $\m{X}_t(\omega)=\m{T}$ for all $t$ and $\omega\in\varOmega$, then eq. (\ref{eq:prod}) becomes $\m{X}^{(t)}(\omega)=\m{T}^t$ for all $\omega\in\varOmega$, so $\m{p}^{(t)}=\m{T}^t\m{p}^{(0)}$ as in the standard DeGroot model.\footnote{To ease notation, we will often omit the argument $\omega$ when referring to $\m{X}_t$, $\m{X}^{(t)}$, and $\m{p}^{(t)}$.} In terms of our primitives, this arises when $\mu$ is a unit mass at $\m{T}\in\varSigma$, i.e., $\varSigma_\mu=\{\m{T}\}$. 

\section{Consensus}\label{sec:consensus}

\subsection{Definitions}
A necessary step for collective intelligence is to first establish whether agents' beliefs can converge to a well-defined limit.  
The next definition presents the convergence used hereafter.
\begin{definition}[Convergence]\normalfont
    A sequence of random nonnegative matrices $\{\m{X}_t\}_{t\geq1}$ is said to be \textit{convergent} if $\underset{t\rightarrow\infty}{\text{lim}}\hspace{0.01in}\m{X}^{(t)}$ exists almost surely.\hfill$\bigtriangleup$
\end{definition}
When $\underset{t\rightarrow\infty}{\text{lim}}\hspace{0.01in}\m{X}^{(t)}$---the limiting product of random interaction matrices---exists a.s. and has \textit{rank one} (i.e., all its rows are identical), the agents are said to have reached \textit{consensus}. 
\par Let $\hat{\varSigma}\subset\varSigma$ denote the set of stochastic matrices whose entries are all strictly positive. The following is a \textit{tail} condition that will play a central role in our model. 
\begin{definition}
   \normalfont A sequence of random nonnegative matrices $\{\m{X}_t\}_{t\geq1}$ is \textit{primitive} if
    \begin{align}
\tag{$\mathscr{C}$}\label{eq:C}\Proba\Bigg(\bigcup_{t=1}^{\infty}\Big\{\m{X}^{(t)}\in\hat{\varSigma}\Big\}\Bigg)>0. 
\end{align}
\end{definition} 
 This is a natural extension of the ``primitive'' condition used for deterministic networks \citep[][Definition 8]{moj10}.\footnote{A nonnegative deterministic matrix $\m{T}$ is said to be primitive if  $\m{T}^t$ is strictly positive for some $t\geq1$.}  
 It holds, for example, if one $\m{X}_\tau$ is strictly positive at some time $\tau$. We analyze and discuss condition \eqref{eq:C} in Sections \ref{sec:fragile} and \ref{sec:C}, respectively. 
Now, for each $t\geq1$, let ${\m{\pi}}_t$  and $\m{r}_t$ be nonzero random vectors in $[0,1]^n$ satisfying  
$$\m{X}^{(t)}\m{r}_t=\m{r}_t,\hspace{0.1in} {\m{\pi}}_t{\m{X}^{(t)}}={\m{\pi}}_t,\text{ and } {\m{\pi}}_t\m{r}_t=1.$$
Since $\m{X}^{(t)}$ is stochastic, its spectral radius is \textit{one} for each $t$, so ${\m{\pi}}_t$ and $\m{r}_t$ denote, respectively, its left and right eigenvectors corresponding to this eigenvalue. 
Here, ${\m{\pi}}_t$ is normalized to a \textit{unit} vector, i.e., its entries sum to one, and $\m{r}_t=\mb{1}$, for all $t$, i.e., the $n$-vector whose entries are all ones. Thus,  the stochastic matrix $\mb{1}{\m{\pi}}_t
$ has rank one---all its rows are identically $\m\pi_t$. 
\subsection{Existence of Consensus}
Our first result establishes that \eqref{eq:C} is sufficient to ensure consensus in dynamic networks. 
\begin{theorem}[Consensus]\label{thm:consensus}
    Suppose \eqref{eq:C} holds. Then, $\{\m{X}_t\}_{t\geq1}$ is convergent and \begin{enumerate}
        \item $\underset{t\rightarrow\infty}{\text{\normalfont lim}}\big(\m{X}^{(t)}-\m{1}{\m{\pi}}_t\big)=\m0$ a.s.
        \item ${\m{\pi}}_t\in[0,1]^n$ converges a.s., as $t\rightarrow\infty$, to a strictly positive random unit vector ${\m{\pi}}$.
    \end{enumerate}
\end{theorem}
Theorem \ref{thm:consensus} is a direct extension of \citet[][Lemma 3]{moj10}. 
Notice that it allows $\{\m{X}_t\}_{t\geq1}$ to be arbitrarily correlated.  
Most importantly, the limit of the left eigenvector ${\m{\pi}}_t$, denoted ${\m{\pi}}=(\pi_1,\dots,\pi_n)\in(0,1)^n$ in Theorem \ref{thm:consensus}.2, is a random vector that collects the long-run \textit{influence} weights of all the agents. The randomness of $\m\pi$ captures the fact that different evolutionary paths of the dynamic network (i.e., different $\omega$'s) may lead to different consensus. Then, for any vector of initial beliefs $\m{p}^{(0)}=({p}^{(0)}_1,\dots,p^{(0)}_n)\in[0,1]^n$, Theorem \ref{thm:consensus} reveals that each agent $i$'s belief $p^{(t)}_i$ will converge a.s., as $t\rightarrow\infty$, to the \textit{same} limit
\begin{align*}
   {p}^{(\infty)}_i= \Big(\underset{t\rightarrow\infty}{\text{lim}}\m{X}^{(t)}\m{p}^{(0)}\Big)_i=\sum_{j\in N}\pi_jp^{(0)}_j.
\end{align*}
Thus, Theorem \ref{thm:consensus} shows that despite the introduction of arbitrary random dynamics in networks, our framework preserves the tractable appeal of the standard DeGroot model.
\subsection{Some Special Cases and Examples}
This section analyzes some dynamic networks. Online Appendix \hyperref[app:ex]{B} explores other examples.
\subsubsection{Unique Consensus}
Since the influence vector $\m\pi$ in dynamic  networks is generally random, characterizing when it is unique is important. Let $\varSigma_{\mu}^\tau$ denote the set of all  $\tau$-length products of matrices in $\varSigma_{\mu}$.
\begin{proposition}\label{thm:uniq}
 Let $\{\m{X}_t\}_{t\geq1}$ be iid and $\m{s}\in[0,1]^n$ be a constant unit vector. Then, $\m{X}^{(t)}$ converges in probability to $\m{1s}$ as $t\rightarrow\infty$ if and only if $\m{1s}$ is the only rank one matrix in
    \begin{align*}
\text{\normalfont closure}\Bigg(\bigcup_{t=1}^{\infty}\varSigma^t_{\mu}\Bigg).
    \end{align*}
\end{proposition}
This result is a weak characterization of the uniqueness of consensus, and it relies solely on our primitives $(\mu,\varSigma_{\mu})$.\footnote{The set $\text{\normalfont closure}\big(\bigcup_{t=1}^{\infty}\varSigma^t_{\mu}\big)$ is the closed ``multiplicative semigroup'' generated by $\varSigma_{\mu}$ (Appendix \hyperref[app:setup]{A.II}).} Next, we illustrate how to apply this insight in Theorem \ref{thm:consensus}.
\begin{corollary}\label{thm:deter}
   In Theorem \ref{thm:consensus}, let $\m{s}$ be a strictly positive constant unit vector  such that $\m{s}\m{X}_t=\m{s}$ holds a.s., for every $t\geq1$. Then, $\underset{t\rightarrow\infty}{\text{\normalfont lim}}\hspace{0.01in}\m{X}^{(t)}=\m{1s}$ a.s. 
\end{corollary}
To predict when all agents have equal influence,  Proposition \ref{thm:uniq} indicates that we can leverage the fact that $\frac{1}{n}\m{11}'$ is the only rank-one bistochastic (or doubly stochastic) matrix. 
\begin{corollary}\label{thm:bi}
   In Theorem \ref{thm:consensus}, let each $\m{X}_t$ be a random bistochastic matrix for every $t\geq1$. Then, as $t\rightarrow\infty$, all 
agents have equal influence, i.e., $\pi_i=1/n$ a.s., for all $i\in N$.
\end{corollary}

\begin{ex}\label{ex:Ann1}\normalfont
  Consider Ann and Bob from Section \ref{sec:prelim}, where $\varSigma_\mu=\big\{\big(\begin{smallmatrix}
    1&0\\0&1
\end{smallmatrix}\big),\big(\begin{smallmatrix}
    1-\epsilon&\epsilon\\\epsilon&1-\epsilon
\end{smallmatrix}\big)\big\}$ and $\epsilon\in(0,1)$. Their encounters can be arbitrarily correlated, e.g., it may be unlikely that Ann and Bob encounter each other two periods in a row. Since these interaction matrices are bistochastic, Corollary \ref{thm:bi} applies, so the influence vector becomes $\m\pi=(1/2,1/2)$. Notice that even with very small amounts of randomness, i.e., $\epsilon\approx0$, Ann and Bob will always reach consensus. Moreover, these results continue to hold even when $\varSigma_\mu=\big\{\big(\begin{smallmatrix}
    0&1\\1&0
\end{smallmatrix}\big),\big(\begin{smallmatrix}
    \epsilon&1-\epsilon\\1-\epsilon&\epsilon
\end{smallmatrix}\big)\big\}$. \hfill$\bigtriangleup$
\end{ex}

\subsubsection{Random Consensus}
 \par When the influence vector $\m\pi$ is actually random, deriving its distribution in closed form becomes critical. The next result provides necessary and sufficient conditions for the distribution of  $\m\pi$ to be \textit{Dirichlet}. For a strictly positive vector $\m{a}=(a_1,\dots,a_n)$, let $D_{\m{a}}$ be the Dirichlet distribution with parameter $\m{a}$. Also, let $G_{\m{a}}$ denote the joint distribution of $n$ independent Gamma random variables each with scale parameter $1$ and shape parameter $a_i$.
\begin{corollary}\label{thm:mic}
  In Theorem \ref{thm:consensus},  let $\{\m{X}_t\}_{t\geq1}$ be iid. Then, $\m\pi\sim D_{\m{\varphi}}$ if and only if \eqref{eq:C} holds and there exists $\m{\varphi}\in\R^n_+$ such that $\m{v}\m{X}_0\sim G_{\m{\varphi}}$, where $\m{v}\sim G_{\m{\varphi}}$ is a vector independent of $\m{X}_0$.
\end{corollary}
This result is due to \citet[][Theorem 4]{mic14}. The next result is \citet[][Theorem 1.2]{letac94}, which gives more insights into the parameter $\m\varphi$ in Corollary \ref{thm:mic}.
\begin{corollary}\label{thm:mic2}
    In Corollary \ref{thm:mic}, suppose each row $(\m{X}_0)_i\sim D_{(\alpha_{i1},\dots,\alpha_{in})}$ is independent, where $r_i=\sum_{j\in N}\alpha_{ij}=\sum_{j\in N}\alpha_{ji}$ holds for all $i\in N$, and $\alpha_{ij}>0$. Then, $\m\varphi=(r_1,\dots,r_n)$.
\end{corollary}
Corollary \ref{thm:mic2} is the only known result in the literature that gives conditions for the distribution of agents' weights to be ``conjugate'' to the distribution of the influence vector. Throughout this paper, we will leverage Corollary \ref{thm:mic2} to illustrate our main results.
\begin{ex}\label{ex:Ann2}
    \normalfont  Suppose that whether or not there is an encounter, Ann and Bob always choose independently (of each other and over time) their weights from a Beta distribution, $\text{Beta}(\alpha,\alpha)$, with  concentration parameter $2\alpha>0$. If $\alpha=1$, then they choose their weights uniformly on $[0,1]$. This setting meets the conditions in Corollary \ref{thm:mic2}, where  $r_1=r_2=2\alpha$, so $\m\varphi=(2\alpha,2\alpha)$. Thus, $\m\pi=(\pi,1-\pi)$ satisfies  $\pi\sim\text{Beta}(2\alpha,2\alpha)$.  As $\alpha$ grows, $\pi$ 
    concentrates around $1/2$ as in Example \ref{ex:Ann1}. Section \ref{sec:med} shows how $\alpha$ affects the speed of convergence.
    \hfill$\bigtriangleup$
\end{ex}

\section{Collective Intelligence}\label{sec:wisdom}
This section characterizes when consensus is \textit{correct} in terms of eigenvector centrality.
\subsection{Setup and Definition}\label{sec:wisdef}
 Let the constant $\gamma\in [0, 1]$ denote the true state of the world. The initial signals $\big\{{p}^{(0)}_i(n)\big\}_{i\leq n}$ and the dynamic network $\{\m{X}_t(n)\}_{t\geq1}$ are drawn from the following joint distribution.\footnote{In this section, we follow \citet{moj10} by explicitly indicating the dependence on $n$.} 
 (i) $\big\{{p}^{(0)}_i(n)\big\}_{i\leq n}$ are independent for each $n$. (ii) Each $p_i^{(0)}(n)\in[0,1]$ is independent of $\{\m{X}_t(n)\}_{t\geq1}$ with mean $\gamma$ and variance of at least $\underline{\upsilon}^2 > 0$. 
 (iii) For all $n$, $\{\m{X}_t(n)\}_{t\geq1}$ is ergodic and satisfies \eqref{eq:C}. 
 Conditions (i) and (ii) are identical to those in \citet[][Section III.A]{moj10}.\footnote{These conditions are not necessary in dynamic networks. Thus, we will prove the main result of this section under weaker conditions that allow signals to be correlated to the network and to also be biased.}  Then, agents' 
 beliefs, $\m{p}^{(t)}(n)$, at $t\geq1$ evolve according to eq. (\ref{eq:evol}),
where, for each $n$, $\m{X}^{(t)}(n)$ is the random product in eq. (\ref{eq:prod}). Now, a large dynamic social network is said to be collectively intelligent (or ``wise'' for short) when the agents' limiting beliefs $\big\{{p}^{(\infty)}_i(n)\big\}_{i\leq n}$ converge jointly in probability to the true state $\gamma$ as defined below.
\begin{definition}
 \normalfont  The sequence  $\big\{\m{X}_{\infty}(n)\big\}_{n\geq1}$ is \textit{collectively intelligent} if, for every $\epsilon>0$,
 \begin{align*}
     \underset{n\rightarrow\infty}{\text{lim}}\hspace{0.03in}\Proba\Big(\underset{i\leq n}{\text{max}}\hspace{0.05in}\Big|{p}^{(\infty)}_i(n)-\gamma\Big|\geq\epsilon\Big)=0.
 \end{align*}
\end{definition}

\subsection{Analysis}
  
Theorem \ref{thm:wisdom} is a characterization of collective intelligence based on eigenvector centrality. 
\begin{theorem}[Collective intelligence]\label{thm:wisdom}
  For each $n$, let $\{\m\pi(n)\}_{n=1}^{\infty}$ be any sequence of random influence vectors from Theorem \ref{thm:consensus}.2. Then, for every $\epsilon>0$, \begin{align*}
        \underset{n\rightarrow\infty}{\text{\normalfont lim}}\hspace{0.03in}\Proba\Bigg(\Bigg|\sum_{i\leq n}\pi_i(n)p_i^{(0)}(n)-\gamma\Bigg|\geq\epsilon\Bigg)=0 \quad\text{if and only if} \quad \E\Big[\underset{i\leq n}{\text{\normalfont max}}\hspace{0.04in}\pi_i(n)\Big]\rightarrow0 \text{ as } n\rightarrow\infty.
    \end{align*}
\end{theorem}

\par In words, Theorem \ref{thm:wisdom} states that, as society grows,
the limiting belief of each agent,
\begin{align*}
    p^{(\infty)}_i(n)=\sum_{j\leq n}\pi_j(n)p_j^{(0)}(n),
\end{align*}
converges in probability to the true state $\gamma$ if and only if the average weight of the most influential agent tends to zero.  
This is a direct extension of \citet[][Lemma 1]{moj10} for dynamic networks. The next result will be useful because it simplifies Theorem \ref{thm:wisdom}. 
\begin{proposition}\label{thm:wis}
    Suppose $\underset{i\leq n}{\text{\normalfont max}}\hspace{0.04in}\E[\pi_i(n)]\rightarrow0\text{ as } n\rightarrow\infty$ and $\text{\normalfont var}(\pi_i(n))=O(1/n^{d})$, for every $i\leq n$ and some constant $d>1$. Then, $\{\m{X}_\infty(n)\}_{n\geq1}$ is collectively intelligent.
\end{proposition}
\subsubsection{Discussion and Intuition}
 \par The next result identifies a large class of dynamic networks that are collectively intelligent.  
\begin{proposition}\label{thm:mic3}
Suppose  $\{\m{X}_t(n)\}_{t\geq1}$ satisfies Corollary \ref{thm:mic}, $\sum_{i\leq n}\varphi_i=O(n^{k})$, for $k\geq1$, and $\varphi_j\big/\sum_{i\leq n}\varphi_i=O(1/n^{m})$, for $m>0$ and all $j$. Then, $\{\m{X}_\infty(n)\}_{n\geq1}$ is collectively intelligent.
\end{proposition}
Proposition \ref{thm:mic3} leverages the conditions in Proposition \ref{thm:wis}, which are illustrated in an example in Section \ref{sec:wisex}. More generally, this is helpful because it clarifies the  conditions for collective intelligence in terms of our primitives. In Corollary \ref{thm:mic}, the influence vector is Dirichlet with parameter $\m\varphi(n)=(\varphi_1,\dots,\varphi_n)$, i.e., $\m\pi(n)\sim D_{\m\varphi(n)}$, where $\sum_{i\leq n}\varphi_i$ measures the ``concentration'' around the mean. For the special case in Corollary \ref{thm:mic2},   $\varphi_i=\sum_{j\leq n}\alpha_{ij}=\sum_{j\leq n}\alpha_{ji}$ and $(\m{X}_t(n))_i\sim_{\text{iid}} D_{(\alpha_{i1},\dots,\alpha_{in})}$ for all $i\leq n$. Here, Proposition \ref{thm:mic3} shows that  $\sum_{i\leq n}\varphi_i=\sum_{i\leq n}\sum_{j\leq n}\alpha_{ij}\rightarrow\infty$ as $n\rightarrow\infty$ is predictive of collective intelligence. Notice that this implies that $\mu(n)$---the distribution  of each $\m{X}_t(n)$---must concentrate around its mean as $n\rightarrow\infty$. To see this,  define $\m{A}(n)=(\alpha_{ij})_{i,j\leq n}$, where $D_{\m{A}(n)}$ denotes a Dirichlet distribution on $n\times n$ stochastic matrices. Here, we have  $\m{X}_t(n)\sim_{\text{iid}}\mu(n)=D_{\m{A}(n)}$, so requiring the sums of the $\alpha_{ij}$'s to be unbounded as $n\rightarrow\infty$ implies that $D_{\m{A}(n)}$ must concentrate around its mean. This and the \textit{balance} equations above---$\sum_{j\leq n}\alpha_{ij}=\sum_{j\leq n}\alpha_{ji}$ for all $i\leq n$---ensure that the limiting distribution of the product $\m{X}^{(t)}(n)$ will concentrate around $\frac{1}{n}\m{1}\m{1}'$ as $n\rightarrow\infty$.
\par To summarize Theorem \ref{thm:wisdom}, collective intelligence requires dynamic networks to obey the following as $n\rightarrow\infty$: (1) the distribution $\mu(n)$ of each $\m{X}_{t}(n)$ should concentrate around its mean $\E_{\mu(n)}[\m{X}_t(n)]$, and (2) $\E_{\mu(n)}[\m{X}_t(n)]$ should behave a lot like a bistochastic matrix.
\subsubsection{Some Numerical Examples}\label{sec:wisex}
 Let's start with a simple warm-up example to build some intuition for collective intelligence. 
 \begin{ex}\normalfont\label{ex:wis1}
 For a constant  $\zeta\in(0,1]$, consider the following dynamic network \begin{align*}
     \m{X}_t(n)=\begin{cases}
         \frac{1}{n}\m{1}\m{1}' \quad &\text{with probability }\zeta\\
         \m{I}_n &\text{with probability }1-\zeta,
     \end{cases}
 \end{align*} for all $t\geq1$, where $\m{I}_n$ is the $n\times n$ identity, and $\{\m{X}_t(n)\}_{t\geq1}$ can be arbitrarily correlated. Then, $\E_{\mu(n)}\big[\m{X}_t(n)\big]=(1-\zeta)\m{I}_n+\zeta\frac{1}{n}\m{1}\m{1}'$, and when $\zeta=1/2$, this mean coincides with the mean in \citet[][Example 1]{iff09}. 
 Since each $\m{X}_t(n)$ is bistochastic, Corollary \ref{thm:bi} applies, so $\pi_i(n)=1/n$ for all $i$. Thus, this dynamic network is collectively intelligent for all $\zeta\in(0,1)$. Notice that as the randomness vanishes (i.e., $\zeta\rightarrow0$), $\m{X}_t(n)\rightarrow\m{I}_n$ a.s. for all $t$ and $n$, so the network would lose the ability to be collectively intelligent. \hfill$\bigtriangleup$
 \end{ex}
The next example shows that the same insights also hold in more complex networks. 
 
  \begin{ex}\normalfont\label{ex:wis2}
  Let $\m{T}(n)$ be the chain/ring network shown below \citep[e.g.,][Fig. 3.2]{touri12}, where each agent $i\leq n$ always adopts the belief of agent $i + 1 \hspace{0.03in} (\text{mod } n)$: 
 \begin{align*}
     \m{T}(n)=\begin{pmatrix}
   0  &1 &0&\dots&0\\
    0&0&1&\ddots&\vdots\\
    \vdots&\ddots&\ddots&\ddots&0\\
    0&\dots&0&0&1\\
    1&0&\dots&0&0
\end{pmatrix}.
 \end{align*}
 When $n=2$, this network is analyzed in  \citet[][Example 1]{moj10}. Since this network does not reach consensus for any $n$, it is never collectively intelligent. Let's now introduce some randomness. At each $t$, let's allow each agent $i$ to independently give to their own belief a weight $x_{i,t}$ uniformly distributed on $[0,1]$. This yields the interaction matrix
 \begin{align*}
\m{X}_t(n)=\begin{pmatrix}
    x_{1,t} &1-x_{1,t} &0&\dots&0\\
    0&x_{2,t}&1-x_{2,t}&\ddots&\vdots\\
    \vdots&\ddots&\ddots&\ddots&0\\
    0&\dots&0&x_{n-1,t}&1-x_{n-1,t}\\
    1-x_{n,t}&0&\dots&0&x_{n,t}
\end{pmatrix},
 \end{align*}
     where $\{\m{X}_t(n)\}_{t\geq1}$ is iid. Notice that $\E_{\mu(n)}\big[\m{X}_t(n)\big]$ is bistochastic for all $t,n$.  Also, $\{\m{X}_t(n)\}_{t\geq1}$ satisfies the conditions in Corollary \ref{thm:mic2} \citep[][Section 4.1]{mic14}. Since $\text{Beta}(1,1)$ is the uniform distribution on $[0,1]$, we have $r_i=2$ for all $i\leq n$, so $\m{\varphi}(n)=(2,\dots,2)\in\R^n_+$, and therefore $\m{\pi}(n)\sim D_{\m{\varphi}(n)}$ in Corollary \ref{thm:mic}, i.e., a Dirichlet distribution with parameter $\m{\varphi}(n)$. By Proposition \ref{thm:mic3}, this dynamic network is collectively intelligent because here $k=m=1$: $$\sum_{i\leq n}\varphi_i=2n=O(n^{k}) \quad\text{and}\quad \frac{\varphi_j}{\sum_{i\leq n}\varphi_i}=\frac{2}{2n}=O(1/n^m) \text{ for all }j\leq n.$$ 
This example highlights that introducing small random dynamics ($x_{i,t}\approx0$) can boost collective intelligence.\footnote{Examples \ref{ex:mckinlay}/\ref{ex:mckinlay2} (Online Appendix \hyperref[app:ex]{B}) show that a complex dynamic network is collectively intelligent.} 
In contrast, when $x_{i,t}\rightarrow0$ for all $i$ and $t$, $\m{X}_{t}(n)$ approaches $\m{T}(n)$, i.e., as random dynamics vanish, the network loses the ability to be collective intelligent. \hfill$\bigtriangleup$
\end{ex}
These insights are consistent with \citeauthor[][]{dynam20}'s (\citeyear[][]{dynam20})  experimental results, who conclude: ``dynamism of the networks has profound effects on the processes taking place on them, allowing networks to become more efficient and enabling them to better adapt to changing environments.'' Notice that $\m{T}(n)$ was sparse in Examples \ref{ex:wis1}--\ref{ex:wis2}. Thus, these examples (and Examples \ref{ex:mckinlay}/\ref{ex:mckinlay2}) indicate that randomness tends to be beneficial when many agents are isolated. Section \ref{sec:med} will show that the opposite happens in ``well-connected'' networks.

\subsection{Undirected Dynamic Networks}
This section explores undirected networks to illustrate that some key insights about collective intelligence in deterministic networks can be extended to dynamic networks. Let $\m{G}_t(n)$ denote a symmetric connected $n\times n$ adjacency matrix of an undirected dynamic network, where $(\m{G}_t(n))_{ij}=1$ indicates that $i$ and $j$ have an undirected link between them at time $t\geq1$, and $(\m{G}_t(n))_{ij}=0$ otherwise. Define $\m{G}(n)=\{\m{G}_t(n)\}_{t\geq1}$. Then, let $d_i(\m{G}(n)) =\sum_{j\leq n}(\m{G}_t(n))_{ij}>0$ be the (random) \textit{degree}---number of neighbors---of agent $i$, for all $t$, i.e., degree is assumed to be stationary. Suppose the interaction matrix at $t$, $\m{X}_t(\m{G}(n))$, is defined as $\big(\m{X}_t(\m{G}(n))\big)_{ij} = (\m{G}_t(n))_{ij}\big/d_i(\m{G}(n))$. Intuitively, $\m{G}_t(n)$ generates a dynamic network of undirected random connections at time $t$, where every agent puts equal weight on all their neighbors at that time. This is a dynamic extension of the deterministic setting studied in \citet[][Section II.C]{moj10}. For all $n$, each entry $i\leq n$ of the random left eigenvector $\m{\pi}(n)$, common to all $\{\m{X}_t(\m{G}(n))\}_{t\geq1}$, in Theorem \ref{thm:consensus}.2 takes the simple form 
$$\pi_i(n)=\frac{d_i(\m{G}(n))}{\sum_{k\leq n}d_k(\m{G}(n))}.$$

 The next result is an extension of \citet[][Corollary 1]{moj10}, which follows from Theorem \ref{thm:wisdom}. It indicates that in undirected networks with random dynamics, the only obstacle to collective intelligence is the disproportionate average popularity of some agents.
\begin{corollary}
In Theorem \ref{thm:wisdom}, let $\{\m{G}(n)\}_{n\geq1}$ be a random sequence of symmetric, connected adjacency matrices. Then, $\big\{\m{X}_{\infty}(\m{G}(n))\big\}_{n\geq1}$ is collectively intelligent if and only if 
\begin{align*}
    \E\Bigg[\underset{i\leq n}{\text{\normalfont max}}\hspace{0.02in}\frac{d_i(\m{G}(n))}{\sum_{k\leq n} d_k(\m{G}(n))}\Bigg]\longrightarrow 0 \quad \text{as}\quad n\rightarrow\infty.
\end{align*}
\end{corollary}

\section{Analysis of Consensus}\label{sec:fragile}

Since consensus is necessary for collective intelligence, identifying its determinants is critical. We seek to identify simple features of dynamic networks that are predictive of consensus. 

\subsection{Characterization of Tail Condition}
\par The following characterization of \eqref{eq:C} will play a central role in the analysis of Section \ref{sec:top}.
\begin{proposition}\label{thm:equiv}
    Suppose $\{\m{X}_t\}_{t\geq1}$ is iid. Then,
   \begin{align*}
    \big\{\m{X}^{(t)}\big\}_{t\geq1} \text{ satisfies }    \eqref{eq:C} \text{ if and only if there exists }t\geq1 \text{ such that }\int_{\varSigma}  c(\sigma)\hspace{0.03in}d\mu^t(\sigma)<1.
    \end{align*}
\end{proposition}
This result is useful because it reveals that determining whether consensus exists in iid networks amounts to examining the distribution of $\m{X}^{(t)}$---the \textit{convolution power} of $\mu$---denoted $\mu^t$ and its support, which depend only on our primitives $(\mu,\varSigma_{\mu})$ (see, Appendices \hyperref[app:setup]{A.II}--\hyperref[app:conv]{A.III}).  The function $c:\varSigma\rightarrow[0,1]$ is continuous, and $c(\sigma)$ is known as the ``contraction coefficient'' of $\sigma\in\varSigma$ (Appendix \hyperref[app:sketch]{A.IV}). The next result sharpens Proposition \ref{thm:equiv} by showing that consensus depends only on the support of the initial interaction matrix $\m{X}_1$. 
\begin{proposition}\label{thm:first}
   Suppose $\{\m{X}_t\}_{t\geq1}$ is iid and $\Proba\big(\text{\normalfont min}_{ij}(\m{X}_1)_{ij}=0\big)<1$. Then, \eqref{eq:C} holds. 
\end{proposition}
 \begin{remark}\normalfont\label{rem:conv}
   Unlike the deterministic case, there are several modes of convergence for the random sequence $\big\{\m{X}^{(t)}\big\}_{t\geq1}$ in dynamic cases. The most important mode is weak convergence because the quantity of interest is the limiting probability measure of the product $\m{X}^{(t)}$. Weak convergence of $\big\{\m{X}^{(t)}\big\}_{t\geq1}$ can be interpreted as convergence of the learning process.  
 \hfill$\bigtriangleup$
\end{remark}
 
\begin{proposition}\label{thm:zero}
   Suppose $\{\m{X}_t\}_{t\geq1}$ is iid, and the initial interaction matrix $\m{X}_1$ has at least $n-1$ strictly positive diagonal entries a.s. Then, $\big\{\m{X}^{(t)}\big\}_{t\geq1}$ converges weakly as $t\rightarrow\infty$. 
\end{proposition}
Proposition \ref{thm:zero} is a much weaker result than  Proposition \ref{thm:first}.  Proposition \ref{thm:zero} indicates that to guarantee convergence of the learning process, it suffices that at least $n-1$ agents initially put some weight on their own beliefs (Remark \ref{rem:diag}). Notice that this condition is tight in the sense that it cannot be relaxed from $n-1$ to $n-2$, due to the case where $n=2$ and  $\mu$ is a unit mass on $\big(\begin{smallmatrix} 0&1\\1&0 \end{smallmatrix}\big)$, which is not convergent \citep[e.g.,][Example 1]{moj10}. 
 \subsection{Social Topology and Consensus}\label{sec:top}
The goal now is to predict \eqref{eq:C} using features of networks that can be easily tested in practice.
\subsubsection{Definitions}
Since consensus is often random in dynamic networks, the quantity of interest is the limiting distribution of the product $\m{X}^{(t)}$ as $t\rightarrow\infty$. This probability measure and its support, denoted ($\psi,{\varSigma_{\psi}}$), are defined below, and Appendix \hyperref[app:conv]{A.III} characterizes them in the iid case.
\begin{definition}\normalfont
    $\m{X}^{(t)}$ converges weakly as $t\rightarrow\infty$ if and only if $\psi$---the weak limit of its probability measure---exists with support ${\varSigma_{\psi}}\subset\varSigma$.  
    Here, $\psi$ is the distribution of $\m{X}^{(\infty)}$. \hfill$\bigtriangleup$
\end{definition}
Consider two equal-sized societies, labeled $X$ and $Y$, and their corresponding interaction matrices are denoted $\{\m{X}_{t}\}_{t\geq1}$ and $\{\m{Y}_{t}\}_{t\geq1}$, respectively. When the limiting distributions of their products exit, they are denoted $\psi_X$ and $\psi_Y$ with support $\varSigma_{\psi_X}$ and $\varSigma_{\psi_Y}$, respectively. 
\begin{definition}[Social topology]\label{def:topo}\normalfont
At any time $t\geq1$, two (distinct) equal-sized societies, with $n\times n$ interaction matrices $\m{X}_{t}$ and $\m{Y}_{t}$, respectively, are said to have the same \textit{social topology} when $(\m{X}_{t})_{ij}>0$ holds if and only if $(\m{Y}_{t})_{ij}>0$ holds, for all $i,j\in N$. \hfill$\bigtriangleup$   
\end{definition}
In graph theory terminology, two societies have the same social topology at time $t$ if their corresponding networks have the same set of ``minimally closed groups'' at time $t$ \citep[][Section A]{moj10}. Alternatively, two societies have the same social topology at time $t$ if their interaction matrices $\m{X}_t$ and $\m{Y}_t$ have the same ``skeleton''---a terminology proposed in \citet{muk05}. For example, $\m{X}_t=\Big(\begin{smallmatrix}
    1-\frac{1}{t+1}&\frac{1}{t+1}\\1&0
\end{smallmatrix}\Big)$ and $\m{Y}_t=\Big(\begin{smallmatrix}
    \frac{2}{3}-\frac{1}{4t}&\frac{1}{3}+\frac{1}{4t}\\1&0
\end{smallmatrix}\Big)$ have the same social topology for all $t\geq1$.  The following event will be useful for our results: $$\mathscr{T}=\big\{\m{X}_1\text{ \normalfont and } \m{Y}_1\text{ \normalfont have the \textit{same} social topology}\big\}.$$
\subsubsection{Analysis}
The next result demonstrates that social topology together with \eqref{eq:C} yield strong predictions.  
\begin{theorem}\label{thm:skel}
Let $\{\m{X}_t\}_{t\geq1}$ be iid, $\{\m{Y}_t\}_{t\geq1}$ be iid, and  $\Proba(\mathscr{T})=1$. Then,\begin{enumerate}
        \item[(i)] $\big\{\m{X}^{(t)}\big\}_{t\geq1}$ converges weakly if and only if $\big\{\m{Y}^{(t)}\big\}_{t\geq1}$ converges weakly. Moreover, there exist two rank one matrices $p\in\varSigma_{\psi_X}$ and $q\in\varSigma_{\psi_Y}$ that have the same social topology.
        \item[(ii)] $\big\{\m{X}^{(t)}\big\}_{t\geq1}$ satisfies \eqref{eq:C} if and only if $\big\{\m{Y}^{(t)}\big\}_{t\geq1}$ satisfies \eqref{eq:C}.
    \end{enumerate} 
\end{theorem}

Theorem \ref{thm:skel} shows that within the class of iid networks, the initial social topology completely characterizes consensus. Since satisfying \eqref{eq:C} implies almost sure convergence to consensus (Theorem \ref{thm:consensus}), Theorem \ref{thm:skel}.(ii) implies Theorem \ref{thm:skel}.(i), but the converse is not necessarily true. The key to prove Theorem \ref{thm:skel}.(ii) is the characterization of \eqref{eq:C} in Propositions \ref{thm:equiv}--\ref{thm:first}.  The proof is technical and novel, so a detailed sketch is presented in Appendix \hyperref[app:sketch]{A.IV}. 
\subsubsection{Discussion and Example}
 Theorem \ref{thm:skel} has two novel implications for social learning in dynamic networks: 
\begin{enumerate}
    \item[1.] \textit{Predictability}: Since consensus is necessary for collective intelligence, the initial interaction matrix can help predict whether a society will be collectively intelligent. Theorem \ref{thm:skel} may also be used by a planner who wishes to design a collectively intelligent society. To see this, consider two societies $X$ and $Y$, where society $X$ is known to be collectively intelligent. Proposition \ref{thm:pred} shows that a planner can make society $Y$ collectively intelligent by simply mirroring (or imitating) the initial social structure of society $X$. 
    \end{enumerate}
    \begin{proposition}\label{thm:pred}
Let $\{\m{X}_t\}_{t\geq1}$ and $\{\m{Y}_t\}_{t\geq1}$ be iid. Suppose society $X$ is collectively intelligent, where $\big\{\m{X}^{(t)}\big\}_{t\geq1}$ satisfies \eqref{eq:C}. If $\Proba(\mathscr{T})=1$, then society $Y$ is also collectively intelligent.   
    \end{proposition}
    \begin{enumerate}
    \item[2.] \textit{Computation}: We can infer whether consensus exists in an iid network by computing the limiting behavior of a deterministic network that happens to the same initial social topology. This is very convenient because determining whether a product of random stochastic matrices converges is an open problem even when $n=2$ (Online Appendix \hyperref[sec:ex2]{B.II}), whereas it is easy in the deterministic case thanks to Markov chain theory. As \citet{krish20} notes, reducing questions about random networks to deterministic calculations is very useful for practitioners because they tend to work with random networks. Example  \ref{ex:comp} illustrates the significance of this computation implication.
\end{enumerate}

\begin{ex}\normalfont\label{ex:comp}
    Suppose the interaction matrices of societies $X$ and $Y$ at $t\geq1$ are, respectively,
    \begin{align*}
        \m{X}_t=\begin{pmatrix}
          x_t &1-x_t\\0&1
        \end{pmatrix}, \text{ and } \quad \m{Y}_t=\begin{pmatrix}
          \frac{1}{2} &\frac{1}{2}\\0&  1
        \end{pmatrix} \text{ a.s.},
    \end{align*}
    where $\{x_t\}_{t\geq1}$ is any iid sequence on $(0,1)$ and society $Y$'s social network is fixed over time as in the standard DeGroot model.   
    Will society $X$ reach consensus? Without using Theorem \ref{thm:skel}, it is not possible to answer this question without probabilistic computations.  We first illustrate that computations are very challenging even though $n=2$, and then will show how Theorem \ref{thm:skel} can simplify the analysis. The product $\m{X}^{(t)}$ can be rewritten explicitly as
\begin{align*}
        \m{X}^{(t)}=\begin{pmatrix}
          \alpha_t &1-\alpha_t\\\beta_t &1-\beta_t  
        \end{pmatrix}= \begin{pmatrix}
          x_t &1-x_t\\0&1  
        \end{pmatrix}\begin{pmatrix}
          x_{t-1} &1-x_{t-1}\\0&1  
        \end{pmatrix}\dots\begin{pmatrix}
          x_1 &1-x_1\\0&1  
        \end{pmatrix},
\end{align*}
which can then be written equivalently as the following random difference equation 
\begin{align*}
    \alpha_t&=(x_t,1-x_t)'(\alpha_{t-1},\beta_{t-1})\\
    \beta_t&=(0,1)'(\alpha_{t-1},\beta_{t-1}),
\end{align*}
for $t\geq2$, where $\alpha_1=x_1$ and $\beta_1=0$ \citep[][eq. (1.2)]{va86}. Since this is a $2\times 2$ case, we know from \citet[][Theorem 1]{va86} that $\{\m{X}_t\}_{t\geq1}$ is convergent if and only if $\mu$ is not concentrated on  $\big\{\big(\begin{smallmatrix}
    0&1\\1&0
\end{smallmatrix}\big),\big(\begin{smallmatrix}
    1&0\\0&1
\end{smallmatrix}\big)\big\}$, which is not the case here. Thus, $\alpha_t-\beta_t$ converges almost surely to zero as $t\rightarrow\infty$, and since $\beta_t=0$ for all $t$, then $\alpha_t\rightarrow0$, so $p=\text{lim}_{t\rightarrow\infty}\m{X}^{(t)}=\big(\begin{smallmatrix}
    0&1\\0&1
\end{smallmatrix}\big)$.
\par In contrast, we could have inferred the above using only deterministic calculations by noticing that $\m{X}_1$ and $\m{Y}_1$ have the same social topology,  $\text{lim}_{t\rightarrow\infty}\m{Y}_1^{t}=p$, and hence $\varSigma_{\psi_Y}=\{p\}$. First, Theorem \ref{thm:skel}.(ii) shows that the limiting matrix in $X$ is not strictly positive since the limiting matrix $p$ in $Y$ is not strictly positive. Second, by Theorem \ref{thm:skel}.(i),  $\psi_X$ exists because $\psi_Y$ exists, and moreover, $\psi_X=\psi_Y=\delta_p$ because $\varSigma_{\psi_Y}$ is a singleton and $p^2=p\in\varSigma_{\psi_Y}$. 
\hfill$\bigtriangleup$
\end{ex}

Theorem \ref{thm:skel} showed that determining whether consensus exists in iid networks is equivalent to locating the zeros in the initial interaction matrix. \citet[][Theorem 1.(i)]{bernd19} shows (without assuming iid) another way through which the initial interaction matrix shapes agents' long-run beliefs. Specifically, they show that the range of possible beliefs shrinks over time. That is, when agents use the DeGroot rule in dynamic networks, their belief adjustments lose \textit{flexibility} the longer they interact with others. As noted in \citet[][Section 7]{bernd19}, this is reminiscent of \textit{anchoring bias}, i.e.,  agents treat their initial opinions of others as ``anchors'' and then proceed to adjust them over time. Thus, the initial structure of a dynamic network plays a central role in shaping long-run beliefs. This suggests that \textit{early} interventions are more likely to be effective because they may exploit the flexibility of initial beliefs. 
Thus, these insights may guide a planner who aims to design and implement policy interventions in  dynamic networks \citep[e.g.,][]{ban19,ban23,ban24}.

\section{Medium-Run Analysis}\label{sec:med}
As \citet{golub17} remark: ``For practical purposes, consensus is often irrelevant unless it is reached reasonably quickly.'' This section therefore analyzes the medium-run behavior of the social learning process to highlight some adverse effects of random dynamics. 
\subsection{Setup}\label{sec:setup}
To compare deterministic and dynamic networks, let's parameterize randomness \textit{locally} around a fixed network $\m{T}$. Define $\m{T}=(T_{ij})_{i,j\in N}$, where $T_{ij}>0$ for all $i,j$, with influence vector $\m{s}=(s_1,\dots,s_n)$. Now, during the transition $t-1\rightarrow t$, the  network $\m{T}$ suffers random perturbations such that it is replaced by a random network $\m{X}_t$. Formally, define the matrix $\m{A}_{\varepsilon}=(\varepsilon s_iT_{ij})_{i,j\in N}$ whose rows and columns are all equal to $(r_1,\dots,r_n)=\varepsilon (s_1,\dots,s_n)$, where $\varepsilon>0$ is a constant. Let  $\m{X}_t\sim D_{\m{A}_{\varepsilon}}$, i.e., each row $(\m{X}_t)_i\sim D_{\varepsilon(s_iT_{i1},\dots,s_iT_{in})}$ is independent across all $i$ and $t$. It follows that $\E_{\mu}[\m{X}_t]=\m{T}$ for all $t$, and the parameter $\varepsilon$ measures the degree of randomness around $\m{T}$, i.e., larger values of $\varepsilon$ lead to smaller fluctuations around $\m{T}$ \citep[see,][p. 425]{letac94}. Notably, this model meets the conditions in Corollary \ref{thm:mic2}, where $\m\varphi=\varepsilon\m{s}$, so the random influence vector satisfies $\m\pi\sim D_{\varepsilon\m{s}}$. As $\varepsilon$ grows, this Dirichlet distribution approaches a unit mass on the influence vector $\m{s}$ of the fixed network $\m{T}$. That is, $D_{\varepsilon \m{s}}$ is a mean-preserving spread of a unit mass on $\m{s}$. Thus, for every $\tau>0$,  $$\Proba\Big(\underset{i\leq n}{\text{max}}\big|\pi_i-s_i\big|\geq\tau\Big)\leq\frac{1}{\varepsilon\tau^2}\Big(1-\sum_{i\leq n}s^2_i\Big),$$ using union bound, Chebyshev inequality, and $\text{var}(\pi_i)=\frac{s_i(1-s_i)}{\varepsilon s_0+1}$, where $s_0:=\sum_{i\leq n}s_i=1$. Now, suppose the fixed network $\m{T}$ is collectively intelligent. Then, $\underset{i\leq n}{\text{max}}\hspace{0.02in}s_i\rightarrow0$ as $n\rightarrow\infty$, so the upper bound above converges to $\frac{1}{\varepsilon\tau^2}>0$. Thus, to ensure that $\m{X}_\infty$ is also collectively intelligent here, we need the randomness to vanish, i.e., $\varepsilon\rightarrow\infty$ as $n\rightarrow\infty$. \textit{--- Interpretation}: Since $\m{T}$ is a strictly positive matrix, the analysis above highlights that introducing random dynamics may prevent collective intelligence when a network is well-connected. This contrasts the conclusion from Section \ref{sec:wisex} where networks were sparse.

\subsection{Speed of Convergence}
Convergence speed of random stochastic matrices is an open problem in general. Here, we study cases where precise rates can be obtained. The network size $n$ will impact the analysis. 
\subsubsection{Simplest Case: $n=2$}\label{sec:2x2}
For any stochastic matrix $\big(\begin{smallmatrix}
    a&1-a\\
    b&1-b
\end{smallmatrix}\big)$, the second largest eigenvalue is $\lambda_2(a,b)=a-b$ \citep[see,][pp. 244--246]{moj08}. Now,  denote the random interaction matrix as $\m{X}_t=\big(\begin{smallmatrix}
    \alpha_t&1-\alpha_t\\
    \beta_t&1-\beta_t
\end{smallmatrix}\big)$, and let $\{\m{X}_t\}_{t\geq1}$ be iid. That is, $\{(\alpha_1,\beta_1),(\alpha_2,\beta_2),\dots\}$ is an iid sequence with distribution $\mu(x,y)=\Proba(\alpha_t\leq x,\beta_t\leq y)$.  Denote the random product as  $\m{X}^{(t)}=(\begin{smallmatrix}
    x_t&1-x_t\\y_t&1-y_t
\end{smallmatrix}\big)$, and to ease notation, let $\big(\begin{smallmatrix}
    a&1-a\\
    b&1-b
\end{smallmatrix}\big)$ be identified by the vector $(a,b)$. Then, for any constant $\phi>0$, define $$t_{\phi}=\text{max}\big\{t\in\mathbb{N}:|x_t-y_t|\geq\phi\big\}$$ as the random variable that counts the number of periods needed to come close to consensus. 
\begin{proposition}\label{thm:conv2x2}
     Let $\{\m{X}_t\}_{t\geq1}$ be an iid sequence, where $n=2$ and $\mu$ is not concentrated on $\{(1,0),(0,1)\}$. Then, as $\phi\rightarrow0$, (i) $\frac{1}{\text{\normalfont log}\hspace{0.02in}\phi}t_{\phi}\rightarrow-I^{-1}_{\mu}$ a.s., and (ii) $\frac{1}{\text{\normalfont log}\hspace{0.02in}\phi}\E[t_{\phi}]\rightarrow-I^{-1}_{\mu}$, where $$I_{\mu}=\int_0^1\int_0^1\text{\normalfont log}\frac{1}{|\lambda_2(x,y)|}\hspace{0.03in}d\mu(x,y).$$
\end{proposition}
\par This result is intuitive. It shows that to compute the convergence time in a dynamic network with two agents, it suffices to compute a $\mu$-average of the convergence time of a fixed network $\m{T}$, which is proportional to $-1/\text{log}|\lambda_2(\m{T})|$ \citep[][p. 132]{moj10}.\footnote{Recall that to achieve an error $\phi>0$, the convergence time of $\m{T}$ is proportional to $-(\text{\normalfont log}\hspace{0.02in}\phi)\big/\text{log}|\lambda_2(\m{T})|$.} 
Averaging occurs because the second largest eigenvalue in dynamic networks is random. 
\par Suppose we further assume that $\alpha_t$ and $\beta_t$ are iid for all $t$, i.e., Ann and Bob choose their weights independently from the same distribution given by $\mu(x,y)=\mu(x)\mu(y)$. Then, $I_{\mu}$ in Proposition \ref{thm:conv2x2} is known as the \textit{logarithmic energy} of the distribution $\mu$. In this special case, the distribution $\mu$ that yields the slowest convergence rate to consensus is not degenerate. 
\begin{corollary}\label{thm:conv2x22}
In Proposition \ref{thm:conv2x2}, suppose $\alpha_t$ and $\beta_t$ are iid for all $t$. Then, as $t\rightarrow\infty$, $\m{X}^{(t)}$ converges most slowly to $\m{1}\m\pi$ when $\mu$ is the arcsine distribution on $[0,1]$. Moreover, the influence vector $\m\pi=(\pi,1-\pi)\in[0,1]^2$ exists a.s., and $\pi$ is uniformly distributed on $[0,1]$.
\end{corollary}

Intuitively, the \textit{arcsine} (or Frostman) distribution has the smallest energy on $[0,1]$ \citep[][Theorem 3.1.(a)]{ull72}, 
and convergence time is inversely proportional to $I_{\mu}$.  $\pi\sim\text{Beta}(1,1)$ holds by Corollary \ref{thm:mic2} since the arcsine distribution is $\text{Beta}(\frac{1}{2},\frac{1}{2})$, so $\m\varphi=(1,1)$. 

\begin{ex}\label{ex:conv}\normalfont We now demonstrate that the effect of random dynamics on the speed of convergence is nuanced even in the simplest case where $n=2$. Consider the random influence vector $\m{\pi}=(\pi,1-\pi)$ in Section \ref{sec:setup}, which satisfies $\pi\sim \text{Beta}(\varepsilon s_i,\varepsilon (1-s_i))$. For ease, suppose $s_i=1/2$  so that the limiting distribution is symmetric $\text{Beta}(\frac{\varepsilon}{2},\frac{\varepsilon}{2})$, which happens when $\m{T}$ is either (i) $\big(\begin{smallmatrix}1/2&1/2\\1/2&1/2\end{smallmatrix}\big)$ or (ii) $\big(\begin{smallmatrix}
    a&1-a\\
    1-a&a
\end{smallmatrix}\big)$ for any $a\in(0,1)$. Let's analyze cases (i) and (ii) to identify settings where random dynamics can be beneficial or detrimental for social learning.
\begin{enumerate}
    \item[(i)] Suppose $\m{T}=\big(\begin{smallmatrix}1/2&1/2\\1/2&1/2\end{smallmatrix}\big)$, i.e., Ann and Bob agree before ever encountering each other. This network is well-connected and hence converges to consensus immediately. Thus, introducing random dynamics $\varepsilon>0$ will trivially slow down the speed of convergence, i.e., $\m{X}^{(t)}$ always converges to $\m{1}\m\pi$ at a slower rate than $\m{T}^{t}$ converges to $\m{1}\m{s}$. In fact, when $\varepsilon=2$, $\m{X}^{(t)}$ will experience its \textit{slowest} convergence rate (see, Corollary \ref{thm:conv2x22}).
    \item[(ii)] Suppose $\m{T}=\big(\begin{smallmatrix}
    a&1-a\\
    1-a&a
\end{smallmatrix}\big)$. For values of $a$ near $0$ or $1$, this network becomes sparse, so it converges very slowly. In such cases, there exist values of $\varepsilon$ for which the dynamic network $\m{X}^{(t)}$ converges at a much faster rate than the deterministic network $\m{T}^{t}$. 
\hfill$\bigtriangleup$ 
\end{enumerate}  
\end{ex}

\subsubsection{General Case: $n\geq2$}\label{sec:conv}
This section analyzes how social structure affects the speed of convergence. \citeauthor{moj12}'s (\citeyear{moj12}) homophily-based analysis does not extend well to dynamic settings due to the random evolution of connections. 
Thus, we propose a probabilistic measure of homophily in terms of how agents can get disconnected. The analysis from the $n=2$ case in Section \ref{sec:2x2} does not generalize for $n>2$, so we will build instead on \citeauthor{doubly13}'s (\citeyear{doubly13}) insights.
\par Consider the sequence
of random graphs $\{G_t\}_{t\geq1}$ defined as $G_t(\omega)=G\big(\m{X}_t(\omega)\big)$, for every $\omega\in\varOmega$. Let $\mathbb{G}_N$ denote the set of all undirected graphs on
the set of agents $N$. For a graph $H\in\mathbb{G}_N$, let $p_{H}$ denote the probability that $H$ occurs in a dynamic network.  Then, the \textit{realized} graph $\mathcal{G}$ is the set of all graphs that occur with strictly positive probability, i.e.,
\begin{align*}
    \mathcal{G}:=\big\{H\in \mathbb{G}_N:p_{H}>0\big\}.
\end{align*}
For a collection of graphs $\mathcal{H}\subseteq\mathbb{G}_N$, we denote by  $\Gamma(\mathcal{H})$ the \textit{accumulation} graph of $\mathcal{H}$, which contains all edges from all graphs in $\mathcal{H}$, i.e., a supergraph of every graph in $\mathcal{H}$ defined as
\begin{align*}
    \Gamma(\mathcal{H}):=\Big(N,\bigcup_{H\in \mathcal{H}}E(H)\Big),
\end{align*}
where $E(G)$ denotes the set of edges of a graph $G$. That is, $\Gamma(\mathcal{H})$ is the ``minimal'' graph.
\begin{definition}\normalfont
  A collection of realized graphs $\mathcal{H}\subseteq \mathcal{G}$ is said to be a \textit{disconnected} collection on $\mathcal{G}$ if its accumulation graph $\Gamma(\mathcal{H})$ is disconnected.   \hfill$\bigtriangleup$
\end{definition}
 In words, a disconnected collection is defined as any collection of realizable
graphs such that the union of all of its graphs yields a disconnected graph. The set of all possible disconnected collections on $\mathcal{G}$ is $\Pi(\mathcal{G}):=\big\{\mathcal{H}\subseteq\mathcal{G}:\mathcal{H}\text{ is a disconnected collection on }\mathcal{G}\big\}$. Then, the key quantity here is the probability of the \textit{most likely} disconnected collection: $$p_{\text{\normalfont max}}=\underset{\mathcal{H}\subseteq\Pi(\mathcal{G})}{\text{max}}\hspace{0.03in}p_{\mathcal{H}},$$ 
where $p_{\mathcal{H}}=\sum_{H\in\mathcal{H}}p_H$, so $p_{\text{\normalfont max}}$ is a worst-case measure of network fragmentation. 
\begin{ex}\normalfont Consider the graph $G=(N,E)$, where $(i,j)\in E$ occurs with probability $p_{ij}$ independently for all other agents. Let $G$ be connected and regular with degree $k=2,\dots,n-1$ and $p_{ij}:=p$. Then, $p_{\text{\normalfont max}}=1-\Proba(\text{agent } i \text{ is isolated})=(1-p)^k$.  \hfill$\bigtriangleup$\end{ex}
The next example illustrates how $p_{\text{\normalfont max}}$ relates to \citeauthor{moj12}'s (\citeyear{moj12}) measures.
\begin{ex}[Islands model]\normalfont
 Consider the following dynamic extension of the \textit{islands} model in \citet[][Section II.C]{moj12}.   There are two islands of equal size $g$. For tractability, assume $\{G_t\}_{t\geq1}$ is iid,  and to simplify $\Pi(\mathcal{G})$, assume each graph has at most one link between the two islands, and each graph has $g-1$ links within each island. Let $p_s:=\Proba\big((G_t)_{ij}=1\big|i,j \text{ in same island}\big)$ and $p_d:=\Proba\big((G_t)_{ij}=1\big|i,j \text{ in different islands}\big)$, so there is homophily when $p_s>p_d$. Then, $p_{\text{\normalfont max}}=1-p_d$, i.e., the probability that the islands are isolated. In the deterministic islands model, \citeauthor{moj12}'s (\citeyear[][eq. (2)]{moj12}) homophily measure is $h^{\text{islands}}=\frac{p_s-p_d}{p_s+p_d}$. Here, $h^{\text{islands}}$ reflects fragmentation as the difference in link probabilities, whereas $p_{\text{\normalfont max}}$ measures fragmentation as the probability of disconnectedness.  Notice that $h^{\text{islands}}$ and $p_{\text{\normalfont max}}$ both capture homophily: as $p_d$ decreases both measures increase. Importantly, Proposition \ref{thm:rate} reveals that $p_{\text{\normalfont max}}$ also governs the speed of convergence. \hfill$\bigtriangleup$
\end{ex}

\begin{proposition}\label{thm:rate}
    Let $\{\m{X}_t\}_{t\geq1}$ be iid and symmetric for all $t\geq1$. If, in addition, every $\m{X}_t$ has strictly positive diagonal entries a.s., then, for every $\epsilon\in(0,1]$,
    \begin{align*}
        -\underset{t\rightarrow\infty}{\text{\normalfont lim}}\hspace{0.03in}\frac{1}{t}\hspace{0.03in}\text{\normalfont log }\Proba\Big(\big\lVert\m{X}^{(t)}-\m{1}\m{1}'/n\big\lVert\geq\epsilon\Big)= \begin{cases}
            +\infty, & \text{\normalfont if }\Pi(\mathcal{G})=\emptyset \\
            |\text{\normalfont log}\hspace{0.03in}p_{\text{\normalfont max}}|, & \text{otherwise}.
        \end{cases}
    \end{align*}
\end{proposition}
A key insight is that when $n>2$, the main driver of convergence speed in dynamic networks is not the second largest eigenvalue but rather 
the 
\textit{Lyapunov exponent} $\Lambda=\underset{t\rightarrow\infty}{\text{ lim}}\big\lVert\m{X}^{(t)}-\m{1}\m{1}'/n\big\lVert^{1/t}$ a.s. \citep[][]{furst60}, where $\lVert.\lVert$ is the spectral norm. Since $|\text{\normalfont log}\hspace{0.03in}p_{\text{\normalfont max}}|$ is defined based on the worst fragmentation, it is a conservative and hence robust measure of homophily in dynamic networks because it is less sensitive to fluctuations in network connectivity. It can also serve as an upper bound for the adverse effect of homophily. 

\subsection{Comparative Statics: deterministic, stationary, iid networks}
We now want to compare different types of dynamic networks.  At one extreme, (1) there is the class of iid networks \citep[][]{iff08}. At the other extreme, (2) there is the class of deterministic networks \citep[][]{moj10}. (3) Our stationary networks therefore serve as an intermediate class of networks between these two extremes. As the interaction matrices become more correlated, our framework approaches (2), whereas when they become less correlated, our framework approaches (1). 
The rest of this section offers some new comparative statics by interpolating between these two extremes.
\begin{itemize}
    \item[(1)] For any deterministic network $\m{T}$, convergence time is proportional to $-1/\text{log}|\lambda_2(\m{T})|$.
    \item[(2)] For any iid network $\{\m{X}_t\}_{t\geq1}$ that satisfies the technical conditions in Assumption \ref{ass:rest} (Section \ref{sec:unique}), convergence time is roughly proportional to $-1/\text{log}|\lambda_2(\E_{\mu}[\m{X}_t])|$. This indicates that iid networks will typically converge faster than deterministic networks---the mean matrix $\E_{\mu}[\m{X}_t]$ has fewer zero entries than any $\m{X}_t$ because it averages nonnegative values and hence more mixing will occur, which then boots convergence speed.
    \item[(3)] For stationary networks, it is perhaps natural to guess that their convergence time will more or less lie somewhere between $-1/\text{log}|\lambda_2(\m{T})|$ and $-1/\text{log}|\lambda_2(\E_{\mu}[\m{X}_t])|$. As the next example shows, this intuition is not incorrect---there are natural settings where the speed of convergence in stationary networks is faster than in iid networks. 
\end{itemize}
 \begin{ex}\normalfont At each $t\geq1$, suppose every agent receives a revision opportunity with probability $\xi\in[0,1]$. When an agent receives a revision opportunity, she samples some of her neighbors, re-evaluates the weights that she assigned to them in the past, and updates these weights according to a random procedure.\footnote{This example can be interpreted as an evolutionary extension of the standard DeGroot learning inspired by the ``sampling best-response dynamics'' literature \citep[e.g.,][]{ryoji23}.} Without a revision opportunity, an agent keeps her last period weights.   
 Suppose all revisions form an iid sequence of random stochastic matrices $\{\m{\Xi}_t\}_{t\geq1}$. Then,  the interaction matrices $\{\m{X}_{t}\}_{t\geq1}$ evolve according to  \begin{align}\label{eq:ar1}
\m{X}_{t}=(1-\xi) \m{X}_{t-1}+\xi\m{\Xi}_{t},\end{align} 
for $t\geq1$, where $\m{X}_0=\m{T}$. This is reminiscent of an autoregressive process of order 1. 
(1) At one extreme, if $\xi=1$, then $\{\m{X}_t\}_{t\geq1}$ is iid (because $\{\m{\Xi}_t\}_{t\geq1}$ is iid). (2) At the other extreme, if $\xi=0$, then $\m{X}_t=\m{T}$ for all $t\geq0$ as in the standard DeGroot model. (3) For any $\xi\in(0,1)$, $\{\m{X}_t\}_{t\geq1}$ is stationary. Here, $\xi$ measures the \textit{persistence} (or stickiness) of weights over time, which allows us to interpolate between the two extreme networks. Notice that, by construction (eq. (\ref{eq:ar1})), the matrices in $\{\m{X}_{t}\}_{t\geq1}$ are $\xi$-mixtures of $\m{T}$ and $\{\m{\Xi}_t\}_{t\geq1}$. Thus, for moderate values of $\xi\in(0,1)$, the speed of convergence in the stationary process $\{\m{X}_{t}\}_{t\geq1}$ in eq. (\ref{eq:ar1}) is much faster than in the deterministic network $\m{T}$ and iid network $\{\m{\Xi}_t\}_{t\geq1}$. 
      \hfill$\bigtriangleup$
 \end{ex}
To summarize, this section has shown that the class of stationary networks is the most conducive for social learning because they are more flexible than deterministic and iid networks. 
 \subsection{Disagreement}\label{sec:disag}
To illustrate how random dynamics can cause disagreement, let's revisit Example \ref{ex:conv}.(ii), where $\m{T}=\big(\begin{smallmatrix}a&1-a\\1-a&a\end{smallmatrix}\big)$, for any constant $0<a<1$. Suppose we introduce the following local randomness around $\m{T}$. Let $\{\m{X}_t\}_{t\geq1}$ be an iid network where $\mu$ puts mass $a$ on $\big(\begin{smallmatrix}1&0\\0&1\end{smallmatrix}\big)$ and $1-a$ on $\big(\begin{smallmatrix}0&1\\1&0\end{smallmatrix}\big)$. Notice that $\E_{\mu}[\m{X}_t]=\m{T}$. Notably, 
\citet[][Section 2, 3rd case]{rao98} show that 
as $t\rightarrow\infty$, $\psi$---the limiting distribution  of $\m{X}^{(t)}$---exists, and it puts equal mass on $\big(\begin{smallmatrix}1&0\\0&1\end{smallmatrix}\big)$ and $\big(\begin{smallmatrix}0&1\\1&0\end{smallmatrix}\big)$, i.e., $\varSigma_{\psi}=\varSigma_{\mu}$ and $\E_{\psi}\big[\m{X}^{(\infty)}\big]=\frac{1}{2}\m{11}'=\m{T}^{\infty}$ for any $0<a<1$.  
\par\noindent--- \textit{Interpretation}: In the long-run, Ann and Bob agree on average as in the deterministic network $\m{T}$ but disagree almost surely due to random dynamics. 
A formal analysis of disagreement requires more technical details, so it is relegated to Online Appendix \hyperref[sec:disagree]{A}. There, we study different degrees of disagreement based on the \textit{cyclicity} of interaction patterns.
\section{Discussions: some intuitions and other remarks}\label{sec:dis}
We conclude with some discussions of our model that may provide insights for future research. Sections \ref{sec:disc}, \ref{sec:disc2}, and \ref{sec:stat} discuss, respectively, our framework, consensus, and stationarity.

\subsection{Framework: microfoundation}\label{sec:disc}
Let's explore a microfoundation of our framework based on the dynamic network formation literature \citep{moj02,moj022}. This literature is concerned with \textit{myopic} agents who form, maintain, and sever links connecting themselves to other agents over time based on the improvement that the resulting network offers them relative to the current network. Instead of forming/severing links, our framework allows agents to change the ``intensity'' of friendships as in \citet{bloch09}. They assume agents have limited attention or resource (e.g., time) to invest in each relationship,  which, in our framework, translates into the restriction that each agent's weights sum to one, i.e., $\sum_{j\in N}(\m{X}_t)_{ij}=1$, for every $i$ and $t$. \par \citet{moj02} propose a general framework where the network evolution can be random according to a stationary Markov chain. 
 Notice that when Conditions (A) and (B) in \citet[][Section 1.4, p. 1549]{hen97} are assumed in our framework,  
 the dynamic network $\{\m{X}_t\}_{t\geq1}$ can also be viewed as a finite stationary ergodic Markov chain on $\varSigma$. \par \citeauthor{moj02}'s (\citeyear{moj02}) framework is based on the idea that agents' payoffs depend on the network of connections among agents. However, they do not impose any restrictions on such payoffs, so let's consider the following setup. Suppose agent $i$'s utility function is 
 \begin{align}\label{eq:util}
     U_{it}\Big(p^{(t)}_i,\{(\m{X}_t)_{ij}\}_{j\in N},\big\{p^{(t-1)}_j\big\}_{j\in N}\Big)=\underbrace{-\alpha_i\Big(p^{(t)}_i-\sum_{j\in N}(\m{X}_t)_{ij}p^{(t-1)}_j\Big)^2}_{(\text{i})}\underbrace{-\beta_i\sum_{j\in N}c^{(t)}_{ij}}_{(\text{ii})},
 \end{align}
 where $\alpha_i\geq0$ and $\beta_i\geq0$ are constants for all $i\in N$. We discuss (i) and (ii) below. 
 \begin{itemize}
     \item[(i)] This is a \textit{conformity}-bias term. It captures the fact that, at every $t$, agents prefer to hold beliefs that align closely with those of their peers. Such a term also appears in the microfoundation of the standard DeGroot model \citep[][p. 521]{golub17}.  In dynamic settings, this form of conformity bias is consistent with \citet{moj022}, where agents also benefit from coordinating their actions.
     \item[(ii)] This is a costly-interaction term. $c^{(t)}_{ij}$ is a function that captures the fact that it may be costly for agent $i$ to maintain a relationship with agent $j$ \citep{bloch09}.
 \end{itemize}
 At every $t$, each agent $i$ chooses $p^{(t)}_i$ that maximizes the utility $U_{it}$ in eq. (\ref{eq:util}). Taking the derivative of $U_{it}$ with respect to $p^{(t)}_i$ and setting it to zero, we obtain $p^{(t)}_i=\sum_{j\in N}(\m{X}_t)_{ij}p^{(t-1)}_j$, for every $t\geq1$, which is precisely the $i$-th component in eq. (\ref{eq:evol}).\footnote{This setup implicitly assumes myopic best-reply dynamics. This is the standard assumption in the DeGroot literature to explain agents' na{\"i}ve learning behavior \citep[e.g.,][p. 521]{golub17}.  } 
 \par \citet{moj02} argue that agents adjust their connections over time based on the payoffs they receive from their current network positions. To incorporate this in our framework, suppose each agent $i$'s weights evolve as follows $(\m{X}_t)_{ij}\propto(\m{X}_{t-1})_{ij}+\eta^{(t)}_i\frac{\partial U_{it}}{\partial (\m{X}_{t-1})_{ij}}$, for every $j\in N$ and $t\geq1$. Here, $\big\{\eta^{(t)}_i\big\}_{t\geq1}$ is a random process that measures how sensitive agent $i$ is to changes in utility due to small adjustments in weights. The standard DeGroot model is recovered, for example, when $\eta^{(t)}_i=0$ or $\frac{\partial U_{it}}{\partial (\m{X}_{t-1})_{ij}}=0$ for all $i,j\in N$ and $t\geq1$.\footnote{\citeauthor{moj02}'s (\citeyear{moj02}) model allows the possibility that networks may evolve randomly due to unintended errors. Consider, for example, \citeauthor{kahn21}'s (\citeyear{kahn21}) \textit{occasion noise}. \citet[][Chapter 7]{kahn21} report that people's judgments are subject to behavioral biases, which lead people to random errors when repeating the same task over time---here the task is to assign weights to neighbors' opinions. They note: ``The variability we observe in judgments of the same problem by the same person is not a fluke observed in a few highly specialized problems: occasion noise affects all our judgments, all the time.''}

\subsection{Consensus: tail condition and special cases}\label{sec:disc2}
\subsubsection{Tail Condition}\label{sec:C}
Our tail condition \eqref{eq:C} is intuitive---it 
requires that all agents eventually pay attention to their own and others' opinions.  
Interestingly, \eqref{eq:C} has yet to be relaxed in random matrix theory. 
 \citet[][Section 1.6]{hen97} anticipated this by noting: ``As far as I know, stationarity and
 \eqref{eq:C} are the weakest conditions that have been considered when dealing with
 random products of positive matrices in order to establish limit theorems.'' Hence, \eqref{eq:C} is the standard assumption, e.g., \citet[][$\text{H}_1$]{kes84}, \citet[][Proposition 2]{cohn93}, \citet[][\eqref{eq:C}]{hen97}, \citet{hen08}, and \citet[][(C2)]{mic14}. 

\par Computing the distribution of the influence vector $\m\pi$ in closed form remains an open problem in the literature (Online Appendix \hyperref[sec:ex2]{B.II}). Even in the simplest case where $n=2$ and the network is iid, all that is known is that the distribution of $\m\pi$ is a solution to an integral equation \citep[][Theorem 1, eq. (2.1)]{va86}, which is only solvable in very few cases. 

\begin{remark}[]\label{rem:allow}\normalfont
     \citet{hen97} assumes $\varSigma$ is the set of \textit{allowable} nonnegative matrices \citep[][Definition 3.1]{seneta81}, i.e., every row and column contains at least one strictly positive entry. For stochastic matrices, only the column constraint is restrictive. The proof of \citet[][Proposition 2]{cohn93} shows that stationarity and \eqref{eq:C} imply that $\{\m{X}_t\}_{t\geq1}$ are allowable almost surely. This restriction is not practically restrictive because it rules out only networks containing agents that nobody (including themselves) listens to. 
    \hfill$\bigtriangleup$
\end{remark}

\subsubsection{A Key Special Case}\label{sec:unique}

The following assumption---due to \citet{iff09}---will be useful. 
\begin{assumption}\label{ass:rest}\normalfont
Set $\varSigma$ and $(\varOmega,\mathscr{F},\Proba,\theta)$, respectively, equal to $\Tilde{\varSigma}$ and $(\Tilde{\varOmega},\Tilde{\mathscr{F}},\Tilde{\Proba},\Tilde{\theta})$ where:
\begin{enumerate}
    \item $\Tilde{\varSigma}=\big\{\text{set of stochastic matrices with strictly positive \textit{diagonal} entries}\big\}$;
    \item $\Tilde{\varOmega}=\big\{\omega=(\omega_1,\omega_2,\dots):\omega_j\in\Tilde{\varSigma}\text{ for all }j\in\mathbb{N}\big\}$;
    \item $\Tilde{\mathscr{F}}=\mathcal{B}\times\mathcal{B}\times\cdots$, where $\mathcal{B}$ is the Borel sigma-algebra on $\Tilde{\varSigma}$;
    \item $\Tilde{\theta}:\Tilde{\varOmega}\rightarrow\Tilde{\varOmega}$ is the \textit{shift map}, i.e., the ergodic map $\Tilde{\theta}(\omega_1,\omega_2,\dots)=(\omega_2,\omega_3,\dots)$. \hfill$\bigtriangleup$ 
\end{enumerate}
\end{assumption}
Under this assumption, eq. (\ref{eq:DGP}) becomes $\m{X}_t(\omega)=\m{X}_0\circ\Tilde{\theta}^{t}(\omega)=\omega_{t+1}$, for $t\geq0$, where $\m{X}_0:\Tilde{\varOmega}\rightarrow\Tilde{\varSigma}$ is the first-coordinate map $\m{X}_0(\omega)=\omega_1$, for any $\omega=(\omega_1,\omega_2,\dots)\in\Tilde{\varOmega}$, and the space $\Tilde{\varOmega}$ is now a bit more intuitive because it is now the set of all paths. Thus, \citeauthor{iff09}'s (\citeyear{iff09}) framework is a special case of our framework when Assumption \ref{ass:rest} is imposed. In this special case, consensus has a simple graph theoretic interpretation: it is equivalent to the
expected graph of the network containing a directed spanning tree. 
\par The most restrictive condition in Assumption \ref{ass:rest} is \ref{ass:rest}.1 (Remark \ref{rem:diag}). It rules out  the standard DeGroot model which allows zeros on the diagonal \citep[][Example 2]{moj10}. It also rules out a large class of settings where interaction matrices have zeros on the diagonal. Popular instances arise when networks are modeled using \textit{directed acyclic graphs},  which are graphs that do not contain self-loops (e.g., citation graphs), so their interaction matrices have zero diagonal entries. 
Thus, our framework complements \citet{iff09} in applied settings where the latter is silent. 
The key takeaway from their main result, \citet[][Theorem 3]{iff09}, is that consensus is completely determined by the average network in the sense that consensus is equivalent to $|\lambda_2(\E_\mu[\m{X}_t])|<1$. In contrast, our framework shows that when their assumptions (Assumption \ref{ass:rest}) are relaxed, the average network no longer suffices for consensus. For example, Section \ref{sec:disag} analyzed a simple dynamic network where $|\lambda_2(\E_\mu[\m{X}_t])|=|1-2a|<1$, yet consensus is never reached for any $a\in(0,1)$. Notice that \eqref{eq:C} is violated in this example because $\varSigma_{\mu}=\big\{\big(\begin{smallmatrix}1&0\\0&1\end{smallmatrix}\big),\big(\begin{smallmatrix}0&1\\1&0\end{smallmatrix}\big)\big\}$, so  $$|\lambda_2(\E_\mu[\m{X}_t])|<1 \text{ does not imply } \eqref{eq:C}.$$ Thus, our condition \eqref{eq:C} serves as a more suitable predictor of consensus in general classes of dynamic networks because it disciplines the tail behavior instead of just the average. 
\begin{remark}\label{rem:diag}\normalfont
    The assumption that $\{\m{X}_t\}_{t\geq1}$ have strictly positive diagonal entries in Assumption \ref{ass:rest}.1 is technically restrictive. This is because, in the iid case, $\big\{\m{X}^{(t)}\big\}_{t\geq1}$ converges weakly  when $\m{X}_1$ has at least $n-1$ strictly positive diagonal entries (see, Proposition \ref{thm:zero}). 
\citet{touri13} show that Assumption \ref{ass:rest}.1 implies a condition called ``strong aperiodicity.'' 
\hfill$\bigtriangleup$
\end{remark}

\subsection{Stationarity}\label{sec:stat}
 Thus far, we have assumed that the dynamical system that generates the interaction matrices is stationary. The next result shows that stationarity is not very restrictive---a unique invariant probability measure exists in a more abstract space under very mild assumptions.

\begin{proposition}\label{thm:stat}
   Let $S$ be a compact metric space, and $\big\{C_t\big\}_{t\geq1}$ be a countable collection of contraction maps from $S$ to itself. Then, there exists a unique invariant probability measure $\mu$ on the Borel subsets of $S$ that satisfies the following fixed-point equation
    \begin{align*}
        \mu(B)=\sum_{t\geq1}\rho_t\hspace{0.03in}\mu\big(C^{-1}_t(B)\big),
    \end{align*}
    for any Borel set $B\subset \varSigma$, where $\sum_{t\geq1}\rho_t=1$ and $\rho_t\geq0$, for all $t$. Moreover, the support of $\mu$ is the unique compact set $S_\mu\subset S$ that satisfies $S_\mu=\text{\normalfont closure}\big(\bigcup_{t\geq1}C_t(S_\mu)\big)$.
\end{proposition}

The next example is a simple illustration of how Proposition \ref{thm:stat} applies in our framework.

\begin{ex}\normalfont
 Let $S=\varSigma$ and $\m{X}_1,\m{X}_2,\dots$ be an iid sequence of random stochastic matrices taking values in the set of contraction maps $\{C_t\}_{t\leq\tau}$  such that $\Proba(\m{X}_1=C_t)=\rho_t\geq0$, for $t=1,\dots,\tau$, $\sum_{t\leq\tau}\rho_t=1$, and any $\tau\geq1$. The support $S_\mu=\varSigma_\mu$ in Proposition \ref{thm:stat}  becomes 
  \begin{align*}
   \varSigma_\mu=\Big\{{\sigma}\in\varSigma: \Proba\big(\m{X}^{(\tau)}\in\mathcal{O}({\sigma})\text{ i.o.}\big)=1,\forall \text{ open set 
 }\mathcal{O}({\sigma})\ni {\sigma}\Big\},
  \end{align*}
which is often referred to as the \textit{attractor} of the contracting system $\{C_t\}_{t\leq\tau}$ and ``i.o.'' stands for ``infinitely often.'' That is, $\varSigma_\mu$ is the smallest closed subset of $\varSigma$ such that $\mu(\varSigma_\mu)=1$.
\hfill$\bigtriangleup$
\end{ex}

\section{Concluding Remarks}\label{sec:conc}

\subsection{Related Literature}\label{sec:lit}
 This paper builds on the na{\"i}ve social learning literature  (or \textit{non}-Bayesian learning),\footnote{In contrast, agents in Bayesian social learning models use Bayes rule to update their beliefs  \citep[][]{bala98,daron11}. These models are considered the ``gold standard'' for
understanding rational learning. However, they assume a lot
of rationality on the part of agents, so they are viewed as models of how people \textit{should} learn rather than how people actually \textit{do} learn. Importantly, Bayesian models are
complex except for simple examples making them prohibitive in real-world settings \citep{golub19}. } which dates back to \citet[][]{french56}. The DeGroot model \citep{degroot74}, named after the statistician Morris H. DeGroot, is the canonical model of decentralized communication among na{\"i}ve agents. It consists of agents who acquire noisy signals of the true state of the world and na{\"i}vely update their beliefs by repeatedly taking (the \textit{same}) weighted averages of their neighbors’ beliefs, i.e., the network is exogenously fixed over time. Aside from its elegance and simplicity, this model is also popular because there is considerable experimental evidence that it best fits the aggregation rules of most people in practice \citep[e.g.,][]{test12,exp15,exp20,test20}.
\par \citet{dem03} apply the DeGroot model in social networks, where bounded rationality (e.g., persuasion bias) is used to justify why agents use the same weights over time. The more recent literature builds on \citeauthor{moj10}'s (\citeyear{moj10}) framework. For instance,  \citet{nonbayes18} propose an axiomatic foundation of the DeGroot model. \citet{banerjee21} extend the DeGroot model to settings where some agents may be initially uniformed.  \citet{ben23} study settings where the state of the world changes over time. \cite{jet23} allow  homophily and influence to co-evolve endogenously. \citet{ban232} use the insights of \citet{fri90} to introduce noisy information transmissions in social networks and study when consensus and disagreement would arise in strategic settings. \citet{nonbayes23} consider applications of updating rules represented by nonlinear opinion aggregators. Notably, all these papers assume that the underlying social network is exogenously fixed over time.
\par \citet{bernd19} rely on a stylized framework to obtain the negative answer noted in the Introduction. Online Appendix \hyperref[sec:disc3]{C} shows 
that their failures of the wisdom of crowds occur for at least one technical reason: they assume the support of the distribution that generates the random interaction matrices is finite. We find that in such cases, there exist examples where even consensus is unlikely to exist. We then show that when this finiteness restriction is relaxed, consensus exists on \textit{average}, irrespective of the network size.

\subsection{Conclusion}\label{sec:conc1}
We have introduced random network dynamics in DeGroot learning to study their effects on social learning. 
We find that random dynamics have double-edged effects on social learning: they can either be beneficial or detrimental depending on social structure. Consistent with experimental evidence, we show that introducing even small amounts of random dynamics can boost collective intelligence.   
Our framework builds on random matrix theory, which allows it to nest most existing models. 
It offers a new spectrum of dynamic networks whose extremes are the deterministic and iid networks. We leverage this to examine the types of dynamic networks are that are most conducive for social learning. We also propose a robust measure of homophily based on the likelihood of the worst possible network fragmentation. 
\par We then show that the initial social structure of a dynamic network plays a key role in shaping long-run beliefs. 
Our results can therefore provide guidance to the literature that relies on insights from the DeGroot model (e.g., centrality measures) to carry out interventions in social networks. Our appendices contain further analyses and details. 
Appendix \hyperref[app:tech]{A} shall serve as a review of key mathematical concepts.  Online Appendix \hyperref[sec:disagree]{A} shows that the presence of random dynamics suffices to explain both consensus and  disagreement.

\phantomsection\label{app:tech}
\section*{Appendix A: Review of Technical Details}

This appendix reviews the key mathematical details that are needed to formalize our framework. The key concept will be the multiplicative semigroup structure of nonnegative matrices. Appendix  \hyperref[app:prelim]{A.I} starts with preliminary definitions, followed by algebraic details about the semigroup and convergence of products of random stochastic matrices in Appendices \hyperref[app:setup]{A.II}--\hyperref[app:conv]{A.III}. Then, Appendix \hyperref[app:sketch]{A.IV} builds on these tools to outline a sketch of Theorem \ref{thm:skel}.

    \phantomsection\label{app:prelim}
    \subsection*{A.I Preliminaries}
    
    Section \ref{sec:dgp} introduces an abstract dynamical system $(\varOmega,\mathscr{F},\Proba,\theta)$  
    and $\varSigma$ as the set of all $n\times n$ stochastic matrices. The only restriction on this dynamical system is ergodicity. This might be very abstract, however, so, to build intuition (e.g., Assumption \ref{ass:rest}), $\varOmega$ can be viewed as a measurable space given by $\varSigma^{\infty}$ whose elements $\omega=\{\omega_t\}_{t\geq1}$ are infinite sequences of elements in $\varSigma$, equipped with the 	sigma-algebra $\mathscr{F}$ generated by coordinate mappings $\omega\mapsto \omega_t$ and the Borel sets of $\varSigma$. The map $\theta$ is often chosen to be an automorphism on  $\varOmega$. The \textit{iterates} of $\theta$ are defined by induction as follows $\theta^0:=\text{Id}$ and $\theta^t:=\theta\circ \theta^{t-1}$, for $t\geq1$.
    
 \par The following definitions are based on \citet[][Chapter 1]{ergo82}. Given a probability measure $\mu$, a measurable map $\theta : \varOmega \rightarrow \varOmega$ is measure-preserving if $\mu(\theta^{-1}(A))=\mu(A)$ for all $A\in\mathscr{F}$ and $\mu$ is said to be \textit{invariant} to $\theta$. A set $A\in\mathscr{F}$ is invariant if $\theta^{-1}(A) = A$. The set of all invariant sets forms a sigma-algebra denoted $\mathscr{F}_{\theta}$. The map $\theta$ is \textit{ergodic} if $\mathscr{F}_{\theta}$ is trivial, i.e., it
contains only sets of measure zero and their complements. A simple example of an ergodic map is the shift map: $\theta(\omega_1,\omega_2,\dots)=(\omega_2,\omega_3,\dots)$ (Assumption \ref{ass:rest}.4). A probability measure $\mu$ invariant with respect to this shift map is said to be \textit{stationary}: $$\mu\big(\{\omega:\omega_{i_1+k}\in A_1,\dots,\omega_{i_{r}+k}\in A_{r}\}\big)=\mu\big(\{\omega:\omega_{i_1}\in A_1,\dots,\omega_{i_{r}}\in A_{r}\}\big),$$ for all $k\geq0$. Another example of an ergodic map is the so-called \textit{circle rotation} defined as $\theta(\omega)=\omega+\alpha \hspace{0.03in}(\text{mod }1)$, where $\varOmega=[0,1)$ is equipped with the Lebesgue measure $m$ (i.e., here $\mu=m$), and $\alpha\in [0,1)$ is an irrational number. More generally, if $f$ is integrable and $\theta$ is measure-preserving, then it is well-known that $f\circ\theta$ is integrable and $\int_{\varOmega}f \hspace{0.03in}d\mu=\int_{\varOmega}f\circ\theta \hspace{0.03in}d\mu$.

    \phantomsection\label{app:setup}
    \subsection*{A.II Probability Measures on Semigroups}
    
   This appendix reviews the semigroup of stochastic matrices. Throughout, we follow the existing literature closely by focusing on the iid case. This is because, as noted earlier, the characterization of the non-iid case remains an open problem in random matrix theory. \par Recall that $\mu$ denotes a probability measure on the Borel subsets of $n\times n$ stochastic matrices, i.e., $\mu(B)=\Proba(\m{X}_t\in B)$, for any Borel set $B\subset\varSigma$ (with
the topology induced by the standard metric on $\R^{n^2}$), and $\varSigma_\mu$ denotes its support. Since the $\m{X}_t$'s are iid, $\Proba(\m{X}^{(t)}\in B)=\mu^{t}(B)$, where, for any $\ell\in\mathbb{N}$, $\mu^{\ell}$ denotes the $\ell$-th \textit{convolution power} of $\mu$. For example, when $\ell=2$, then $\mu^{2}=\mu*\mu$, where $*$ denotes the convolution product of any pair of probability measures.  Equivalently, $\mu^{t}=\overbrace{\mu*\dots*\mu}^{t\text{-times}}$ can be defined recursively for all $t\geq1$ as follows $$\mu^{t+1}(B)=\int_{\varSigma} \mu^{t}\big(B\sigma^{-1}\big)\hspace{0.03in}d\mu(\sigma),$$ 
where $B\sigma^{-1}=\{x\in \varSigma:x\sigma\in B\}$, for any Borel set $B\subset\varSigma$ \citep[see,][Section 2]{muk97}.\footnote{For any two probability measures $\mu,\nu$ on $\varSigma$, $\mu*\nu(B)=\int_{\varSigma} \mu(B\sigma^{-1})\hspace{0.03in}d\nu(\sigma)=\int_{\varSigma} \nu(\sigma^{-1}B)\hspace{0.03in}d\mu(\sigma)$, where given the above notation, $\sigma^{-1}B=\{x\in \varSigma:\sigma x\in B\}$, for any Borel set $B\subset\varSigma$.} 
The product measure on $\varOmega$ can therefore be denoted as $\Proba_{\mu}$.  More importantly, if the starting point of the process---$\m{X}_0$ in eq. (\ref{eq:DGP})---lies in $\varSigma_\mu$,\footnote{A standard assumption is that the process starts at the identity element of $\varSigma$ (adjoined if need be).} then it is well known that the sequence of random products $\big\{\m{X}^{(t)}\big\}_{t\geq1}$ will $\Proba_{\mu}$-a.s. never leave the \textit{closed multiplicative semigroup} of $n\times n$ stochastic matrices generated by $\varSigma_\mu$ defined as
\begin{align*}
\text{closure}\Bigg(\bigcup_{t=1}^{\infty}\varSigma^t_{\mu}\Bigg)=\text{closure}\Bigg(\bigcup_{t=1}^{\infty}\varSigma_{\mu^t}\Bigg),
\end{align*}
which holds by independence,\footnote{For any two probability measures $\mu$ and $\nu$, $\varSigma_{\mu*\nu}=\text{closure}(\varSigma_{\mu}\cdot\varSigma_{\nu})$, where $\cdot$ is matrix multiplication.} where $\varSigma_{\mu^t}=\text{closure}\big(\big\{\m{X}^{(t)}:\m{X}_\ell\in\varSigma_{\mu},1\leq \ell\leq t\big\}\big)$ and $\varSigma_\mu=\big\{\sigma\in\varSigma:\mu(\mathcal{O}(\sigma))>0, \forall\text{ open set }\mathcal{O}(\sigma)\ni \sigma\big\}$. It therefore suffices to set
\begin{align}\label{eq:sigma}
\varSigma=\text{closure}\Bigg(\bigcup_{t=1}^{\infty}\varSigma^t_{\mu}\Bigg)
\end{align}
\citep[see,][eq. (5)]{rosen65}, in which case, $\varSigma$ is said to be \textit{generated} by $\varSigma_\mu$. That is, when $\varSigma$ is defined as in eq. (\ref{eq:sigma}), $\Proba\big(\m{X}^{(t)}\notin \varSigma \text{ for some }t\geq1\big)=0$. 
The restriction in eq. (\ref{eq:sigma}) is standard in the literature on random products of nonnegative matrices \citep[e.g.,][]{rosen65,muk87,mukherjea91,muk97}. 
\par Since we are in the iid case, the product $\big\{\m{X}^{(t)}\big\}_{t\geq1}$ is synonymous to a random walk whose state space is the semigroup $\varSigma$. That is, each $\mu$ on $\varSigma$ defines a Markov process on $\R^n$ via random iterations as follows: if we are at a point $\m{p}\in\R^n$ and  select a matrix $\sigma\in\varSigma$ according to $\mu$, we would move to $\m{q}=\sigma\m{p}\in\R^n$. The set $\varSigma$ can also be described as the support of the probability measure $\sum_{t=1}^{\infty}\frac{1}{2^t}\mu^t$, i.e., any open set in $\varSigma$ has $\mu^t$-measure for some $t$. Moreover, the subsemigroup $\hat{\varSigma}$, the set of strictly positive stochastic matrices, is an ideal of $\varSigma$, i.e., for any $\hat{\sigma}\in\hat{\varSigma}$ and $\sigma\in\varSigma$, their product satisfies $\sigma\hat{\sigma}\in \hat{\varSigma}$. Consequently, $\hat{\varSigma}$ is stochastically closed for the random walk $\big\{\m{X}^{(t)}\big\}_{t\geq1}$. It also holds that when $\varSigma_\mu$ is a countable set, then so is $\varSigma$. Since the matrices are stochastic, $\varSigma$ is a compact Hausdorff topological semigroup (with respect to standard matrix multiplication), so the literature typically assumes $\varSigma$ is second countable \citep[see,][]{muk87,mukherjea91}.\footnote{This assumption is not restrictive. For instance, let $M_n$ denote the set of $n\times n$ real matrices (i.e., all entries are real numbers). Under matrix multiplication, this set forms a semigroup. If we define a norm on $M_n$ according to the Euclidean distance on $\R^{n^2}$ (i.e., the Frobenius norm), then $M_n$ becomes a locally compact second-countable topological space with jointly continuous multiplication. }

\phantomsection\label{app:conv}
\subsection*{A.III Weak Convergence of Convolutions}

\par We are mainly interested in the convergence properties of the product $\m{X}^{(t)}$ as $t\rightarrow\infty$. This product converges weakly (or in distribution) whenever $\mu^{t}$ converges weakly as $t\rightarrow\infty$. Specifically, $\m{X}^{(t)}$ converges weakly if and only if there exists a probability measure $\psi$ that is the weak limit of the sequence $\{\bar{\mu}_t\}_{t}$ as $t\rightarrow\infty$, where $\bar{\mu}_t$ denotes the Ces\`aro average $$\bar{\mu}_t=\frac{1}{t}\sum_{\ell=1}^t\mu^\ell,$$
\citep[e.g.,][Theorem 2.1]{mukherjea91}. Algebraically, $\psi$ is the unique solution to the convolution equation $\psi=\psi*\mu=\mu*\psi$ (and $\psi=\psi*\psi)$. Let ${\varSigma_{\psi}}$ be the support of $\psi$. It is well-known that ${\varSigma_{\psi}}$ coincides with the (completely simple) \textit{kernel} of $\varSigma$, denoted $K$:\footnote{Suppose $\varSigma$ contains a rank-one matrix $\sigma$. On one hand, since $\sigma$ is a stochastic matrix and has identical rows, it forms by itself a minimal left ideal $\{\sigma\}=\varSigma \sigma$. On the
other hand, $\sigma\varSigma$ is the set of all stochastic matrices of rank one and hence is the minimal two-sided ideal of $\varSigma$. This set is by definition the kernel of $\varSigma$.}
\begin{align}\label{eq:kernel}
    K=\Big\{y\in\varSigma:\text{rank}(y)\leq\text{rank}(\sigma),\forall \sigma\in\varSigma\Big\},
\end{align}
so $K={\varSigma_{\psi}}$ is the support of $\psi$. Eq. (\ref{eq:kernel}) states that the elements in ${\varSigma_{\psi}}$ are all the stochastic matrices in $\varSigma$ with minimal rank \citep[][Proposition 2.3]{clark65,mukherjea91}. To fix ideas, the next example illustrates how to find the kernel $K$ in very simple cases.
\begin{ex}\normalfont\label{ex:2x2id}
    Let $n=2$, so the semigroup $\varSigma$ consists of stochastic matrices of the form $\big(\begin{smallmatrix}
    a&1-a\\b&1-b
\end{smallmatrix}\big)$, where $a,b\in[0,1]$. There is a one-to-one correspondence between such stochastic matrices and the points $(a,b)$ in the unit square. Then, the product of any two matrices $(a,b)$ and $(a',b')$, respectively, corresponds to the point $\big(aa'+b'(1-a'),ba'+b'(1-b)\big)$.\footnote{Intuitively, multiplication from the left with $(a,b)$ always moves a point $({x},{y})$ in the same direction with step length being a multiple of $x-y$.} The kernel $K$ of $\varSigma$ (eq. (\ref{eq:kernel})) in this case is therefore isomorphic to the diagonal $a=b$. If in addition, $a=1-b$, i.e., $\varSigma$ consists of bistochastic matrices, then $\varSigma$ is a compact abelian semigroup and hence  the kernel becomes the singleton $\big\{\big(\begin{smallmatrix}
    1/2&1/2\\1/2&1/2
\end{smallmatrix}\big)\big\}$. 
Alternatively, suppose $\mu$ puts all its mass on only two points $(a,a+d)$ and $(a',a'+d)$, where $d < 1/3$. Then, $K$ is a Cantor-like compact subset of the diagonal $\big\{(x,x):\forall x\in[0,1]\big\}$.\hfill$\bigtriangleup$
\end{ex}
Thus, when the minimal rank of $\varSigma$ is one, it indicates that agents will reach (a random) consensus when their beliefs converge. Since each stochastic matrix in ${\varSigma_{\psi}}$ has identical rows under consensus, ${\varSigma_{\psi}}$ can be viewed as a subset of $\R^{n-1}$ \citep{mukherjea02}, and hence the influence vector $\m\pi=(\pi_1,\dots,\pi_n)$ is understood to be distributed according to $\psi$. 

\begin{remark}
 \normalfont  A necessary and sufficient condition for the existence of $\psi$ is: $\underset{t\rightarrow\infty}{\text{lim inf}}\hspace{0.03in}\varSigma^t_{\mu}\neq \emptyset$ \citep[see,][Theorem 2.1]{mukherjea91}, where this set is defined as $$\underset{t\rightarrow\infty}{\text{lim inf}}\hspace{0.03in}\varSigma^t_{\mu}=\Big\{\sigma\in\varSigma:\forall\text{ open set }\mathcal{O}(\sigma)\ni \sigma, \exists k\in\mathbb{N}\text{ s.t. }t\geq k\implies\mathcal{O}(\sigma)\cap\varSigma^t_{\mu}\neq\emptyset\Big\}.$$
If $\varSigma$ does not contain a rank-one matrix, $\psi$ may not exist \citep[][Remark 1]{mukherjea02}. When $\varSigma_\mu$ contains only invertible matrices and $\varSigma$ contains a rank-one matrix, \citet[][Theorem 3]{muk97} shows that $\psi$ can only be \textit{one} of the following:
\begin{enumerate}
    \item discrete, i.e., $\psi(E)=1$ for some countable set $E$;
   \item continuous and singular with respect to the $n-1$ dimensional Lebesgue measure on ${\varSigma_{\psi}}$ (viewed as a subset of $\R^{n-1}$) denoted $m_{n-1}$;
   \item absolutely continuous with respect to $m_{n-1}$. \hfill$\bigtriangleup$
\end{enumerate}
\end{remark}
\phantomsection\label{app:finite}
The remark above has therefore completely characterized  ($\psi,\varSigma_{\psi}$) in the iid case. When the rank of the matrices in $\varSigma_{\psi}$ (the kernel of $\varSigma$) is $n$, $\varSigma_{\psi}$ is said to be \textit{cancellative}, which makes it a group, and therefore, $\varSigma=\varSigma_{\psi}$ is a compact group of $n\times n$ stochastic matrices of \textit{full rank}, hence $\varSigma$ is also finite \citep[][Section 1]{muk07}. In this case, convergence results of the convolution sequence $\{\mu^t\}_{t\geq1}$ are well-known (see, Lemma \ref{thm:finite}).
\phantomsection\label{app:fail}

\phantomsection\label{app:sketch}
\subsection*{A.IV Sketch of Theorem \ref{thm:skel}}
 
The proof of Theorem \ref{thm:skel} will rely on the algebraic structure of the kernel $K$ in eq. (\ref{eq:kernel}). As noted earlier, Theorem \ref{thm:skel}.(ii) implies Theorem \ref{thm:skel}.(i), so we will prove the former in the main text and the latter in Online Appendix \hyperref[thm21]{D} since the latter is mostly based on the results in \citet{muk05}. We outline the key steps below.

\begin{enumerate}
    \item \citet[][Section 9.2: Remarks]{hen97} presents two equivalent formulations of \eqref{eq:C} in the iid case. Specifically, we focus on the following equivalence (see, Proposition \ref{thm:equiv})
    \begin{align}\label{eq:equiv}
        \eqref{eq:C} \text{ holds if and only if there exists }t\geq1 \text{ such that }\int  c(\sigma)\hspace{0.03in}d\mu^t(\sigma)<1,
    \end{align}
    where $c(.)$ is a continuous function on $\varSigma$ \citep[][Lemma 10.8]{hen97}. By \citet[][Lemma 10.6.(iii)]{hen97}, $c(\sigma)\leq1$ for any $\sigma\in\varSigma$, and $c(\sigma)<1$ for all $\sigma\in\hat{\varSigma}$. 
    \item Eq. (\ref{eq:equiv}) indicates that, to understand condition \eqref{eq:C}, we only need to characterize the convolution power $\mu^t$ and its support for large $t$. Specifically, when  eq. (\ref{eq:equiv}) holds, \eqref{eq:C} is satisfied, so Theorem \ref{thm:consensus} indicates that $\m{X}^{(t)}$ converges almost surely, and hence its distribution $\mu^t$ converges weakly to $\psi$ with support $\varSigma_{\psi}=K$ in eq. (\ref{eq:kernel}). As noted earlier, $K$ is the kernel of $\varSigma$ in eq. (\ref{eq:sigma}), which contains matrices in $\varSigma$ with the minimal rank. Importantly, $K$ is a \textit{completely simple} semigroup \citep[e.g.,][Section 2]{muk87},  and therefore it is topologically \textit{isomorphic} to (i.e., can be identified with) the so-called \textit{Rees-Suschkewitsch} (or simply \textit{Rees}) product decomposition
    \begin{align}\label{eq:rees}
      K\cong  W \times G \times Z,
    \end{align}
    where  $W=\mathbb{I}(K\vartheta)$, $G=\vartheta K\vartheta$, and $Z=\mathbb{I}(\vartheta K)$, with $\vartheta$ being an \textit{idempotent} element in $K$, i.e., $\vartheta\in\mathbb{I}(K)$ or $\vartheta^2=\vartheta\in K$, so $W$ and $Z$ are the sets of idempotent elements  in $K\vartheta$ and $\vartheta K$, respectively. For notation, $\mathbb{I}(A)$ denotes the set of all idempotents in $A$. The multiplication in $W\times G\times Z$ is given by $(w,\sigma,z)(\hat{w},\hat{\sigma},\hat{z})=(w,\sigma(z\hat{w})\hat{\sigma},\hat{z})$. Importantly, since $\varSigma$ in eq. (\ref{eq:sigma}) is compact, then so is $K$, and hence the group factor $G$ in eq. (\ref{eq:rees}) must be finite \citep[][Section 4]{muk03}.\footnote{This follows from the fact that any compact group of $n\times n$ nonnegative matrices of rank $\eta$ is isomorphic to a subgroup of permutations $\{1,\dots,\eta\}$ and therefore must be finite \citep[e.g.,][]{muk86}.}
    \item Since we have two societies, we aim to show that society $X$ satisfies eq. (\ref{eq:equiv}) (i.e., \eqref{eq:C}) if and only if society $Y$ satisfies eq. (\ref{eq:equiv}). We therefore need to relate $\mu^t_X$ and $\mu^t_Y$ as well as $\varSigma_{\psi_X}$ and $\varSigma_{\psi_Y}$ using an \textit{isomorphism}. We will achieve this by using the fact that $\Proba(\mathscr{T})=1$. However, $\varSigma_{\psi_X}$ and $\varSigma_{\psi_Y}$ are arbitrary large sets (since they are compact), so constructing an isomorphism between them is very challenging. Instead, we will use the fact that $G_X$ and $G_Y$---the group factors in eq. (\ref{eq:rees}), for society $X$ and $Y$---are finite. We will show that Suppose $\Proba(\mathscr{T})=1$. Then, matrices in $\varSigma_{\psi_X}$ and $\varSigma_{\psi_Y}$ have the same rank $\eta$ (Lemma \ref{thm:rank}), which will help prove that $G_X$ and $G_Y$ are isomorphic (Lemma \ref{thm:factor}). The fact that these two groups are finite means that we can show that they are isomorphic by showing that they are both isomorphic to the same subgroup of permutations on $\{1,\dots,\eta\}$, where $\eta$ is the rank of the stochastic matrices in $\varSigma_{\psi_X}\cup\varSigma_{\psi_Y}$.
    \item We will then use the isomorphism between $G_X$ and $G_Y$ to show that their \textit{maximal homomorphic group images} $G_X/Q_X$ and $G_Y/Q_Y$ \citep{stoll51}, respectively, are also isomorphic (Lemma \ref{thm:maxfactor}). Here, $Q_X\subset G_X$ denotes the smallest (compact) normal subgroup of $G_X$ such that $W_XZ_X\subset Q_X$ and similarly for $Q_Y$. Then, $G_X/Q_X$ denotes the factor group (i.e., quotient group) of $G_X$ by $Q_X$. We can then define a map $\varPhi_X:\varSigma_X\rightarrow G_X/Q_X$ such that $\varPhi_X(x)=pxp\hspace{0.02in}Q_X$, where $p$ is an idempotent element in $\varSigma_{\psi_X}$. That is, $\varPhi_X$ maps the element $x\in \varSigma_X$ onto the coset of $Q_X$ containing $pxp$. Then, $\varPhi_X$ is a (surjective) continuous \textit{homomorphism} (with quotient topology on $G_X/Q_X$) (see, after 5.), which defines a probability measure $\Tilde{\mu}_X$ on $G_X/Q_X$:
\begin{align*}
    \Tilde{\mu}_X(B)=\mu_X\big(\varPhi_X^{-1}(B)\big),
\end{align*}
for any Borel set $B\subset G_X/Q_X$.\footnote{That is, if $U$ is an $\varSigma_X$-valued random variable with distribution $\mu_X$, then $\varPhi_X(U)$ is
a random variable with values in $G_X/Q_X$ and distribution $\Tilde{\mu}_X$.} Then, the sequence $\mu^t_X$ on $\varSigma_X$  converges weakly  if and only if the sequence $\Tilde{\mu}^t_X$
on $G_X/Q_X$ converges weakly (Lemma \ref{thm:homom}).
    \item Finally, we will then use the isomorphism between $G_X/Q_X$ and $G_Y/Q_Y$ to show that the supports $\varSigma_{\Tilde{\mu}_X}$ and $\varSigma_{\Tilde{\mu}_Y}$ are also isomorphic (Lemma \ref{thm:support}). Since $G_X/Q_X$ is a finite group, $\Tilde{\mu}^t_X$ converges weakly to the uniform probability measure on the closed subgroup generated by $\varSigma_{\Tilde{\mu}_X}$ (Lemma \ref{thm:finite}). Then, under  $\Proba(\mathscr{T})=1$, $\Tilde{\mu}^t_X$ converges weakly if and only if $\Tilde{\mu}^t_Y$ converges weakly (Lemma \ref{thm:iff}), which combined with Lemma \ref{thm:homom} establishes Theorem \ref{thm:skel}.(i). This equivalence will then be used to show that $\big\{\m{X}^{(t)}\big\}_{t\geq1}$ satisfies eq. (\ref{eq:equiv}) if and only if $\{\m{Y}^{(t)}\}_{t\geq1}$ satisfies eq. (\ref{eq:equiv}), which establishes Theorem \ref{thm:skel}.(ii).
\end{enumerate}

We now show that $\varPhi_j$ is a continuous homomorphism, for each society $j\in\{X,Y\}$ as noted in part 4. above \citep[see,][Section 2]{muk03}, so, for ease, we suppress the subscript $j$ in what follows. For any $x,y\in\varSigma$, let $\vartheta x=a\sigma b$ and
$y\vartheta=\hat{a}\hat{\sigma}\hat{b}$, where $\vartheta^2=\vartheta\in K$, $a,\hat{a}\in W$, $b,\hat{b}\in Z$, and $\sigma,\hat{\sigma}\in G$, for $K\cong W\times G\times Z$ in eq. (\ref{eq:rees}). 
Then, 
\begin{align*}
    \vartheta (xy)\vartheta =\vartheta \big(\sigma(b\hat{a})\hat{\sigma}\big)\hat{b}\vartheta &=\vartheta a\big(\sigma(b\hat{a})\hat{\sigma}\big)\vartheta \\
    &=\sigma(b\hat{a})\hat{\sigma}\in \sigma Q. Q.\hat{\sigma}Q\\
    &=(\vartheta x\vartheta )Q.(\vartheta y\vartheta )Q,
\end{align*}
so $\varPhi(xy)=(\vartheta xy\vartheta )Q=(\vartheta x\vartheta )Q.(\vartheta y\vartheta )Q=\varPhi(x)\varPhi(y)$ using the normality of $Q$, and hence $\varPhi$ is a homomorphism. With the quotient topology on $G/Q$, the continuity
of $\varPhi$ follows since it is the composition of the canonical homomorphism from $G$ to
$G/Q$ and the continuous map $s\mapsto \vartheta s\vartheta$ from $\varSigma$ to $G$. We refer to \citet{muk03} for further details.

\phantomsection\label{app:proofs}
\section*{Appendix B: Proofs}

\subsection*{Proof of Theorem \ref{thm:consensus}}
Theorem \ref{thm:consensus} is a special case of \citet[][Theorem 1.(i)--(ii).(a)]{hen97} applied to random stochastic matrices. We therefore refer to \citet[][Section 4]{hen97} for a proof.

\subsection*{Proof of Proposition \ref{thm:uniq}}
Let $K$ denote the kernel of $\text{\normalfont closure}\big(\bigcup_{t=1}^{\infty}\varSigma^t_{\mu}\big)$---the closed semigroup generated by $\varSigma_{\mu}$. $K$ contains \textit{all} the matrices in this semigroup with the minimal rank (Appendix \hyperref[app:conv]{A.III}: eq. (\ref{eq:kernel})).  
\begin{lemma}\label{thm:weak1}
When there exists a rank one matrix in $\text{\normalfont closure}\big(\bigcup_{t=1}^{\infty}\varSigma^t_{\mu}\big)$, the convolution sequence $\{\mu^t\}_{t\geq1}$ converges weakly to a probability measure with support $K$.
\end{lemma}

To ease notation in the proofs of Lemma \ref{thm:weak1} and Proposition \ref{thm:uniq}, define $S:=\text{\normalfont closure}\big(\bigcup_{t=1}^{\infty}\varSigma^t_{\mu}\big)$.
\begin{proof}[Proof of Lemma \ref{thm:weak1}]
    Since $S$ is compact, $\{\mu^t\}_{t\geq1}$ is tight, so the sequence $\big\{\frac{1}{t}\sum_{\ell=1}^t\mu^\ell\big\}_{t\geq1}$ of average convolution powers of $\mu$ converges weakly to a probability measure $\psi$ with support $K$ such that $\mu*\psi=\psi*\mu=\psi$ \citep[][Theorem 1.1]{muk07}.
    \par By \citet[][Theorem 2.1]{muk87}, $\underset{t\rightarrow\infty}{\text{lim}}\mu^t(\mathcal{O})=1$ holds for any open set $\mathcal{O}\supset K$, so if $\nu$ is ever a weak limit of $\mu^t$ it must be that   $\varSigma_{\nu}\subset K$. Then, $\nu*\psi=\psi*\nu=\psi$ implies that, for any Borel subset $B\subset S$, 
    \begin{align*}
        \psi(B)=\psi*\nu(B)=\int_K\psi\Big(\big\{x\in S:x\sigma\in B\big\}\Big) \hspace{0.03in}d\nu(\sigma)=\nu(B)
    \end{align*}
    because $x\sigma=\sigma$ holds when the rank in $K$ is one. Thus, $\{\mu^t\}_{t\geq1}$ converges weakly to $\psi$.
\end{proof}

\begin{proof}[Proof of Proposition \ref{thm:uniq}] $(\Longrightarrow)$: Suppose $\m{X}^{(t)}$ converges in probability to $\m{1s}$. Then, $\mu^t$ converges weakly to a probability measure $\psi$ with support $\varSigma_{\psi}=\{\m{1s}\}$. It is well-known that $\varSigma_{\psi}=K$---the kernel of $S$---which contains \textit{all} the matrices from this semigroup with minimal rank. Thus, $\varSigma_{\psi}=\{\m{1s}\}$ implies that $\m{1s}$ must be the \textit{only} rank one matrix in $S$.
\par $(\Longleftarrow)$: Suppose $\m{1s}$ is a rank one matrix in the closed semigroup $S$. Then, by Lemma \ref{thm:weak1}, $\mu^t$ must converge weakly to a probability measure $\psi$ with support $\varSigma_{\psi}=K$. 
Since $\m{1s}$ is also assumed to be the \textit{only} rank one matrix in $S$, we have $K=\{\m{1s}\}$, so $\psi$ is a unit mass on $\m{1s}$. Thus, $\m{X}^{(t)}$ converges in probability to $\m{1s}$. 
\end{proof}

\subsection*{Proof of Corollaries \ref{thm:deter} and \ref{thm:bi}}

\begin{proof}
---Corollary \ref{thm:deter}: This follows directly by applying simple properties of eigenvectors.

\par\noindent---Corollary \ref{thm:bi}: By Theorem \ref{thm:consensus}, the limiting interaction matrix is a strictly positive random stochastic  matrix of rank one. Since the $\m{X}_t$'s are all bistochastic, the limiting interaction matrix is also bistochastic (because bistochastic matrices are closed under multiplication). Since the only $n\times n$ bistochastic matrix of rank one is $\frac{1}{n}\mb{1}\mb{1}'$, it must be the limiting interaction matrix because it is strictly positive. Thus, each agent's influence is $\pi_i=1/n$ for all $i$.\end{proof}

\subsection*{Proof of Theorem \ref{thm:wisdom}}
Consider the following weakening of Assumptions (i) and (ii) stated in Section \ref{sec:wisdef}.
\begin{assumption}\label{ass:wisdom1}\normalfont The joint distribution of $\{{p}^{(0)}_i(n)\}_{i\leq n}$ and $\m\pi(n)$ satisfies the following:
    \begin{enumerate}
        \item $\sum_{i<j}^n\Big|\text{cov}\Big({p}^{(0)}_i(n),{p}^{(0)}_j(n)\Big|\m\pi(n)\Big)\Big|$ is uniformly bounded above as $n\rightarrow\infty$.
        \item $\E\big[{p}^{(\infty)}_i(n)\big|\m\pi(n)\big]=\gamma$, for all $i$ and $n$.
        \item ${p}^{(\infty)}_i(n)$ is not a constant, for all $i$ and $n$. \hfill$\bigtriangleup$
    \end{enumerate}
\end{assumption}
Let's first check that these assumptions are weaker than Assumptions (i) and (ii) in Section \ref{sec:wisdef}. (i) If all initial signals $\{{p}^{(0)}_i(n)\}_{i\leq n}$ are independent and are jointly independent of the dynamic network $\{\m{X}_{t}\}_{t\geq1}$, then Assumption \ref{ass:wisdom1}.1 holds since $\text{cov}\big({p}^{(0)}_i(n),{p}^{(0)}_j(n)|\pi(n)\big)=\text{cov}\big({p}^{(0)}_i(n),{p}^{(0)}_j(n)\big)=0$ for all $i\neq j$ and every $n$. (ii) If each initial signal is conditionally unbiased for $\gamma$, i.e., $\E\big[p^{(0)}_i(n)\big]=\gamma$ for all $i\leq n$, and is independent of $\{\m{X}_{t}\}_{t\geq1}$, then Assumption \ref{ass:wisdom1}.2 holds because $\E\big[{p}^{(\infty)}_i(n)\big|\m\pi(n)\big]=\sum_{i}\pi_i(n)\E[p_i^{(0)}(n)]=\gamma$ for all $i$ and $n$. Assumption \ref{ass:wisdom1}.3 also holds under (i) and (ii). Recall that Assumption (iii) is always maintained. 
\par We start with a lemma that will simplify the proof. Let $Z(n)=\sum_{i}\pi_i(n)p_i^{(0)}(n)$, and recall that $\m\pi(n)=\big(\pi_1(n),\dots,\pi_n(n)\big)\in(0,1)^n$ for each $n$ by Theorem \ref{thm:consensus}. 
Also, recall that for any arbitrary random variables $\{Y_i\}_{i=1}^n$, $V$, and arbitrary constants $\{a_i\}_{i=1}^n$, the following holds $\text{var}\big(\sum_{i=1}^na_iY_i\big|V\big)=\sum_{i=1}^na_i^2\text{var}(Y_i|V)+2\sum_{i<j}a_ia_j\text{cov}(Y_i,Y_j|V)$, where $\text{cov}(Y_i,Y_j|V)=\E[Y_iY_j|V]-\E[Y_i|V]\E[Y_j|V]$, for all $i\neq j$.
To ease notation, define $\pi^*(n):=\underset{i\leq n}{\text{\normalfont max}}\hspace{0.04in}\pi_i(n)$, and we suppress the argument $n$ in the statement and proof of Lemma \ref{thm:unbias}.
\begin{lemma}\label{thm:unbias}
    Suppose Assumption \ref{ass:wisdom1} holds.  Then,
    \begin{enumerate}
        \item $\E[Z|\m\pi]=\gamma$, for all $n$;
        \item $\sum_{i}\text{\normalfont var}\big(\pi_ip_i^{(0)}\big|\m\pi\big)\leq \overline{\upsilon}^2\pi^*$, for all $i\neq j$ and $n$ and some finite constant $\overline{\upsilon}^2\geq0$.
        \item $\sum_{i<j}\text{\normalfont cov}\big(\pi_ip_i^{(0)},\pi_jp_j^{(0)}\big|\m\pi\big)\leq \overline{\nu}\pi^*$ as $n\rightarrow\infty$ for some finite constant $\overline{\nu}\geq0$. 
    \end{enumerate}
\end{lemma}
\begin{proof}[Proof of Lemma \ref{thm:unbias}]
To simplify the notation, we omit the argument $n$ in this proof. 
\par For part 1., it follows by Assumption \ref{ass:wisdom1}.2 because (by definition) $p_i^{(\infty)}(n)=Z(n)$. 

  \par For part 2.,
    \begin{align*}
\sum_i\text{var}\big(\pi_ip_i^{(0)}\big|\m\pi\big)=\sum_i\pi_i^2\text{var}\big(p_i^{(0)}\big|\m\pi\big)\leq \pi^*\sum_i\pi_i\text{var}\big(p_i^{(0)}\big|\m\pi\big)\leq \overline{\upsilon}^2\pi^*\sum_i\pi_i=\overline{\upsilon}^2\pi^*,
    \end{align*}
    where the first inequality holds since $\pi^*\geq\pi_{i}$ a.s., for all $i$ and $n$. Here, $\overline{\upsilon}^2$ denotes the uniform upper bound on all the conditional variances $\big\{\text{var}\big(p_i^{(0)}\big|\m\pi\big)\big\}_{i\leq n}$, which is finite because all the signals are bounded random variables.
    \par For part 3.,
    \begin{align*}
\sum_{i<j}\text{cov}\Big(\pi_ip_i^{(0)},\pi_jp_j^{(0)}\Big|\m\pi\Big)=\sum_{i<j}\pi_i\pi_j\text{cov}\big(p_i^{(0)},p_j^{(0)}\big|\m\pi\big)\leq \pi^*\sum_{i<j}\pi_j\Big|\text{cov}\big(p_i^{(0)},p_j^{(0)}\big|\m\pi\big)\Big|\leq \overline{\nu}\pi^*,
    \end{align*}
    where $\overline{\nu}$ denotes the uniform upper bound on the sum $\sum_{i<j}\big|\text{cov}(p_i^{(0)},p_j^{(0)}|\m\pi)\big|$ as $n\rightarrow\infty$, which exists almost surely by Assumption \ref{ass:wisdom1}.1.  
\end{proof}

We can now prove the following generalization of Theorem \ref{thm:wisdom}.
\begin{theorem}\label{thm:wisdom2}
  For each $n$, let $\{\m\pi(n)\}_{n=1}^{\infty}$ be a sequence of arbitrarily random influence vectors from Theorem \ref{thm:consensus}.2, and suppose Assumption \ref{ass:wisdom1} holds. Then, for any $\epsilon>0$, \begin{align*}
        \underset{n\rightarrow\infty}{\text{\normalfont lim}}\hspace{0.03in}\Proba\Big(\Big|\sum_{i\leq n}\pi_i(n)p_i^{(0)}(n)-\gamma\Big|\geq\epsilon\Big)=0 \quad\text{if and only if} \quad \E\big[\underset{i\leq n}{\text{\normalfont max}}\hspace{0.04in}\pi_i(n)\big]\rightarrow0 \text{ as } n\rightarrow\infty.
    \end{align*}
\end{theorem}

\begin{proof}[Proof of Theorem \ref{thm:wisdom2}]

First, suppose $\E\big[\underset{i\leq n}{\text{\normalfont max}}\hspace{0.04in}\pi_i(n)\big]\rightarrow0$ as $n\rightarrow\infty$,  then 
\begin{align*}
    \text{var}\big(Z(n)\big)&=\E\Big[\text{var}\big(Z(n)\big|\m\pi(n)\big)\Big]+\text{var}\Big(\underbrace{\E\big[Z(n)\big|\m\pi(n)\big]}_{=\gamma}\Big)\\
    &=
    \E\Bigg[\text{var}\Big(\sum_{i}\pi_i(n)p_i^{(0)}(n)\Big|\m\pi(n)\Big)\Bigg]\\
    &=\E\Bigg[\sum_{i}\text{var}\Big(\pi_i(n)p_i^{(0)}(n)\Big|\m\pi(n)\Big)+2\sum_{i<j}\text{cov}\Big(\pi_i(n)p_i^{(0)}(n),\pi_j(n)p_j^{(0)}(n)\Big|\m\pi(n)\Big)\Bigg]\\
    &=\E\Bigg[\sum_{i}\text{var}\Big(\pi_i(n)p_i^{(0)}(n)\Big|\m\pi(n)\Big)\Bigg]+2\E\Bigg[\sum_{i<j}\text{cov}\Big(\pi_i(n)p_i^{(0)}(n),\pi_j(n)p_j^{(0)}(n)\Big|\m\pi(n)\Big)\Bigg]\\
    &\leq (\overline{\upsilon}^2+2\overline{\nu})\hspace{0.04in}\E\big[\underset{i\leq n}{\text{\normalfont max}}\hspace{0.04in}\pi_i(n)\big]\underset{n\rightarrow\infty}{\longrightarrow}0.
\end{align*}
The first equality is the law of total variance. The second equality holds since $\E[Z(n)|\m\pi(n)]=\gamma$ for all $n$ by Lemma \ref{thm:unbias}.1, so $\text{var}(\gamma)=0$ since $\gamma$ is a constant. The third equality is the definition of conditional variance. The inequality holds by applying both Lemmas \ref{thm:unbias}.2 and \ref{thm:unbias}.3 for the first and second term, respectively, where $\overline{\upsilon}^2$ and $\overline{\nu}$ are nonnegative finite constants (Lemma \ref{thm:unbias}). 
By the law of iterated expectation, Lemma \ref{thm:unbias}.1 implies that $\E[Z(n)]=\E\big[\sum_{i}\pi_i(n)p_i^{(0)}(n)\big]=\gamma$ for all $n$. Then, for every $\epsilon>0$, Chebyshev's inequality yields
\begin{align*}
    \Proba\Bigg(\Bigg|\sum_{i\leq n}\pi_i(n)p_i^{(0)}(n)-\gamma\Bigg|\geq\epsilon\Bigg)\leq \frac{\text{var}\big(Z(n)\big)}{\epsilon^2}\underset{n\rightarrow\infty}{\longrightarrow}0.
\end{align*}
The converse follows similar steps as in \citet[][Lemma 1]{moj10}. Suppose (taking a subsequence if necessary) that $\E\big[\underset{i\leq n}{\text{\normalfont max}}\hspace{0.04in}\pi_i(n)\big]\rightarrow\hat{\pi}>0$ as $n\rightarrow\infty$. By Assumption \ref{ass:wisdom1}.3, this implies that $Z(n)=\sum_{i}\pi_i(n)p_i^{(0)}(n)$ is not a constant for all $n$. Thus, there exists a constant $\delta > 0$ such that $\text{var}\big(Z(n)\big) > \delta$ for all $n$, and hence $Z(n)$ cannot converge in probability to $\gamma$. This continues to hold even when $\pi_i(n)$ is a constant for all $i$ and $n$ \citep[see,][Lemma 1]{moj10}. 
\end{proof}

\subsection*{Proof of Proposition \ref{thm:wis}}
The proof of this result is elementary, so it is relegated to Online Appendix \hyperref[app:oproof]{D}.
\subsection*{Proof of Proposition \ref{thm:mic3}}
\begin{proof}
Let $\alpha_0(n) := \sum_{i=1}^n \alpha_i$. For every $j\leq n$, $\E[\pi_j(n)]=\frac{\alpha_j}{\alpha_0(n)}=O(1/n^m)$ and, for large $n$,
$$\text{var}(\pi_j(n)) = \frac{\alpha_j(\alpha_0(n) - \alpha_j)}{\alpha_0(n)^2(\alpha_0(n) + 1)} \sim \frac{\alpha_j}{\alpha_0(n)^2}.$$
Since $\alpha_0(n) = O(n^k)$ and $\frac{\alpha_j}{\alpha_0(n)} = O(1/n^m)$, for $k\geq1$, $m>0$, so every $j\leq n$, there exist constants $C_1, C_2 > 0$ such that
$\alpha_j \leq \frac{C_2}{n^m} C_1 n^k = C_1 C_2 n^{k-m}.$ Then, for every $j\leq n$,
$$\text{var}(\pi_j(n)) \leq \frac{C_1 C_2 n^{k-m}}{(C_1 n^k)^2} = \frac{C_2}{C_1 n^{k+m}}.$$
Collective intelligence holds by Proposition \ref{thm:wis} since $\text{var}(\pi_j(n)) = O(1/n^{d})$, $d:=k + m > 1$. 
\end{proof}

\subsection*{Proof of Propositions \ref{thm:equiv}, \ref{thm:first}, and \ref{thm:zero}}
The equivalence in Proposition \ref{thm:equiv} can be found in \citet[][Section 9.2: Remarks]{hen97}. Proposition \ref{thm:first} was famously proved by  \citet[][Proposition 2.2]{letac94}. Proposition \ref{thm:zero}
is due to \citet[][Theorem 1]{muk97}.

\subsection*{Proof of Theorem \ref{thm:skel}.(ii)}

We recall that under the (surjective) continuous homomorphisms $\varPhi_X$ and $\varPhi_Y$ in Appendix \hyperref[app:sketch]{A.IV}, the support of $\Tilde{\mu}_X$ is $\varSigma_{\Tilde{\mu}_X}=\varPhi_X(\varSigma_{\mu_X})=\{pxp\hspace{0.02in}Q_X:x\in \varSigma_{\mu_X}\}$, and  the support of $\Tilde{\mu}_Y$ is  $\varSigma_{\Tilde{\mu}_Y}=\varPhi_Y(\varSigma_{\mu_Y})=\{qyq\hspace{0.02in}Q_Y:y\in \varSigma_{\mu_Y}\}$ \citep[][Section 3]{muk03}. Then, it follows that $(\Tilde{\mu}_X)^t=\Tilde{\mu}^t_X$ and  $(\Tilde{\mu}_Y)^t=\Tilde{\mu}^t_Y$, for all $t\geq1$.\footnote{If $V$ and $P$ are independent random variables with distribution $\Tilde{\mu}_j$, then $\Tilde{\mu}^2_j$, the law of $\varPhi_j(VP)$, is equal to the law $(\Tilde{\mu}_j)^2$ of
$\varPhi_j(V)\varPhi_j(P)$, for each society $j\in\{X,Y\}$ \citep[][Section 3]{muk03}.} Moreover, the supports of the limiting distributions $\psi_X$ and $\psi_Y$ denoted, respectively,
 $\varSigma_{\psi_X}$ and $\varSigma_{\psi_Y}$ are the kernels of the closed semigroups $\varSigma_X=\text{closure}\big(\bigcup_{t\geq1}\varSigma^t_{\mu_X}\big)$ and $\varSigma_Y=\text{closure}\big(\bigcup_{t\geq1}\varSigma^t_{\mu_Y}\big)$ (see, eq. (\ref{eq:sigma})), respectively, which are compact sets (since they consist of stochastic matrices). Thus,  $\varSigma_{\psi_X}$ and $\varSigma_{\psi_Y}$ are completely simple semigroups (see, eq. (\ref{eq:rees})) and therefore \begin{align}\label{eq:reesx}
     \varSigma_{\psi_X}\cong W_X\times G_X\times Z_X \text{ and } \varSigma_{\psi_Y}\cong W_Y\times G_Y\times Z_Y
 \end{align}  are, respectively, their Rees-Suskewitsch product decomposition in eq. (\ref{eq:rees}), where for a fixed idempotent element $p\in \varSigma_{\psi_X}$, $W_X=\mathbb{I}(\varSigma_{\psi_X}p)$, $G_X=p\varSigma_{\psi_X}p$, $Z_X=\mathbb{I}(p\varSigma_{\psi_X})$, and similarly for society $Y$, i.e., for a fixed idempotent element $q\in \varSigma_{\psi_Y}$, $W_Y=\mathbb{I}(\varSigma_{\psi_Y}q)$, $G_Y=q\varSigma_{\psi_Y}q$, $Z_Y=\mathbb{I}(q\varSigma_{\psi_Y})$. Then, the following lemmas will be useful.
\begin{lemma}\label{thm:rank}
    Suppose $\Proba(\mathscr{T})=1$. Then, the minimal rank of the matrices in $\varSigma_{\psi_X}$ is
the same as that of the matrices in $\varSigma_{\psi_Y}$. Furthermore, there exist rank one matrices
$p$ in $\varSigma_{\psi_X}$ and $q$ in $\varSigma_{\psi_Y}$ such that $p$ and $q$ have the same basis and the same social topology.\footnote{Eq. (\ref{eq:basis}) in Online Appendix \hyperref[app:oproof]{D} defines the \textit{basis} of a nonnegative matrix.}
\end{lemma}

\begin{lemma}\label{thm:factor}
    Suppose $\Proba(\mathscr{T})=1$. Then, the group factors $G_X$ and $G_Y$ in
the Rees-Suskewitsch product representations of $\varSigma_{\psi_X}$ and $\varSigma_{\psi_Y}$ are isomorphic. 
\end{lemma}

\begin{lemma}\label{thm:maxfactor}
    Suppose $\Proba(\mathscr{T})=1$. Then,  the maximal homomorphic group images $G_X/Q_X$ and $G_Y/Q_Y$
are isomorphic.
\end{lemma}

\begin{lemma}\label{thm:support}
    Suppose $\Proba(\mathscr{T})=1$. Then,  the supports $\varSigma_{\Tilde{\mu}_X}$ and $\varSigma_{\Tilde{\mu}_Y}$ are isomorphic.
\end{lemma}

Lemmas \ref{thm:rank}--\ref{thm:support} will be proved in Online Appendix \hyperref[thm21]{D} by following \citet{muk05}. The next result is a special case of  \citet[][Theorem]{muk03}.
\begin{lemma}\label{thm:homom}
    The convolution sequence $\mu^t_j$ on $\varSigma_j$ converges weakly  if and only if the convolution sequence $\Tilde{\mu}^t_j$  on $G_j/Q_j$ converges weakly, for each society $j\in\{X,Y\}$.
\end{lemma}

The next result is well-known  \citep[e.g.,][Remark 3]{chak01}.

\begin{lemma}\label{thm:finite}
    Let $H$ be a finite group and $\nu$ be a probability measure on $H$. Then, the convolution sequence $\nu^t$ converges weakly if and only if there exists an integer $\tau\geq1$ such that the support of $\nu^\tau$ is the subgroup generated by the support of $\nu$. In this case, $\nu^t$ converges weakly to the uniform probability measure on the subgroup generated by the support of $\nu$.
\end{lemma}

If $\langle \varSigma_\nu\rangle$ denotes the subgroup of a finite group $H$ generated by the support of $\nu$, denoted $\varSigma_\nu\subset H$,\footnote{If $G$ is a group, $V$ is a subset of $G$, and $\{S_{\ell}<G:\ell\in I\}$ is the set of all subgroups of $G$ each one containing $V$ as a subset, then $\langle V\rangle:=\bigcap_{\ell \in I}S_{\ell}$ is the subgroup of $G$ generated by $V$.} then Lemma \ref{thm:finite} states that $\nu^t\rightarrow 1\big/\big|\langle\varSigma_\nu\rangle\big|$ weakly as $t\rightarrow\infty$. A sufficient condition for this is that the greatest common divisor of the orders of the elements in $\varSigma_\nu$ is one. 
\par The next result together with Lemma \ref{thm:rank} constitute a proof of Theorem \ref{thm:skel}.(i).

\begin{lemma}\label{thm:iff}
   Suppose $\Proba(\mathscr{T})=1$. Then,  the convolution sequence $\mu^t_X$ converges weakly if and only the convolution sequence $\mu^t_Y$ converges weakly.
\end{lemma}

\begin{proof}[Proof of Theorem \ref{thm:skel}.(ii)]
Let $\Proba(\mathscr{T})=1$. $(\Longrightarrow)$: we want to show that if $\big\{\m{X}^{(t)}\big\}_{t\geq1}$ satisfies \eqref{eq:C} then so does $\{\m{Y}^{(t)}\}_{t\geq1}$. To this end, assume $\big\{\m{X}^{(t)}\big\}_{t\geq1}$ satisfies \eqref{eq:C}. Then, Proposition \ref{thm:equiv} indicates that we can use the fact that eq. (\ref{eq:equiv}) is equivalent to \eqref{eq:C}. More specifically, this means that there exits a $\tau\geq1$ such that the inequality below holds
  \begin{align}\label{eq:equivX}
    \int c_X(\sigma)\hspace{0.03in}d\Tilde{\mu}^\tau_X\big(\varPhi_X(\sigma)\big)=  \int c_X(\sigma)\hspace{0.03in}d\mu^\tau_X(\sigma)<1.
  \end{align}
  where the equality above follows from the fact that $\Tilde{\mu}_X(B)=\mu_X(\varPhi^{-1}_X(B))$, for any Borel set $B\subset G_X/Q_X$, with support $\varSigma_{\Tilde{\mu}_X}=\varPhi_X(\varSigma_{\mu_X})\subset G_X$ \citep[see,][Section 3]{muk03}, and recalling from part 4. of the sketch (Appendix \hyperref[app:sketch]{A.IV}) that $\varPhi_X(x)=pxp\hspace{0.02in}Q_X$ is a continuous homomorphism, for any $x\in \varSigma_{X}$ and a fixed idempotent element $p\in \varSigma_{\psi_X}$. This means that $p$ is the identity of $G_X$, and by eq. (\ref{eq:equivX}), $p$ is strictly positive.
\par Since eq. (\ref{eq:equivX}) implies a.s. convergence of $\big\{\m{X}^{(t)}\big\}_{t\geq1}$ (Theorem \ref{thm:consensus}), it also implies that its convolution sequence $\mu^t_X$ converges weakly, and therefore so does $\Tilde{\mu}^t_X$ by Lemma \ref{thm:homom}. Since the group factor $G_X$ in the Rees-Suskewitsch product representation of $\varSigma_{\psi_X}$ in eq. (\ref{eq:reesx}) is a finite group \citep[][Section 4]{muk03}, we know from Lemma \ref{thm:finite} that there exists an integer $s\geq1$ such that $\Tilde{\mu}^s_X$ is indistinguishable from the uniform probability measure on the subgroup (of $G_X/Q_X$) generated by $\varSigma_{\Tilde{\mu}_X}=\varPhi_X(\varSigma_{\mu_X})$. 
Then,  $\Tilde{\mu}^t_Y$ also converges weakly by Lemma \ref{thm:iff}, so applying Lemma \ref{thm:finite} (since $G_Y/Q_Y$ is a finite group), there exists an integer $r\geq1$ such that $\Tilde{\mu}^r_Y$ is indistinguishable from the uniform probability measure on the subgroup (of $G_Y/Q_Y$) generated by $\varSigma_{\Tilde{\mu}_Y}=\varPhi_Y(\varSigma_{\mu_Y})$, denoted $\langle\varSigma_{\Tilde{\mu}_Y}\rangle$. By Lemma \ref{thm:rank}, there exists an idempotent matrix $q\in\varSigma_{\psi_Y}$ with the same social topology as $p$, so $q$ is strictly positive, and since it is the identity of $G_Y$, $qQ_Y$ is the identity of $\langle\varSigma_{\Tilde{\mu}_Y}\rangle$, hence $qQ_Y\in \langle\varSigma_{\Tilde{\mu}_Y}\rangle$. Define  $\Tilde{\mu}_Y(D)=\mu_Y(\varPhi^{-1}_Y(D))$, for any Borel set $D\subset G_Y/Q_Y$, where $\varPhi_Y(y)=qyq\hspace{0.02in}Q_Y$ is a (surjective) continuous homomorphism, for any $y\in \varSigma_{Y}$. 
This shows that 
there exists a  $T=\text{max}\{\tau,s,r\}$ such that $\Tilde{\mu}^T_Y(\varPhi_Y(q))=\Tilde{\mu}^T_Y(qQ_Y)>0$, so the following inequality holds
\begin{align*}
\int c_Y(\sigma)\hspace{0.03in}d\mu^T_Y(\sigma)=\int c_Y(\sigma)\hspace{0.03in}d\Tilde{\mu}^T_Y(\varPhi_Y(\sigma))<1
\end{align*} because $c_Y(q)<1$ (since $q$ is strictly positive). The equality above holds by definition of $\Tilde{\mu}_Y$ and $\varPhi_Y$.
This completes $(\Longrightarrow)$: there exists a $T\geq1$ such that $\int c_Y(\sigma)\hspace{0.03in}d\mu^T_Y(\sigma)<1$ and hence $\{\m{Y}^{(t)}\}_{t\geq1}$ satisfies \eqref{eq:C}. Reversing the role of $X$ and $Y$ yields the other direction $(\Longleftarrow)$. 
\end{proof}

\subsection*{Proof of Proposition \ref{thm:pred}}
\begin{proof}
 Since society $X$ satisfies  \eqref{eq:C},   its limiting support, $\varSigma_{\psi_X}$, must consist of strictly positive interaction matrices of rank one, so the influence vector $\m\pi^X(n)$ is a strictly positive unit vector (Theorem \ref{thm:consensus}.2).  We know from Theorem \ref{thm:skel}.(ii) that society $Y$ must satisfy  \eqref{eq:C} whenever society $X$ satisfies \eqref{eq:C} and $\Proba(\mathscr{T})=1$. This means that  $\varSigma_{\psi_Y}$ must also contain strictly positive matrices of rank one and $\m\pi^Y(n)$ must also be a strictly positive unit vector. 

   \par Since society $X$ is collectively intelligent,  $\E\big[\underset{i\leq n}{\text{max}}\hspace{0.03in}\pi^X_i(n)\big]\rightarrow0$ as $n\rightarrow\infty$ (Theorem \ref{thm:wis}). As noted above, the positivity of  $\m\pi^Y(n)$ must match the positivity of $\m\pi^X(n)$, so it must be that $\E\big[\underset{i\leq n}{\text{max}}\hspace{0.03in}\pi^Y_i(n)\big]\rightarrow0$ as $n\rightarrow\infty$. Thus, society $Y$ is also collectively intelligent.
\end{proof}

\subsection*{Proofs of Proposition \ref{thm:conv2x2} and Corollary \ref{thm:conv2x22}}
A proof of Proposition \ref{thm:conv2x2} can be found in \citet[][Theorem 3]{va86} by recalling the definition $\lambda_2(x,y)=x-y$. Similarly, Corollary \ref{thm:conv2x22} follows from \citet[][Corollary]{va86}.

\subsection*{Proof of Proposition \ref{thm:rate}}
Proposition \ref{thm:rate} is a special case of \citet[][Theorem 6]{doubly13}, when $d_k:=\epsilon$ for all $k$. The symmetry assumption can be relaxed \citep[e.g., see,][Theorem 3]{convrate12}.
\subsection*{Proof of Proposition \ref{thm:stat}}
Our proof builds on the techniques in \citet[][Chapter 3.3]{fractal98}. For notation, let $\mathcal{P}(S)$ denote the set of all probability measures on the Borel subsets of $S$. Let $d$ denote a metric associated with the metric space $S$, which we denote as the pair $(S,d)$. Since each map $C_t$ is assumed to be a contraction on $S$, then for all $x,y\in S$, $d\big(C_t(x),C_t(y)\big)\leq \kappa_td(x,y)$, where $0<\kappa_t<1$ for all $t\in\mathcal{T}$ and define $\overline{\kappa}=\text{sup}\{\kappa_t:t\in\mathcal{T}\}$ for a countably infinite index set $\mathcal{T}$.
\par Let $\{\rho_t\}_{t\in \mathcal{T}}$ be a discrete probability measure on $\mathcal{T}$ and
let $\tau$ be a $\mathcal{T}$-valued random variable distributed according to $\{\rho_t\}_{t\in \mathcal{T}}$. Since $S$ is compact, there exists a $\alpha>0$ such that $d(x,y)<\alpha$ for all $x,y\in S$, so define the set
\begin{align*}
    \mathcal{F}_{\alpha}=\Big\{h:S\rightarrow\R:|h(x)-h(y)|<d(x,y),|h(x)|<\alpha,\forall x,y\in S\Big\}.
\end{align*}
Then, for any $\mu_1,\mu_2\in\mathcal{P}(S)$, define on $\mathcal{P}(S)$ the so-called \textit{Monge-Kantorovich} metric
\begin{align*}
    \nu(\mu_1,\mu_2)=\text{sup}\Big\{\Big|\int h\hspace{0.03in}d\mu_1-\int h\hspace{0.03in}d\mu_2\Big|:h\in\mathcal{F}_{\alpha}\Big\},
\end{align*}
and hence $(\mathcal{P}(S),\nu)$ is a complete metric space. Moreover, let $\mathcal{C}(S)$ denote the class of all closed subsets of $S$. Then, for any $A,B\subset S$ define on $\mathcal{C}(S)$  the \textit{Hausdorff} metric 
\begin{align*}
    \nu(A,B)=\text{inf}\Big\{s>0:A\subseteq Q_s(B) \text{ and }B\subseteq Q_s(A)\Big\},
\end{align*}
where $Q_s(F)=\big\{y\in S:\exists x\in F,d(x,y)<s\big\}$ for any $F\subset S$, and hence $(\mathcal{C}(S),\nu)$ is a complete metric space. Given all the above, we are now ready to prove Proposition \ref{thm:stat}.
\begin{proof}
    Consider the map $\varLambda:\mathcal{P}(S)\rightarrow\mathcal{P}(S)$ defined such that for any Borel subset $B$ of $S$:
    \begin{align}\label{eq:mu}
\varLambda(\mu(B))=\sum_{t\in\mathcal{T}}\rho_t\hspace{0.02in}\mu\big(C_t^{-1}(B)\big).
    \end{align}
    Now, fix a point $z\in S$, then for any $h\in \mathcal{F}_{\alpha}$ and $\mu_1,\mu_2\in\mathcal{P}(S)$, we have
    \begin{align*}
        \nu\big(\varLambda(\mu_1),\varLambda(\mu_2)\big)&=\Big|\int h\hspace{0.03in}d\varLambda(\mu_1)-\int h\hspace{0.03in}d\varLambda(\mu_2)\Big|\\
        &=\sum_{t\in\mathcal{T}}\rho_t\Big(\int\Big[h\circ C_t(x)-h\circ C_t(z)\Big]d\mu_1(x)-\int\Big[h\circ C_t(x)-h\circ C_t(z)\Big]d\mu_2(x)\Big)\\
        &\leq \Big|\sum_{t\in\mathcal{T}}\rho_t\kappa_t\Big(\int h_t\hspace{0.03in}d\mu_1-\int h_t\hspace{0.03in}d\mu_2\Big)\Big|\\
        &\leq \E[\kappa_{\tau}]\hspace{0.03in}\nu(\mu_1,\mu_2),
    \end{align*}
    where $h_t(x):=\frac{1}{\kappa_t}\big(h\circ C_t(x)-h\circ C_t(z)\big)$ is used to establish the first inequality above, and the fact that $h_t\in \mathcal{F}_{\alpha}$ for all $t\in \mathcal{T}$ is used for the last inequality. Then, since $\overline{\kappa}<1$,  $\E[\kappa_{\tau}]<1$, and hence $\varLambda$ is a contraction. Thus, we can apply the contraction mapping theorem to establish that there exists a unique fixed point $\mu\in\mathcal{P}(S)$ that satisfied eq. (\ref{eq:mu}). 
    \par For the support $S_{\mu}$, consider the function $\varGamma:\mathcal{C}(S)\rightarrow\mathcal{C}(S)$ such that for any $B\in \mathcal{C}(S)$
    \begin{align}\label{eq:support}
\varGamma(B)=\text{closure}\Bigg(\bigcup_{t\in\mathcal{T}}C_t(B)\Bigg).
    \end{align}
 Then, for any sets $A,B\in \mathcal{C}(S)$, it follows that $\nu(\varGamma(A),\varGamma(B))\leq \overline{\kappa}\hspace{0.03in}\nu(A,B)$. Since $\overline{\kappa}<1$, the map $\varGamma$ is a contraction. By the contraction mapping theorem and since $(\mathcal{C}(S),\nu)$ is complete, there exists a unique compact subset $V\in \mathcal{C}(S)$ that satisfies eq. (\ref{eq:support}). The support of any probability measure $\mu$ on the Borel subsets of $S$ is defined as $S_\mu=\big\{\sigma\in S:\mu(\mathcal{O}(\sigma))>0,\forall \text{open set }\mathcal{O}(\sigma)\ni \sigma\big\}$. That is, $S_\mu$ is the smallest closed subset $D$ of $S$ such that $\mu(D)=1$. Given eq. (\ref{eq:mu}), such a set must be a Borel subset of $S$ that satisfies eq. (\ref{eq:support}), and since we have shown that there exists a unique set that does so, it must be that $S_\mu=V$. 
\end{proof}

\begin{singlespace}
\bibliography{ref}

\begin{thebibliography}{}

\bibitem[Acemoglu et~al., 2011]{daron11}
Acemoglu, D., Dahleh, M.~A., Lobel, I., and Ozdaglar, A. (2011).
\newblock Bayesian learning in social networks.
\newblock {\em The Review of Economic Studies}, 78(4):1201--1236.

\bibitem[Almaatouq et~al., 2020]{dynam20}
Almaatouq, A., Noriega-Campero, A., Alotaibi, A., Krafft, P., Moussaid, M., and Pentland, A. (2020).
\newblock Adaptive social networks promote the wisdom of crowds.
\newblock {\em Proceedings of the National Academy of Sciences}, 117(21):11379--11386.

\bibitem[Bajovi{\'c} et~al., 2013]{doubly13}
Bajovi{\'c}, D., Xavier, J., Moura, J.~M., and Sinopoli, B. (2013).
\newblock Consensus and products of random stochastic matrices: Exact rate for convergence in probability.
\newblock {\em IEEE Transactions on Signal Processing}, 61(10):2557--2571.

\bibitem[Bajovic et~al., 2012]{convrate12}
Bajovic, D., Xavier, J., and Sinopoli, B. (2012).
\newblock Products of stochastic matrices: Exact rate for convergence in probability for directed networks.
\newblock In {\em 2012 20th Telecommunications Forum (TELFOR)}, pages 883--886. IEEE.

\bibitem[Bala and Goyal, 1998]{bala98}
Bala, V. and Goyal, S. (1998).
\newblock Learning from neighbours.
\newblock {\em The review of economic studies}, 65(3):595--621.

\bibitem[Banerjee et~al., 2024a]{ban23}
Banerjee, A., Breza, E., Chandrasekhar, A.~G., Duflo, E., Jackson, M.~O., and Kinnan, C. (2024a).
\newblock Changes in social network structure in response to exposure to formal credit markets.
\newblock {\em Review of Economic Studies}, 91(3):1331--1372.

\bibitem[Banerjee et~al., 2021]{banerjee21}
Banerjee, A., Breza, E., Chandrasekhar, A.~G., and Mobius, M. (2021).
\newblock Naive learning with uninformed agents.
\newblock {\em American Economic Review}, 111(11):3540--74.

\bibitem[Banerjee et~al., 2024b]{ban24}
Banerjee, A., Chandrasekhar, A.~G., Dalpath, S., Duflo, E., Floretta, J., Jackson, M.~O., Kannan, H., Loza, F.~N., Sankar, A., Schrimpf, A., et~al. (2024b).
\newblock Selecting the most effective nudge: Evidence from a large-scale experiment on immunization.
\newblock {\em forthcoming, Econometrica}.

\bibitem[Banerjee et~al., 2019]{ban19}
Banerjee, A., Chandrasekhar, A.~G., Duflo, E., and Jackson, M.~O. (2019).
\newblock Using gossips to spread information: Theory and evidence from two randomized controlled trials.
\newblock {\em The Review of Economic Studies}, 86(6):2453--2490.

\bibitem[Banerjee and Compte, 2024]{ban232}
Banerjee, A. and Compte, O. (2024).
\newblock Consensus and disagreement: Information aggregation under (not so) naive learning.
\newblock {\em forthcoming, Journal of Political Economy}.

\bibitem[Bloch and Dutta, 2009]{bloch09}
Bloch, F. and Dutta, B. (2009).
\newblock Communication networks with endogenous link strength.
\newblock {\em Games and Economic Behavior}, 66(1):39--56.

\bibitem[Brandts et~al., 2015]{exp15}
Brandts, J., Giritligil, A.~E., and Weber, R.~A. (2015).
\newblock An experimental study of persuasion bias and social influence in networks.
\newblock {\em European Economic Review}, 80:214--229.

\bibitem[Breza et~al., 2019]{golub19}
Breza, E., Chandrasekhar, A., Golub, B., and Parvathaneni, A. (2019).
\newblock {Networks in economic development}.
\newblock {\em Oxford Review of Economic Policy}, 35(4):678--721.

\bibitem[Cerreia-Vioglio et~al., 2024]{nonbayes23}
Cerreia-Vioglio, S., Corrao, R., and Lanzani, G. (2024).
\newblock Dynamic opinion aggregation: long-run stability and disagreement.
\newblock {\em Review of Economic Studies}, 91(3):1406--1447.

\bibitem[Chakraborty and Mukherjea, 2014]{muk14}
Chakraborty, S. and Mukherjea, A. (2014).
\newblock Limit distributions of random walks on stochastic matrices.
\newblock {\em Proceedings-Mathematical Sciences}, 124:603--612.

\bibitem[Chakraborty and Rao, 1998]{rao98}
Chakraborty, S. and Rao, B. (1998).
\newblock Covolution powers of probabilites on stochastic matrices of order 3.
\newblock {\em Sankhy{\=a}: The Indian Journal of Statistics, Series A}, pages 151--170.

\bibitem[Chakraborty and Rao, 2001]{chak01}
Chakraborty, S. and Rao, B. (2001).
\newblock Convolution powers of probabilities on stochastic matrices.
\newblock {\em Journal of Theoretical Probability}, 14(2):599--603.

\bibitem[Chamayou and Letac, 1994]{letac94}
Chamayou, J.-F. and Letac, G. (1994).
\newblock A transient random walk on stochastic matrices with dirichlet distributions.
\newblock {\em The Annals of Probability}, pages 424--430.

\bibitem[Chandrasekhar et~al., 2020]{test20}
Chandrasekhar, A.~G., Larreguy, H., and Xandri, J.~P. (2020).
\newblock Testing models of social learning on networks: Evidence from two experiments.
\newblock {\em Econometrica}, 88(1):1--32.

\bibitem[Chatterjee and Seneta, 1977]{seneta77}
Chatterjee, S. and Seneta, E. (1977).
\newblock Towards consensus: Some convergence theorems on repeated averaging.
\newblock {\em Journal of Applied Probability}, 14(1):89--97.

\bibitem[Chetty and Hendren, 2015]{raj15}
Chetty, R. and Hendren, N. (2015).
\newblock The impacts of neighborhoods on intergenerational mobility: Childhood exposure effects and county-level estimates.
\newblock {\em Harvard University and NBER}, 133(3):1--145.

\bibitem[Clark, 1965]{clark65}
Clark, W.~E. (1965).
\newblock Remarks on the kernel of a matrix semigroup.
\newblock {\em Czechoslovak Mathematical Journal}, 15(2):305--310.

\bibitem[Cohn et~al., 1993]{cohn93}
Cohn, H., Nerman, O., and Peligrad, M. (1993).
\newblock Weak ergodicity and products of random matrices.
\newblock {\em Journal of Theoretical Probability}, 6:389--405.

\bibitem[Conlon et~al., 2025]{frank25}
Conlon, J.~J., Mani, M., Rao, G., Ridley, M., and Schilbach, F. (2025).
\newblock Not learning from others.
\newblock {\em working paper}.

\bibitem[Corazzini et~al., 2012]{test12}
Corazzini, L., Pavesi, F., Petrovich, B., and Stanca, L. (2012).
\newblock Influential listeners: An experiment on persuasion bias in social networks.
\newblock {\em European Economic Review}, 56(6):1276--1288.

\bibitem[Cornfeld et~al., 1982]{ergo82}
Cornfeld, I., Fomin, S., and Sinai, Y.~G. (1982).
\newblock {\em Ergodic Theory}.
\newblock Springer, New York.

\bibitem[Cureg and Mukherjea, 2007]{muk07}
Cureg, E. and Mukherjea, A. (2007).
\newblock Convergence of convolution powers of a probability measure on d$\times$ d stochastic matrices and a cyclicity condition.
\newblock {\em Sankhy{\=a}: The Indian Journal of Statistics}, pages 289--303.

\bibitem[Dasaratha, 2020]{krish20}
Dasaratha, K. (2020).
\newblock Distributions of centrality on networks.
\newblock {\em Games and Economic Behavior}, 122:1--27.

\bibitem[Dasaratha et~al., 2023]{ben23}
Dasaratha, K., Golub, B., and Hak, N. (2023).
\newblock Learning from neighbours about a changing state.
\newblock {\em Review of Economic Studies}, 90(5):2326--2369.

\bibitem[DeGroot, 1974]{degroot74}
DeGroot, M.~H. (1974).
\newblock Reaching a consensus.
\newblock {\em Journal of the American Statistical association}, 69(345):118--121.

\bibitem[DeMarzo et~al., 2003]{dem03}
DeMarzo, P.~M., Vayanos, D., and Zwiebel, J. (2003).
\newblock Persuasion bias, social influence, and unidimensional opinions.
\newblock {\em The Quarterly journal of economics}, 118(3):909--968.

\bibitem[Dhar and Mukherjea, 1997]{muk97}
Dhar, S. and Mukherjea, A. (1997).
\newblock Convergence in distribution of products of iid nonnegative matrices.
\newblock {\em Journal of Theoretical Probability}, 10:375--393.

\bibitem[Edgar, 1998]{fractal98}
Edgar, G.~A. (1998).
\newblock {\em Integral, Probability, and Fractal Measures}.
\newblock Springer Science \& Business Media.

\bibitem[Erd{\"o}s, 1939]{erdo39}
Erd{\"o}s, P. (1939).
\newblock On a family of symmetric bernoulli convolutions.
\newblock {\em American Journal of Mathematics}, 61(4):974--976.

\bibitem[French, 1956]{french56}
French, J.~R. (1956).
\newblock A formal theory of social power.
\newblock {\em Psychological review}, 63(3):181.

\bibitem[Friedkin and Johnsen, 1990]{fri90}
Friedkin, N.~E. and Johnsen, E.~C. (1990).
\newblock Social influence and opinions.
\newblock {\em Journal of mathematical sociology}, 15(3-4):193--206.

\bibitem[Friedkin and Johnsen, 1999]{fri99}
Friedkin, N.~E. and Johnsen, E.~C. (1999).
\newblock Social influence networks and opinion change.
\newblock {\em Advances in Group Processes}, 16:1–29.

\bibitem[Furstenberg and Kesten, 1960]{furst60}
Furstenberg, H. and Kesten, H. (1960).
\newblock Products of random matrices.
\newblock {\em The Annals of Mathematical Statistics}, 31(2):457--469.

\bibitem[Galton, 1907]{galton907}
Galton, F. (1907).
\newblock Vox populi.
\newblock {\em Nature}, 75(1949):450--451.

\bibitem[Golub and Jackson, 2010]{moj10}
Golub, B. and Jackson, M.~O. (2010).
\newblock Naive learning in social networks and the wisdom of crowds.
\newblock {\em American Economic Journal: Microeconomics}, 2(1):112--49.

\bibitem[Golub and Jackson, 2012]{moj12}
Golub, B. and Jackson, M.~O. (2012).
\newblock How homophily affects the speed of learning and best-response dynamics.
\newblock {\em The Quarterly Journal of Economics}, 127(3):1287--1338.

\bibitem[Golub and Sadler, 2016]{golub17}
Golub, B. and Sadler, E. (2016).
\newblock Learning in social networks.
\newblock In {\em The Oxford Handbook of the Economics of Networks}. Oxford University Press.

\bibitem[Grimm and Mengel, 2020]{exp20}
Grimm, V. and Mengel, F. (2020).
\newblock Experiments on belief formation in networks.
\newblock {\em Journal of the European Economic Association}, 18(1):49--82.

\bibitem[Hennion, 1997]{hen97}
Hennion, H. (1997).
\newblock Limit theorems for products of positive random matrices.
\newblock {\em The Annals of Probability}, pages 1545--1587.

\bibitem[Hennion and Herv{\'e}, 2008]{hen08}
Hennion, H. and Herv{\'e}, L. (2008).
\newblock Stable laws and products of positive random matrices.
\newblock {\em Journal of Theoretical Probability}, 21:966--981.

\bibitem[H{\"o}gn{\"a}s and Mukherjea, 2003]{muk03}
H{\"o}gn{\"a}s, G. and Mukherjea, A. (2003).
\newblock Maximal homomorphic group image and convergence of convolution sequences on a semigroup.
\newblock {\em Journal of Theoretical Probability}, 16:847--854.

\bibitem[H{\"o}gn{\"a}s and Mukherjea, 2005]{muk05}
H{\"o}gn{\"a}s, G. and Mukherjea, A. (2005).
\newblock Products of iid random nonnegative matrices: Their skeletons and convergence in distribution.
\newblock {\em Sankhy{\=a}: The Indian Journal of Statistics}, pages 615--633.

\bibitem[Huang et~al., 2019]{bernd19}
Huang, J.-P., Heidergott, B., and Lindner, I. (2019).
\newblock Na{\"\i}ve learning in social networks with random communication.
\newblock {\em Social Networks}, 58:1--11.

\bibitem[Jackson, 2008]{moj08}
Jackson, M. (2008).
\newblock {\em Social and Economic Networks}.
\newblock Princeton University Press.

\bibitem[Jackson, 2019]{moj19}
Jackson, M.~O. (2019).
\newblock {\em The human network: How your social position determines your power, beliefs, and behaviors}.
\newblock Vintage.

\bibitem[Jackson et~al., 2023]{moj23}
Jackson, M.~O., Nei, S.~M., Snowberg, E., and Yariv, L. (2023).
\newblock The dynamics of networks and homophily.
\newblock Technical report, National Bureau of Economic Research.

\bibitem[Jackson and Watts, 2002a]{moj02}
Jackson, M.~O. and Watts, A. (2002a).
\newblock The evolution of social and economic networks.
\newblock {\em Journal of economic theory}, 106(2):265--295.

\bibitem[Jackson and Watts, 2002b]{moj022}
Jackson, M.~O. and Watts, A. (2002b).
\newblock On the formation of interaction networks in social coordination games.
\newblock {\em Games and Economic Behavior}, 41(2):265--291.

\bibitem[Kahneman et~al., 2021]{kahn21}
Kahneman, D., Sibony, O., and Sunstein, C.~R. (2021).
\newblock {\em Noise: A flaw in human judgment}.
\newblock Hachette UK.

\bibitem[Katz et~al., 2017]{leader17}
Katz, E., Lazarsfeld, P.~F., and Roper, E. (2017).
\newblock {\em Personal influence: The part played by people in the flow of mass communications}.
\newblock Routledge.

\bibitem[Katz et~al., 2001]{mto01}
Katz, L.~F., Kling, J.~R., and Liebman, J.~B. (2001).
\newblock Moving to opportunity in boston: Early results of a randomized mobility experiment.
\newblock {\em The quarterly journal of economics}, 116(2):607--654.

\bibitem[Kesten and Spitzer, 1984]{kes84}
Kesten, H. and Spitzer, F. (1984).
\newblock Convergence in distribution of products of random matrices.
\newblock {\em Zeitschrift f{\"u}r Wahrscheinlichkeitstheorie und Verwandte Gebiete}, 67:363--386.

\bibitem[Lo and Mukherjea, 1991]{mukherjea91}
Lo, C.-C. and Mukherjea, A. (1991).
\newblock Convergence in distribution of products of d$\times$ d random matrices.
\newblock {\em Journal of mathematical analysis and applications}, 162(1):71--91.

\bibitem[McKinlay, 2014]{mic14}
McKinlay, S. (2014).
\newblock A characterisation of transient random walks on stochastic matrices with dirichlet distributed limits.
\newblock {\em Journal of Applied Probability}, 51(2):542--555.

\bibitem[Molavi et~al., 2018]{nonbayes18}
Molavi, P., Tahbaz-Salehi, A., and Jadbabaie, A. (2018).
\newblock A theory of non-bayesian social learning.
\newblock {\em Econometrica}, 86(2):445--490.

\bibitem[Mukherjea, 1986]{muk86}
Mukherjea, A. (1986).
\newblock Completely simple semigroups of matrices.
\newblock In {\em Semigroup Forum}, volume~33, pages 405--429. Springer.

\bibitem[Mukherjea, 1987]{muk87}
Mukherjea, A. (1987).
\newblock Convergence in distribution of products of random matrices: a semigroup approach.
\newblock {\em Transactions of the American Mathematical Society}, 303(1):395--411.

\bibitem[Mukherjea and Nakassis, 2002]{mukherjea02}
Mukherjea, A. and Nakassis, A. (2002).
\newblock On the continuous singularity of the limit distribution of products of iid d$\times$ d stochastic matrices.
\newblock {\em Journal of Theoretical Probability}, 15:903--918.

\bibitem[Polanski and Vega-Redondo, 2023]{jet23}
Polanski, A. and Vega-Redondo, F. (2023).
\newblock Homophily and influence.
\newblock {\em Journal of Economic Theory}, 207:105576.

\bibitem[Rosenblatt, 1965]{rosen65}
Rosenblatt, M. (1965).
\newblock Products of independent identically distributed stochastic matrices.
\newblock {\em Journal of Mathematical Analysis and Applications}, 11:1--10.

\bibitem[Sawa and Wu, 2023]{ryoji23}
Sawa, R. and Wu, J. (2023).
\newblock Statistical inference in evolutionary dynamics.
\newblock {\em Games and Economic Behavior}, 137:294--316.

\bibitem[Seneta, 1981]{seneta81}
Seneta, E. (1981).
\newblock Non-negative matrices and markov chains.
\newblock {\em Springer Series in Statistics}.

\bibitem[Solomyak, 1995]{solomyak95}
Solomyak, B. (1995).
\newblock On the random series $\sum\pm\lambda^n$ (an erdos problem).
\newblock {\em Annals of Mathematics}, pages 611--625.

\bibitem[Stoll, 1951]{stoll51}
Stoll, R. (1951).
\newblock Homomorphisms of a semigroup onto a group.
\newblock {\em American Journal of Mathematics}, 73(2):475--481.

\bibitem[Surowiecki, 2005]{wis05}
Surowiecki, J. (2005).
\newblock {\em The wisdom of crowds}.
\newblock Anchor.

\bibitem[Tahbaz-Salehi and Jadbabaie, 2008]{iff08}
Tahbaz-Salehi, A. and Jadbabaie, A. (2008).
\newblock A necessary and sufficient condition for consensus over random networks.
\newblock {\em IEEE Transactions on Automatic Control}, 53(3):791--795.

\bibitem[Tahbaz-Salehi and Jadbabaie, 2009]{iff09}
Tahbaz-Salehi, A. and Jadbabaie, A. (2009).
\newblock Consensus over ergodic stationary graph processes.
\newblock {\em IEEE Transactions on automatic Control}, 55(1):225--230.

\bibitem[Touri, 2012]{touri12}
Touri, B. (2012).
\newblock {\em Product of random stochastic matrices and distributed averaging}.
\newblock Springer Science \& Business Media.

\bibitem[Touri and Nedi{\'c}, 2013]{touri13}
Touri, B. and Nedi{\'c}, A. (2013).
\newblock Product of random stochastic matrices.
\newblock {\em IEEE Transactions on Automatic Control}, 59(2):437--448.

\bibitem[Ullman, 1972]{ull72}
Ullman, J. (1972).
\newblock On the regular behaviour of orthogonal polynomials.
\newblock {\em Proceedings of the London Mathematical Society}, 3(1):119--148.

\bibitem[Van~Assche, 1986]{va86}
Van~Assche, W. (1986).
\newblock Products of 2$\times$ 2 stochastic matrices with random entries.
\newblock {\em Journal of applied probability}, 23(4):1019--1024.

\bibitem[Weld et~al., 2015]{ai15}
Weld, D.~S., Lin, C.~H., and Bragg, J. (2015).
\newblock Artificial intelligence and collective intelligence.
\newblock {\em Handbook of collective intelligence}, pages 89--114.

\end{thebibliography}
\bibliographystyle{apalike}
\end{singlespace}

  \clearpage

\phantomsection\label{app:online}
\begin{center}
{\LARGE\bf Online Appendix: \par ``Collective Intelligence in Dynamic Networks''}
\end{center}
The online appendix is organized as follows. Online Appendix \hyperref[sec:disagree]{A} shows how long-run disagreement can be explained in our framework. Online Appendix \hyperref[app:ex]{B} explores other examples of dynamic networks.
Online Appendix \hyperref[sec:disc3]{C} relates our positive results to \citeauthor{bernd19}'s (\citeyear{bernd19}) negative results about collective intelligence. Online Appendix \hyperref[app:oproof]{D} collects the omitted proofs. 

\phantomsection\label{sec:disagree}
\section*{Online Appendix A: Disagreement}
The main analysis of this paper focused on consensus and social learning. However, the leading criticism of standard DeGroot learning is that it can explain consensus but not \textit{disagreement}.\footnote{\citet[][p. 3]{fri99} note on DeGroot literature: ``These initial formulations described the formation of group consensus, but did not provide an adequate account of settled patterns of disagreement.''}  
The most popular way to introduce disagreement was pioneered by \citet{fri90}.
They assume a very specific source of disagreement: agents are \textit{stubborn}---they keep putting disproportionate weight on their own initial beliefs \citep[e.g.,][Example 8.3.8]{moj08}. In fact, \citet{frank25} find experimental evidence that people tend to systematically underweight equally-relevant information that their peers discovered.
\par This appendix shows that stubbornness can also be captured in our dynamic model. We show that introducing random network dynamics suffices to explain both consensus and (arbitrary forms of) disagreement. To this end, let $\{\m{X}_t\}_{t\geq1}$ be iid so that we can build on the tools in Appendix \hyperref[app:conv]{A.III}. We start by defining the degree of disagreement in a society as the \textit{minimal rank} of all the interaction matrices  in $\varSigma$---the closed semigroup generated by the support of $\mu$, denoted $\varSigma_{\mu}$. This set is formally defined in Appendix \hyperref[app:tech]{A}: eq. (\ref{eq:sigma}).
\begin{definition}\label{def:rank}\normalfont
    The \textit{degree} of disagreement in a society is $\eta=\text{min}\big\{\text{rank}(\sigma):\forall\sigma\in\varSigma\big\}$.\hfill$\bigtriangleup$
\end{definition}
Importantly, the parameter $\eta\in\{1,\dots,n\}$ in Definition \ref{def:rank} is also the rank of \textit{all} the matrices in $\varSigma_\psi$---the limiting support of the random product $\m{X}^{(t)}$---which coincides with the ``kernel'' of $\varSigma$ (Appendix \hyperref[app:tech]{A}: eq. (\ref{eq:kernel})). At one extreme, $\eta=1$ (first degree) is trivial because it corresponds to consensus. When $\eta=2$, there are two distinct groups of agents whose limiting beliefs agree within each group but disagree with the limiting beliefs of those outside the group. When $\eta=3$, there are three such groups, and so on for $\eta>3$. At the other extreme, $\eta=n$ ($n^\text{th}$ degree) means total disagreement, i.e., no two agents' limiting beliefs ever coincide, hence $\varSigma=\varSigma_\psi$ must be a finite set (Appendix \hyperref[app:finite]{A.III}). Thus, as $\eta$ increases from $1$ (consensus) to $n$ (total disagreement), our framework captures how societies can get \textit{fragmented} into increasingly larger subgroups of agents with disagreeable limiting beliefs.
  
  \par In general, the characterization of disagreement in random dynamic networks is an open problem. There exists, however, special cases where strong predictions can be made. To see this, we first introduce a useful definition due to \citet[][p. 169]{rao98}.
\begin{definition}[Cyclicity]\label{def:cyclic}\normalfont
   The support of $\mu$, denoted $\varSigma_\mu$, is said to be \textit{cyclic} with respect to a set $\{A_1,\dots,A_m\}$ if, for each $\sigma\in\varSigma_\mu$, $\sum_{\ell\in A_{s+1}}(\sigma)_{i\ell}=1$ for $i\in A_{s}$ and $s=1,\dots,m$, where the $A_j$'s, for $j=1,\dots,m$ are all pairwise disjoint subsets of $N=\{1,\dots,n\}$ such that $\bigcup_{j=1}^mA_j$ may or may not equal $N$ and $A_{m+j}:=A_j$ for all $j=1,\dots,m$. \hfill$\bigtriangleup$
\end{definition}
\textit{Cyclicity} can be viewed as a generalization of periodicity \citep[][Definition 2]{moj10}. For instance, let $n=2$, so if $\varSigma_\mu$ is cyclic with respect to $A_1=\{1\}$ and $A_2=\{2\}$ (Definition \ref{def:cyclic}), then it must be that $\varSigma_\mu=\big\{\big(\begin{smallmatrix}
    0&1\\1&0
\end{smallmatrix}\big)\big\}$, which is periodic. When $\eta=2$, the definition above simplifies to the fact that there exist two disjoint subsets $A_1$ and $A_2$ of $N$ such that $\sum_{\ell\in A_{2}}(\sigma)_{i\ell}=1$ for $i\in A_{1}$ and $\sum_{\ell\in A_{1}}(\sigma)_{i\ell}=1$ for $i\in A_{2}$. The next result uses this to characterize  \textit{second-degree} disagreement $(\eta=2)$ in terms of the cyclicity of $\varSigma_\mu$. 
\begin{proposition}\label{thm:cyclic}
    Suppose $\{\m{X}_t\}_{t\geq1}$ is iid  and $\eta=2$. Then, $\big\{\m{X}^{(t)}\big\}_{t\geq1}$ converges weakly as $t\rightarrow\infty$ if and only if $\varSigma_\mu$ is not cyclic.
\end{proposition}
This result shows that a society reaches second-degree disagreement whenever $\varSigma_\mu$ is not cyclic, 
which is due to \citet[][Theorem 2.1]{muk07}. When $\eta=2$ and agents' beliefs $\m{p}^{(t)}=\m{X}^{(t)}\m{p}^{(0)}$ converges weakly to $\m{p}^{(\infty)}$, all matrices in the limiting support $\varSigma_\psi$ will have rank two, i.e., $\text{rank}(\sigma)=\eta=2$, for all $\sigma\in\varSigma_\psi$. 
 An intuitive example of second-degree disagreement is the two-party divide observed in US Congress, where Republicans tend to agree (i.e., vote on the same side of a bill) with their fellow Republicans but disagree with Democrats, and vice versa \citep[see,][Figure 7.4(b)]{moj19}.
 The next example illustrates how our framework can capture \citeauthor{fri90}'s (\citeyear{fri90}) insights about stubbornness. 
\begin{ex}\label{ex:disag}\normalfont
    Let $n=3$, $\varSigma_{\mu}=\{h_{\kappa},g\}$, where $h_{\kappa}=\Big(\begin{smallmatrix}
        0&0&1\\
        0&0&1\\
        \kappa &1-\kappa&0
    \end{smallmatrix}\Big)$, $g=\Big(\begin{smallmatrix}
        0&1&0\\
        1&0&0\\
        0&0&1
    \end{smallmatrix}\Big)$, for some constant $\kappa\in[0,1]$. 
    Then, when $\{\m{X}_t\}_{t\geq1}$ are iid, the limiting distribution $\psi$ exists with support $\varSigma_{\psi}=\big\{h_{\kappa},h_{1-\kappa},q_{\kappa},q_{1-\kappa}\big\}$, where  $q_{\kappa}=\Big(\begin{smallmatrix}
        \kappa&1-\kappa&0\\
       \kappa&1-\kappa&0\\
        0 &0&1
    \end{smallmatrix}\Big)$ \citep[][Section 4]{muk97}. For instance, if $\kappa=1/2$, then $\varSigma_{\psi}=\{h_{1/2},q_{1/2}\}$, and hence $\psi\{h_{1/2}\}=1/2$. More generally, if $\mu\{h_{\kappa}\}=r$, then $\psi\{h_{\kappa}\}+\psi\{h_{1-\kappa}\}=\psi\{q_{\kappa}\}+\psi\{q_{1-\kappa}\}=1/2$, where $\psi\{h_{\kappa}\}=\psi\{q_{\kappa}\}=\frac{1}{2(2-r)}$, for $r\in(0,1)$. Here, \eqref{eq:C} is not satisfied because $\varSigma_{\psi}$ contains matrices that are not strictly positive. Instead, $\varSigma_{\psi}$ consists of rank-two matrices, i.e., $\eta=2$, and it follows that agents 1 and 2 will reach consensus but always disagree with agent 3. It can also be seen (by $g\in\varSigma_\mu$) that long-run disagreement arises here because agent 3 tends to be stubborn.\hfill$\bigtriangleup$
\end{ex}

Extending Proposition \ref{thm:cyclic} for $\eta>2$ will not work---\citet[][Section 3]{muk07} show that cyclicity does not characterize weak convergence when $\eta>2$. In the next result, we therefore use a different approach to characterize total disagreement ($\eta=n$).
\begin{proposition}
    Suppose $\{\m{X}_t\}_{t\geq1}$ is iid  and $\eta=n$. Then, $\big\{\m{X}^{(t)}\big\}_{t\geq1}$ converges weakly if and only if there exists a $\tau\geq1$ such that the support of $\m{X}^{(\tau)}$ is the finite set $\varSigma_\psi=\varSigma$. When the limiting distribution $\psi$ exists, it is the uniform probability measure on $\varSigma_\psi$.
\end{proposition}
This result is a special case of Lemma \ref{thm:finite} because when $\eta=n$, $\varSigma_\psi=\varSigma$ is finite (Appendix \hyperref[app:finite]{A.III}). As noted after the statement of Lemma \ref{thm:finite}, a sufficient condition for $\psi$ to exist when $\eta=n$ is that the greatest common divisor of the orders of the elements in $\varSigma$ is one. 
\par\noindent--- \textit{Interpretation}: Lemma \ref{thm:finite} shows that total disagreement is very special: when agents reach total disagreement in the limit, it must be that the agents have totally disagreed at every $t$. 
 \par We also consider the predictability of disagreement. The next result follows directly from Theorem \ref{thm:skel}.(i), so we use the language of societies $X$ and $Y$ introduced in Section \ref{sec:fragile}.
\begin{corollary}
In Theorem \ref{thm:skel}.(i), choose any $k\in\{1,\dots,n\}$. Then, Society $X$ reaches $k^{\text{\normalfont th}}$-degree disagreement if and only if Society $Y$ reaches $k^{\text{\normalfont th}}$-degree disagreement.
\end{corollary}
A key step to prove Theorem \ref{thm:skel}.(i)  (Lemma \ref{thm:rank}) shows that the interaction matrices in the limiting supports $\varSigma_{\psi_X}$ and $\varSigma_{\psi_Y}$ have the \textit{same} rank, i.e., $\eta_X=\eta_Y(=k)$, when $\Proba(\mathscr{T})=1$. Thus, the initial social topology (Definition \ref{def:topo}) also reveals disagreement in iid networks.

\phantomsection\label{app:ex}
\section*{Online Appendix B: Other Numerical Examples}
\phantomsection\label{sec:ex}
\subsection*{B.I Setup of Numerical Examples}
There are two categories of dynamic/time-dependent interaction matrices: they can either be deterministic or random. We illustrate simple examples in each category.

\textit{-- Deterministic setting}: \citet[][]{seneta77} characterize the convergence of products of deterministic time-dependent interaction matrices. For example, suppose $(\m{X}_{t})_{ij}\rightarrow1/n$ deterministically, for all $i,j\in N$, as $t\rightarrow\infty$.
A more sophisticated example of deterministic interaction matrices is due to \citet[][eq. (5)]{dem03}, where for all $t$ \begin{align}\label{eq:dem}
    \m{X}_t=(1-\lambda_t)\m{I}_n+\lambda_t\m{T},
\end{align} where $\lambda_t\in(0,1]$ is deterministic for all $t$. When $\lambda_t=1$ for all $t$, we recover the standard DeGroot model. When $\lambda_t=\big(1+\frac{k}{1+k}t\big)^{-1}$, for a constant $k$, \citet[][footnote 19]{dem03} interpret this as capturing  the agents having ``increased self-confidence''. 
\par \textit{-- Random setting}. We start with simple $2\times2$ examples, then move to the $n\times n$ case.
\begin{ex}[$2\times2$]\label{ex:2x2}\normalfont 
  Consider two dynamic networks: (1) when the matrix $\m{X}_0$ is either
 \begin{align*}
      \begin{pmatrix}
          x &1-x\\0&1  
        \end{pmatrix} \quad \text{\normalfont or} \quad
     \begin{pmatrix}
          1&0\\1-x &x 
        \end{pmatrix}
 \end{align*}
 with equal probability for a constant $x\in(0,1)$ and the $\m{X}_t$'s are iid with $\m{X}_0$ for $t\geq1$, and an interpretation is given shortly below. (2) For each $t\geq1$, let the matrix $\m{X}_t$ be
 \begin{align*}
     \m{X}_t=\begin{pmatrix}
          a_t&1-a_t\\b_t &1-b_t 
        \end{pmatrix},
 \end{align*}
 where $x^{-1}a_t\sim\text{Bern}(p_a)$ and $x^{-1}b_t\sim\text{Bern}(p_b)$ are Bernoulli random variables with (distinct) parameters $p_a,p_b\in(0,1)$ for all $t$ and a constant $x\in(0,1/2]$.\footnote{That is, $a_t\sim p_a\delta_x+(1-p_a)\delta_0$ and $b_t\sim p_b\delta_x+(1-p_b)\delta_0$, for all $t$, where $\delta_z$ is the unit mass at any $z$. } 
 \hfill$\bigtriangleup$
\end{ex}

 \begin{ex}[$n\times n$]\label{ex:mckinlay}\normalfont 
 
The following example is due to \citet[][Section 4.1]{mic14}. 
It is similar to the dynamic network in Section \ref{sec:wisex} but allows correlations across rows:
      \begin{align*}
\m{X}_0=\begin{pmatrix}
    x &1-x &0&\dots&0\\
    0&\mathds{1}_{x\geq\frac{1}{2}}&1-\mathds{1}_{x\geq\frac{1}{2}}&\ddots&\vdots\\
    \vdots&\ddots&\ddots&\ddots&0\\
    0&\dots&0&\mathds{1}_{x\geq\frac{1}{2}}&1-\mathds{1}_{x\geq\frac{1}{2}}\\
    1-\mathds{1}_{x\geq\frac{1}{2}}&0&\dots&0&\mathds{1}_{x\geq\frac{1}{2}}
\end{pmatrix},
 \end{align*}
 where $x\sim\mathcal{U}(0,1)$ and the $\m{X}_t$'s are iid with $\m{X}_0$ for $t\geq1$. 
This network can be interpreted as follows. At time $t\geq0$, agent 1, acting as a ``leader'' (or ``first mover''), allocates a uniform proportion of her attention to agent 2. If this proportion is greater than $1/2$, then all other agents pay no attention to their peers. If instead the proportion is less than or equal to $1/2$, then each other agent $i\neq1$ pays full attention to her neighboring
peer $i + 1\hspace{0.03in} (\text{mod } n)$. This example captures the so-called ``two-step communication'' process where some agents may operate as ``opinion leaders'' and others as ``opinion followers'' \citep[see,][]{leader17}.\hfill$\bigtriangleup$
 \end{ex}

 \phantomsection\label{sec:ex2}
\subsection*{B.II Analysis of Numerical Examples}

This appendix revisits all the numerical examples above to study their limiting properties.

\begin{ex}[]\label{ex:unif2}
    \normalfont When $(\m{X}_{t})_{ij}\underset{}{\rightarrow}1/n$ for all $i,j\in N$, as $t\rightarrow\infty$, then it is easy to see that consensus is reached 
     \citep[e.g.,][Theorem 5's Corollary]{seneta77}. When instead $\m{X}_t=(1-\lambda_t)\m{I}_n+\lambda_t\m{T}$ (eq. (\ref{eq:dem})), \citet[][Theorem 1]{dem03} show that consensus is reached when $\sum_{t=0}^{\infty}\lambda_{t}=\infty$ and $\m{T}$ is strongly connected.
    \hfill$\bigtriangleup$ 
\end{ex}

\begin{ex}[$2\times2$, revisited]\label{ex:2x2rev}
    \normalfont 
\par (1) The limiting distribution $\psi$ exists and $\pi\sim F$ denotes agent 1's influence. The distribution $F$ (with respect to $\psi$) uniquely solves the functional equation
    $F(\pi)=\frac{1}{2}F\big(\frac{\pi}{x}\big)+\frac{1}{2}F\big(\frac{\pi-(1-x)}{x}\big),$
 where $F(0)=0$, $F(1)=1$ \citep[][eq. (1.2)]{mukherjea02}. 
 Solving this equation is an open problem known as Erd{\"o}s problem \citep[][]{erdo39}. Some special cases are known. \citet{solomyak95} finds that $F$ is absolutely continuous for almost all $x\in(1/2, 1)$. When $x=1/2$, $F=\mathcal{U}[0,1]$. For $x<1/2$, $F$ is continuous singular, e.g., when  $x=1/3$, $F$ is the Cantor distribution ($\varSigma_{\psi}$ is the Cantor set). This shows that it is not even clear when $F$ is discrete, continuous singular, or absolutely continuous. This remark dates back to \citet[][eq. (47)]{rosen65}, and yet it remains an open problem. \par (2) Here, $\varSigma_\mu=\big\{v_{0,0},v_{0,x},v_{x,0},v_{x,x}\big\}$ and each element has probability (according to $\mu$) $p_{00},p_{01},p_{10},p_{11}$, respectively, where $p_{10}+p_{11}=p_a$,  $p_{01}+p_{11}=p_b$, $v_{s,r}=\big(\begin{smallmatrix}
    s&1-s\\r&1-r
\end{smallmatrix}\big)$, and define $v_s:=v_{s,s}$. Then, $\psi$ exists and $\text{lim}_{t\rightarrow\infty}\m{X}^{(t)}=\big(\begin{smallmatrix}
        \pi&1-\pi\\
        \pi&1-\pi
    \end{smallmatrix}\big)$ exists, so if  $x=1$, $\pi\in\{0,1\}$, then $\psi\{v_{0}\}=\frac{p_{00}(1-p_{10})+p_{11}p_{01}}{(1-p_{10})^2-p_{01}^2}$. When $x\in(0,1/2]$, $\varSigma_\psi\subset\big\{v_{\pi}:\pi\in[0,x]\big\}$, where for $\pi\in\{0,1\}$, $\psi\{v_{0}\}=\frac{p_{00}}{1-p_{10}}$, $\psi\{v_{x}\}=\frac{p_{11}(1-p_{10})+p_{00}p_{01}}{1-p_{10}}$, and for $\pi\in(0,x)$ with $\psi\{v_\pi\}>0$, $\psi\{v_{\pi}\}=p_{10}\psi\{v_{\pi/x}\}+p_{01}\psi\{v_{1-\pi/x}\}$ \citep[see,][Section 2]{muk14}.
  \hfill$\bigtriangleup$
\end{ex}

\begin{ex}[$n\times n$, revisited]\label{ex:mckinlay2}\normalfont
In  Example \ref{ex:mckinlay}, notice that Corollary \ref{thm:mic2} cannot be applied because the rows are correlated.  
 However, this dynamic network surprisingly has the same limiting distribution as the second example in Section \ref{sec:wisex} \citep[see,][Section 4.1,  Example 2]{mic14}. It therefore follows by Corollary \ref{thm:mic} that the distribution of the influence vector satisfies $\m{\pi}\sim D_{\m{\varphi}}$, i.e., a Dirichlet distribution with parameter $\m{\varphi}=(2,\dots,2)\in\R^n_+$. Thus, this dynamic network is also collectively intelligent by Proposition \ref{thm:mic3}.  
\hfill$\bigtriangleup$
\end{ex}

A general theme in all the above examples is that random dynamics in social networks renders the computation of consensus and the limiting distribution of the influence vector very challenging (even when $n=2$). This is an active research area in random matrix theory. 

\phantomsection\label{sec:disc3}
\section*{Online Appendix C: Robustness of Collective Intelligence}

\citet{bernd19} show, using examples and simulations, that the wisdom of crowds may fail to hold even when no agent has influence as random networks grow. In particular, they consider the case where there is a finite collection of independent random influence matrices. The examples in their Section 6 focus specifically on the case where the interaction matrix is a weighted average of two non-wise matrices. 
In what follows, we show that their restriction to a finite collection of matrices dramatically limits the possibility of even reaching consensus. 

\par Suppose the limiting distribution $\psi$ exists in the sense of Remark \ref{rem:conv}. We then ask: ``What is the probability that the agents reach consensus?'' That is, we seek to understand how likely it is that $n\geq2$ agents would reach consensus when their beliefs are known to converge. The next result answers this question when the influence matrices are iid and $\varSigma$ is finite, so that $\varSigma_{\mu}$ necessarily contains a finite number of elements as in \citet{bernd19}.  
Let $\phi_{n}:=\Proba\big(R_{n,m}=1\big)\in[0,1]$, where $R_{n,m}$ denotes the rank of the $m$-th $n\times n$ matrix in $\varSigma$, for any $m=1,\dots,k$. 
Also, $|.|$ stands for cardinality.
\begin{proposition}\label{thm:prob}
Suppose the limiting distribution $\psi$ of the iid sequence $\big\{\m{X}^{(t)}\big\}_{t\geq1}$ exists and $|\varSigma|=k<\infty$. Then, the probability that $n$ agents reach consensus is
\begin{align}\label{eq:prob}
    \sum_{j=0}^{k-1}\binom{k}{j}\big(1-\phi_n\big)^j\phi_n^{k-j}\in[0,1].
\end{align}

\end{proposition}
The rank of matrices appears in this result for a profound reason, which we elaborate in details in Appendix \hyperref[app:conv]{A.III}: when a product of iid random nonnegative matrices converges in distribution, the support $\varSigma_{\psi}$ of the limiting distribution $\psi$ consists of all the matrices in $\varSigma$ that have \textit{minimal} rank.
Then, the next example illustrates that the probability of reaching consensus may vanish as the number of agents $n$ increases.
\begin{ex}\normalfont
    Let $k=2$, then the probability of consensus is $\phi_n\big(2-\phi_n\big)$ (eq. (\ref{eq:prob})). Suppose the rank $R_{n,m}$ of the $m$-th matrix in $\varSigma$ is uniformly distributed on $\{1,\dots,n\}$, for all $m=1,2$, so each of its values has probability $1/n$. Thus, $\phi_n=1/n$ such that consensus is reached with probability $(2-1/n)/n$, and therefore, this probability vanishes as $n\rightarrow\infty$.\hfill$\bigtriangleup$
\end{ex}

Although the above restriction to $k=2$ might seem extreme, \citet{bernd19} assume the same in their examples to show the failure of the wisdom of crowds. We will now consider what happens when the size of the set $\varSigma$ is unrestricted. The next result shows that when the agents' beliefs converge, consensus can be reached on \textit{average} irrespective of $n$.  
\begin{proposition}\label{thm:avg}
Suppose the limiting distribution $\psi$ of the iid sequence $\big\{\m{X}^{(t)}\big\}_{t\geq1}$ exists, $\psi(\hat{\varSigma})>0$, and $|\varSigma|$ is unrestricted. Then, $\int_{\varSigma} \sigma\hspace{0.03in}d\psi(\sigma)$ has rank one, for all $n$. 
\end{proposition}
 This result states that when the size of $\varSigma$  is unrestricted and agents' beliefs converge, then consensus is reached on average for any $n$. 
 Intuitively, this result indicates that the randomness in social networks needs to be sufficiently \textit{rich} to ensure a society can reach consensus. The assumptions in Proposition \ref{thm:avg} are easily satisfied, e.g., when there exists an integer $r\geq1$ such that the convolution power satisfies $\mu^r(\hat{\varSigma})>0$, then $\psi$ exists and $\psi(\hat{\varSigma})>0$ \citep[see,][eq. (1)]{muk14}.

\phantomsection\label{app:oproof}
\section*{Online Appendix D: Proofs}

\subsection*{Proof of Proposition \ref{thm:wis}}
\begin{proof}
Let $M_n := \max_{i \leq n} \pi_i(n)$. Then, by union bound, for any $\epsilon > 0$:
\begin{align*}
\mathbb{P}(M_n > \epsilon) = \mathbb{P}\left(\bigcup_{i=1}^n {\pi_i(n) > \epsilon}\right) \
\leq \sum_{i=1}^n \mathbb{P}(\pi_i(n) > \epsilon).
\end{align*}
Since $\max_i \mathbb{E}[\pi_i(n)] \to 0$ for every $i\leq n$,  we have $\mathbb{E}[\pi_i(n)] < \epsilon/2$ for large $n$. Thus,
\begin{align*}
\mathbb{P}(\pi_i(n) > \epsilon) \leq \mathbb{P}\big(\big|\pi_i(n) - \mathbb{E}[\pi_i(n)]\big| > \varepsilon/2\big) \
\leq \frac{4\text{var}(\pi_i(n))}{\epsilon^2}
\end{align*}
using Chebyshev inequality. 
Now, since $\text{var}(\pi_i(n))=O(n^{-d})$, for some $d>1$, there exists a constant $C$ such that $\text{var}(\pi_i(n))\leq C/n^d$. Then, as $n\rightarrow\infty$, we have
\begin{align*}
\mathbb{P}(M_n > \epsilon) \
\leq \sum_{i=1}^n \mathbb{P}(\pi_i(n) > \epsilon)\leq n \frac{4C}{\epsilon^2 n^d} = \frac{4C}{\epsilon^2 n^{d-1}} \to 0 .
\end{align*}
Since $M_n$ is uniformly bounded on $[0,1]$ and $M_n\rightarrow0$ in probability, then $\E[M_n]\rightarrow0$ as $n\rightarrow0$ (by dominated convergence theorem). By Theorem \ref{thm:wisdom}, collective intelligence holds.
\end{proof}

\subsection*{Proof of Proposition \ref{thm:prob}}
\begin{proof}
    When the limiting distribution $\psi$ exists, its support satisfies $\varSigma_{\psi}=K$, where $$K=\big\{{y}\in\varSigma:\text{rank}({y})\leq\text{rank}({\sigma}), \forall{\sigma}\in\varSigma\big\}$$
  is the kernel of $\varSigma$, so its elements are the $n\times n$ stochastic matrices in $\varSigma$ with minimal rank (see, eq. (\ref{eq:kernel})). Now, let $R_{n,m}$ denote the rank of the $m$-th $n\times n$ stochastic matrix in $\varSigma$, for $m=1,\dots,k$. Then, agents reach consensus when $\text{min}_mR_{n,m}=1$ such that the limiting support $\varSigma_{\psi}$ contains only rank-one stochastic matrices. Moreover, $\text{min}_mR_{n,m}$ is a first order statistics that takes values in $\{1,2,\dots n\}$, where zero is excluded because the matrices are stochastic. Since there are $k<\infty$ such matrices in $\varSigma$ and these matrices are iid, then so are their ranks $\{R_{n,m}\}_{m=1}^k$. Thus, the probability of reaching consensus coincides with the probability that the first order statistics equals 1, i.e., $\Proba(\text{min}_mR_{n,m}=1)$, which is given by
  \begin{align*}
      \sum_{j=0}^{k-1}\binom{n}{j}\Big[(1-{\phi}_n)^j{\phi}_n^{k-j}-(1-{\phi}_n+p_n)^j({\phi}_n-p_n)^{k-j}\Big],
  \end{align*}
  where $p_n=\Proba(R_{n,m}= 1)$ and $\phi_n=\Proba(R_{n,m}\leq1)$, and since $R_{n,m}$ cannot take values below 1 (i.e., the matrices are stochastic), $p_n={\phi}_n$, so the expression above simplifies to eq. (\ref{eq:prob}).
\end{proof}

\subsection*{Proof of Proposition \ref{thm:avg}}

This result follows directly by combining  \citet[][Lemmas 9.2 and 9.3]{hen97} for the special case of random stochastic matrices.

\phantomsection\label{thm21}
\subsection*{Proof of Theorem \ref{thm:skel}.(i)}
The proof of Theorem \ref{thm:skel}.(i) is technical and first appeared in \citet{muk05}. However, \citet{muk05}  mix results for stochastic matrices with those of nonnegative matrices, so the goal of this appendix is to collect a simplified version of the proof for stochastic matrices. We aim to show that, when $\Proba(\mathscr{T})=1$, $\big\{\m{X}^{(t)}\big\}_{t\geq1}$ converges weakly if and only if $\big\{\m{Y}^{(t)}\big\}_{t\geq1}$  converges weakly, which is equivalent to showing that the corresponding convolution sequence $\mu^t_X$ converges weakly if and only if $\mu^t_Y$ converges weakly.  As outlined in the sketch (Appendix \hyperref[app:sketch]{A.IV}), it suffices to prove Lemmas \ref{thm:rank}--\ref{thm:support} respectively.
\par Let $\mu_X$ and $\mu_Y$ be probability measures on the Borel subsets of $n\times n$ stochastic
matrices. Let $\varSigma_X=\text{closure}\big(\bigcup_{t\geq1}\varSigma^t_{\mu_X}\big)$ and $\varSigma_Y=\text{closure}\big(\bigcup_{t\geq1}\varSigma^t_{\mu_Y}\big)$ be, respectively, the
compact multiplicative semigroups of stochastic matrices generated by the
supports $\varSigma_{\mu_X}$ and $\varSigma_{\mu_Y}$, respectively (eq. (\ref{eq:sigma})). For any $\sigma\in\varSigma_X\cup \varSigma_Y$, denote by $\overline{\sigma}$ the idempotent (identity) element in the group of limit points $\{\sigma^t:t\geq1\}$.\footnote{For any stochastic matrix $\sigma$, the set of limit points of the sequence $\sigma^t$ coincides with the set $\bigcap_{k=1}^{\infty}\text{closure}\big(\{\sigma^t:t\geq k\}\big)$, which forms a group. The identity of this group is idempotent, denoted $\overline{\sigma}$.} A useful property that we will need is that any idempotent nonnegative matrix $p$ uniquely defines a partition 
\begin{align}\label{eq:basis}
    \mathscr{B}_{\eta}=\{T,C_1,\dots,C_{\eta}\}
\end{align}
of the set $N=\{1,\dots,n\}$ called the \textit{basis} of $p$ with rank $\eta$ \citep[see,][]{muk86}. An element $i$ in $N$ belongs to $T$ if and only if either the $i$-th row of $p$ is a zero row or $i$-th column of $p$ is a zero column or both. For each $j=1,\dots,\eta$, the $C_j\times C_j$ block of $p$ is a matrix of rank one whose entries are all strictly positive. For $j,k=1,\dots,\eta$ with $j\neq k$, the $C_j\times C_k$ block of $p$ is an all zero block. In our context, $p$ is a stochastic matrix, so each $C_j\times C_j$ block of $p$ is a strictly positive stochastic matrix with rank one.
\par Our definition of \textit{social topology} in Definition \ref{def:topo} is identical to that of \textit{skeleton} in \citet[][Assumption 1]{muk05}. We start by following \citet[][Remark 1]{muk05} to show that $\Proba(\mathscr{T})=1$ extends to $\varSigma_X$ and $\varSigma_Y$, where we recall that
\begin{align*}
    \mathscr{T}=\big\{\m{X}_1\text{ \normalfont and } \m{Y}_1\text{ \normalfont have the same social topology}\big\}.
\end{align*}
To see this, define the event $$\mathscr{A}=\Big\{\m{X}_1\in\varSigma_{\mu_X}, \m{Y}_1\in\varSigma_{\mu_Y}, \m{X}_1\text{ \normalfont and } \m{Y}_1 \text{ \normalfont have the same social topology}\Big\}$$
then $\Proba(\mathscr{A})=1$ whenever $\Proba(\mathscr{T})=1$ holds. Define the sets $C=\m{X}_1(\mathscr{A})$ and $D=\m{Y}_1(\mathscr{A})$. Then, $\text{closure}(C)\subset\varSigma_{\mu_X}$, $\text{closure}(D)\subset\varSigma_{\mu_Y}$, and $\mu_X(\text{closure}(C))=1=\mu_Y(\text{closure}(D))$. It therefore follows that $\text{closure}(C)=\varSigma_{\mu_X}$ and $\text{closure}(D)=\varSigma_{\mu_Y}$. This then means that for any $x\in\varSigma_{\mu_X}$, there exists a sequence $x_t\in\varSigma_{\mu_X}$ such that $x_t\rightarrow x$ as $t\rightarrow\infty$, and $y_t\in\varSigma_{\mu_Y}$ such that $y_t$ and $x_t$ have the same social topology for each $t$, and for some $y\in\varSigma_{\mu_Y}$, $y_t\rightarrow y$ as $t\rightarrow\infty$. We next show that this property extends to $\varSigma_X=\text{closure}\big(\bigcup_{t\geq1}\varSigma^{t}_{\mu_X}\big)$ and $\varSigma_Y=\text{closure}\big(\bigcup_{t\geq1}\varSigma^{t}_{\mu_Y}\big)$. Let $x\in \varSigma_X$, then there  exists $x_{t_k}\in\varSigma^{t_k}_{\mu_X}$, for some subsequence $\{t_k\}$, such that $x_{t_k}\rightarrow x$ as $t\rightarrow\infty$. For a fixed $k$, define $x_{t_k}=z_{t_k}\dots z_2z_1$, where $z_j\in\varSigma_{\mu_X}$ for all $j=1,\dots,t_k$. Then, there exist sequences $\big\{u_j\in\varSigma_{\mu_X}\big\}_{j=1}^{t_k}$ and $\big\{v_j\in\varSigma_{\mu_Y}\big\}_{j=1}^{t_k}$, where $u_j$ and $v_j$ have the same social topology for each $j=1,\dots,t_k$, and $\lVert z_j-u_j\lVert<\epsilon_k$,\footnote{Here, let $\lVert.\lVert$ denote the standard Euclidean norm.} for some small enough $\epsilon_k$ such that $\big\lVert z_{t_k}\dots z_2z_1-u_{t_k}\dots u_2u_1 \big\lVert<1/k$. That is, given any $x\in\varSigma_{\mu_X}$, there exists $x_{t_k}\in\varSigma_{X}$ and $y_{t_k}= v_{t_k}\dots v_2v_1\in\varSigma_{Y}$ such that, for each $k$, $x_{t_k}$ and $y_{t_k}$ have the same social topology and $x_{t_k}\rightarrow x$. The same property continues to hold when we interchange the role of $X$ and $Y$ above. We are now ready to prove Lemma \ref{thm:rank}. 
\begin{proof}[Proof of Lemma \ref{thm:rank}]
Let $p\in\varSigma_{\psi_X}$ be an idempotent element with basis $\mathscr{B}_{\eta}=\{T,C_1,\dots,C_{\eta}\}$ in eq. (\ref{eq:basis}) such that among all idempotent elements in $\varSigma_{\psi_X}$, $p$ has the minimum number of zero columns. First, suppose $p\in \bigcup_{t\geq1}\varSigma^t_{\mu_X}$. Then, there exists some $q\in \bigcup_{t\geq1}\varSigma^t_{\mu_Y}$  such that $p$ and $q$ have the same social topology. Thus, for $i,j=1,\dots,\eta$, each $C_i\times T$ block, each $C_i\times C_j$ block for $i\neq j$, and the $T\times T$ block of $q$ is an all zero block, and each $C_i\times C_i$ block of $q$ is a strictly positive stochastic matrix. The same is true for all powers of $q$ and therefore, for the element $\overline{q}$, an idempotent element in $\varSigma_Y$. Since $\overline{q}|_{C_i\times C_i}=\text{lim}\big[\overline{q}|_{C_i\times C_i}\big]^t$, for $i=1,\dots,\eta$, then $\overline{q}|_{C_i\times C_i}$ has rank one. It also follows that $q|_{T\times C_i}$ is strictly positive, for some $i$, then the same is true for $\overline{q}$. Thus, $p$ and $\overline{q}$ have the same basis and also the same social topology.
\par Now, suppose $p^2=p\in \varSigma_{\psi_X}-\bigcup_{t\geq1}\varSigma^t_{\mu_X}$ with basis $\mathscr{B}_{\eta}$. Without loss of generality, we can assume that there are elements $p_t\in \bigcup_{t\geq1}\varSigma^t_{\mu_X}$ such that $p_t\rightarrow p$ as $t\rightarrow\infty$ for each $i=1,\dots,\eta$, $p_t|_{C_i\times C_i}$ is strictly positive for each $t$, for all $j\in N=\{1,\dots,n\}$ and $k\in T$, $(p_t)_{jk}<\frac{1}{2n^2}$, and $p_t|_{T\times C_i}$ is strictly positive for each $t$ whenever $p|_{T\times C_i}$ is so. For each $t\geq1$, let $\mathcal{G}_t$ be the group of limit points of $\{p^m_t:m\geq1\}$. Then, $\overline{p}_t$ is the identity of $\mathcal{G}_t$, so choose $y_t\in \mathcal{G}_t$ such that $y_t(p_t\overline{p}_t)=(p_t\overline{p}_t)y_t\overline{p}_t$. But since $y_t\overline{p}_t=y_t=\overline{p}_ty_t$ and $p_t\overline{p}_t=\overline{p}_tp_t$, we have 
\begin{align}\label{eq:a1}
    y_tp_t=p_ty_t=\overline{p}_t.
\end{align}
    Let $\mathscr{B}'_s=\{T',C'_1,\dots,C'_s\}$ be the basis of $\overline{p}_t$. Note that $\overline{p}_t\in \varSigma_X$, but $p\in \varSigma_{\psi_X}$, and hence $s\geq \eta$.  We aim to show that $p$ and $\overline{p}_t$ have the same basis and the same social topology. To see this, let $j\in C_i'$ for $i=1,\dots,s$ and $k\in T$. Then, $$0=(p_t\overline{p}_t)_{jk}=(\overline{p}_tp_t)_{jk}=\sum_{m\in C'_i}(\overline{p}_t)_{jm}(p_t)_{mk}$$
    which implies that $p_t|_{C'_i\times T'}$ is an all zero block. It is known that for $y_t$ in eq. (\ref{eq:a1}), there is an associated permutation $\pi$ of $\{1,\dots,s\}$ such that the $C'_i\times C'_j$ of $y_t$ is a block with all entries strictly positive or zero accordingly as $\pi(i)=j$ or $\pi(i)\neq j$. We may also notice that $y_tp_t=\overline{p}_t$ and hence, for $u\in C'_k$, we have
    \begin{align*}
        1=\sum_{j\in C'_k}(\overline{p}_t)_{uj}=\sum_{\ell\in C'_{\pi(k)}}(y_t)_{u\ell}\sum_{j\in C'_k}(p_t)_{\ell j},
    \end{align*}
    which implies that $p_t|_{C'_{\pi(k)}\times C'_{k}}$ is a stochastic matrix, whenever $k\in\{1,\dots,s\}$, and therefore, $\sum_{j\in C'_k}(e_t)_{\ell j}=1$, for $\ell\in C'_{\pi(k)}$. We will now show that the permutation $\pi$ corresponding to $y_t$ must be the identity permutation. If this is not the case, then there exists some $j\in\{1,\dots,s\}$ such that $j\neq \pi(j)$, and consequently $\pi(\pi(j))\neq \pi(j)$. Let $\pi(j)=i$ and $\pi(k)=j$, so $i\neq j$ and $j\neq k$. It follows that $p_t|_{C'_i\times C'_i}$ as well as $p_t|_{C'_j\times C'_j}$ is an all zero block. Thus, $C'_i\cup C'_j\subset T$, because of our choice of $p_t$, and $C'_i\times C'_j\subset T\times T$. However, this contradicts the fact that $p_t|_{C'_i\times C'_j}$ is a stochastic matrix, since for $k_1,k_2\in T$, $(p_t)_{k_1k_2}<\frac{1}{2n^2}$, because of the way $p_t$ was chosen. Thus, $\pi$ must be the identity permutation, and therefore, $y_t$ must be the element $\overline{p}_t$ \citep[see,][]{muk86} and this means that, for $k\geq1$, $\overline{p}_tp^k_t=p^k_t\overline{p}_t=\overline{p}_t$. It follows that for any $y\in \mathcal{G}_t$, $\overline{p}_ty=y\overline{p}_t=\overline{p}_t$ so that $\mathcal{G}_t$ is a singleton and $\text{lim}_{k\rightarrow\infty}p^k_t=\overline{p}_t=\overline{p}_t^2$. Also, since $p_t\overline{p}_t=\overline{p}_tp_t=\overline{p}_t$, $p_t$ must be of the form
   \begin{table}[hbt!]
   \centering
\begin{tabular}{c|c|c|c}
       & $T'$ & $C'_1$     & $C'_2$     \\ \hline
$T'$   &      &            &            \\ \hline
$C'_1$ & 0    & stochastic & 0          \\ \hline
$C'_2$ & 0    & 0          & stochastic
\end{tabular}
\end{table}
\par Recall that $p_t|_{T\times T}$ is strictly sub-stochastic and thus, no $C'_i$ can be completely contained in $T$, and no $C_i$ can intersect two different $C'_j$ (because of the way $p_t$ was chosen). Thus, each $C'_j$ is completely contained in $C_i\cup T$, for some $i$ ($i=i(j),$ i.e., $i$ depending on $j$), and $i(j)\neq i(k)$ for $j\neq k$. It follows that the ranks $r=s$, $T'\subset T$, and each $C_i$ is a subset of some $C'_j$. Now, $p$, by choice, has the minimum number of zero columns among all idempotent elements of $\varSigma_{\psi_X}$, which means that $T'=T$, and consequently, $C_i$ is equal to some $C'_j$. That is, $p$ and $\overline{p}_t$ have the same basis, and also must have the same social topology since $p_t\overline{p}_t=\overline{p}_tp_t=\overline{p}_t$. 
\par Now, since $p_t\in \bigcup_{m\geq1}\varSigma^m_{\mu_X}$ and $\Proba(\mathscr{T})=1$, there must exist $q_t\in \bigcup_{m\geq1}\varSigma^m_{\mu_Y}$ such that $q_t$ and $p_t$ have the same social topology for each $t$. Thus, $q_t$, for each $t$, must be of the form
   \begin{table}[hbt!]
   \centering
\begin{tabular}{m{0.5cm}|m{0.5cm}|m{4cm}|m{4cm}}
       & $T$ & $C_1$     & $C_2$   \\ \hline
$T$   &      &            &            \\ \hline
$C_1$ & 0    & {stochastic  (and all entries strictly positive)} & 0          \\ \hline
$C_2$ & 0    & 0          & stochastic (and all entries strictly positive)
\end{tabular}
\end{table}

\par We see that each row in the $q_t|_{T\times T}$ block is less than 1 since the same is true for $p_t$. It follows (using standard Markov chain theory) that since $q_t|_{C_i\times C_i}$ is a strictly positive stochastic matrix, then $\overline{q}_t$ must also have the same basis $\mathscr{B}_{\eta}=\{T,C_1,\dots,C_{\eta}\}$ and also have the same social topology as $p$. Thus, the minimal rank of the matrices in $\varSigma_Y$ is at most $\eta$. Reversing the roles of $X$ and $Y$ yields the same conclusion.
\end{proof}

\begin{proof}[Proof of Lemma \ref{thm:factor}]
    By Lemma \ref{thm:rank}, there exist idempotent elements $p\in \varSigma_{\psi_X}$ and $q\in \varSigma_{\psi_Y}$ which have the same rank $\eta$ and the same basis $\mathscr{B}_{\eta}=\{T,C_1,\dots,C_{\eta}\}$. For any $x\in \varSigma_{\psi_X}$, there exists $x_t\in \bigcup_{m\geq1}\varSigma^m_{\mu_X}$ such that $x_t\rightarrow x$. Then, $px_tp\rightarrow pxp$ as $t\rightarrow\infty$. Now, choose $t$ sufficiently large so that whenever $(pxp)_{ij}>0$, then $(px_tp)_{ij}>0$. That is, for large $t$, $pxp$ and $px_tp$ (both elements in  $G_X$) must correspond to the same permutation $\pi$ on $\{1,\dots,\eta\}$ in the following sense: the block $pxp|_{C_i\times C_j}$ is strictly positive if and only if $j=\pi(i)$.
    \par Since $G_X$ is finite, the correspondence $\sigma\mapsto \pi$ from $G_X$ to the group of permutations on $\{1,\dots,\eta\}$ is an isomorphism. This means that there are elements $y_t\in \bigcup_{m\geq1}\varSigma^m_{\mu_Y}$ such that, for each $t$, $px_tp$ and $qy_tq$ have the same social topology so that the element $pxp\in \varSigma_{\psi_X}$ and $qyq\in \varSigma_{\psi_Y}$ (when $y$ is a limit point of the sequence $y_t$) must correspond to the same permutation $\pi$ on $\{1,\dots,\eta\}$. Reversing the role of $G_X$ and $G_Y$, it follows that the groups $G_X$ and $G_Y$ are isomorphic.\end{proof}

\begin{proof}[Proof of Lemma \ref{thm:maxfactor}]
Recall the continuous homomorphism $\varPhi_X$ and that $G_X=p\varSigma_{\psi_X}p$ and $G_Y=q\varSigma_{\psi_Y}q$ are isomorphic by Lemma \ref{thm:factor}. Also, recall the Rees-–Suskewitsch decompositions $\varSigma_{\psi_X}\cong W_X\times G_X\times Z_X$ and $\varSigma_{\psi_X}\cong W_Y\times G_Y\times Z_Y$ in eq. (\ref{eq:reesx}).  Notice that $Z_XW_X\subset G_X$ and $Z_YW_Y\subset G_Y$. It therefore suffices to show that $Z_XW_X$ and $Z_YW_Y$ correspond to the same set of permutations under the standard isomorphism from the groups $G_X$ and $G_Y$ to the group of permutations on $\{1,\dots,\eta\}$, then it would follow that the factor groups $G_X/Q_X$ and $G_Y/Q_Y$ are also isomorphic, where $Q_X$ and $Q_Y$ are the smallest normal subgroups containing $Z_XW_X$ and $Z_YW_Y$, respectively. To show this, let $x_1\in Z_X$ and $x_2\in W_X$. Then, exist $x_{2t},x_{1t}\in \bigcup_{m\geq1}\varSigma^m_{\mu_X}$ with $x_{2t}\rightarrow x_2$ and $x_{1t}\rightarrow x_1$ such that whenever $(x_2)_{ij}>0$, $(x_{2t})_{ij}>0$, and whenever $(x_1)_{ij}>0$, $(x_{1t})_{ij}>0$.
    \par Now, choose $y_{2t},y_{1t}\in \bigcup_{m\geq1}\varSigma^m_{\mu_Y}$ such that, for each $t$, $x_{2t}$ and $y_{2t}$ have the same social topology, and $x_{1t}$ and $y_{1t}$ have the same social topology. Let $\sigma_t$ be the identity in the group of limit points of  $\big\{(q y_{1t} )^m:m\geq1\big\}$, and $h_t$ be the identity in the group of limit points of $\big\{(q y_{2t} )^m:m\geq1\big\}$. Then, $q\sigma_t=\sigma_t$ and $h_tq=h_t$. Choose $M$ and $L$ sufficiently large so that $[(q y_{1t} )^M]_{1j}>0$ whenever $(\sigma_t)_{ij}>0$, and $[(q y_{2t} )^L]_{1j}>0$ whenever $(h_t)_{ij}>0$. 
    \par Note that the elements $p(px_{2t})^M(x_{1t}p)^Lp$ and $q(qy_{2t})^M(x_{2t}q)^Lq$ have the same skeleton and, therefore, represent the group elements in $G_X$ and $G_Y$ that correspond to the same permutation on $\{1,\dots,\eta\}$. Moreover, $\big[p(px_{2t})^M(x_{1t}p)^Lp\big]_{ij}>0$  whenever $\big[p(px_{2})^M(x_{1}p)^Lp\big]_{ij}=(x_{2}x_1)_{ij}>0$. Similarly,  $\big[q(qy_{2t})^M(y_{1t}q)^Lq\big]_{ij}>0$  whenever $\big[q\sigma_th_t q]_{ij}=(\sigma_th_t)_{ij}>0$, and thus, both represent the same group element $\sigma_th_t$. Also, we may notice that $\sigma^2_t=\sigma_t=q\sigma_t\in q\varSigma_{\psi_Y}$ so that $\sigma_t\in Z_Y$ and $h^2_t=h_t=h_tq\in \varSigma_{\psi_Y}q$ so that $h_t\in Z_Y$. Thus, with every element $x_2x_1\in Z_XW_X$, we have associated an element $\sigma_th_t\in Z_YW_Y$ that corresponds to the same permutation on $\{1,\dots,\eta\}$. Reversing the roles of $X$ and $Y$, it follows that the sets $Z_XW_X$ and $Z_YW_Y$ both correspond to the same set of permutation under the standard isomorphism from the groups $G_X$ and $G_Y$ to the group of permutations on $\{1,\dots,\eta\}$. 
\end{proof}
    \begin{proof}[Proof of Lemma \ref{thm:support}]
        For any Borel set $B\subset G_X/Q_X$, we can define the probability measure $\Tilde{\mu}_X$ by $\Tilde{\mu}_X(B)=\mu_X\big(\varPhi^{-1}_X(B)\big)$, where $\varPhi_X:\varSigma_X\rightarrow G_X/Q_X$ is a continuous homomorphism defined by $\varPhi_X(x)=pxp\hspace{0.02in} Q_X$. We define $\varPhi_Y$ and $\Tilde{\mu}_Y$ similarly. Then, the supports of $\Tilde{\mu}_X$ and $\Tilde{\mu}_Y$ become $\varSigma_{\Tilde{\mu}_X}=\big\{pxp\hspace{0.02in}Q_X:x\in \varSigma_{\mu_X}\big\}$ and $\varSigma_{\Tilde{\mu}_Y}=\big\{qyq\hspace{0.02in}Q_Y:y\in \varSigma_{\mu_Y}\big\}$, respectively. 
        \par We may notice that for $x\in \varSigma_{\mu_X}$, there exists $y\in\varSigma_{\mu_Y}$ such that $x$ and $y$  (and therefore $pxp$ and $qyq$) have the same social topology. It follows from Lemma \ref{thm:maxfactor} that the cosets $pxp\hspace{0.02in}Q_X$ and $qyq\hspace{0.02in}Q_Y$ correspond to the same set of permutations under the isomorphism from the groups $G_X$ and $G_Y$ to the group of permutations on $\{1,\dots,\eta\}$. Similarly, for each $y\in \varSigma_{\mu_Y}$, there exists $x\in \varSigma_{\mu_X}$ such that a similar conclusion is again true for the cosets $qyq\hspace{0.02in}Q_Y$ and $pxp\hspace{0.02in}Q_X$. It therefore follows that, under the isomorphism from $G_X/Q_X$ to $G_Y/Q_Y$ (Lemma \ref{thm:maxfactor}), $\varSigma_{\Tilde{\mu}_Y}$ is the isomorphic image of $\varSigma_{\Tilde{\mu}_X}$.
    \end{proof}
\begin{proof}[Proof of Lemma \ref{thm:iff}]
    By Lemma \ref{thm:homom}, we know that $\mu^t_X$ (respectively, $\mu^t_Y$) converges weakly if and only if $\Tilde{\mu}^t_X$ (respectively, $\Tilde{\mu}_Y$) converges weakly. By Lemma \ref{thm:support}, the supports $\varSigma_{\Tilde{\mu}_X}$ and $\varSigma_{\Tilde{\mu}_Y}$ are isomorphic. Thus, $\Tilde{\mu}^t_X$ converges weakly if and only if $\Tilde{\mu}^t_Y$ converges weakly.
\end{proof}

\end{document}